\documentclass[11pt]{article}
\usepackage{citesort,latexsym,graphics,color}

\newcommand{\tr}{{\mathrm{tr}}}

\newcounter{Diagrams}
\Alph{Diagrams}
\newcounter{TLDiagrams}
\Alph{TLDiagrams}
\newcounter{cancellation}
\addtocounter{cancellation}{1}

\newtheorem{Diag}{}[Diagrams]
\newtheorem{DC}[Diag]{\{}
\newtheorem{DC-B}[Diag]{$|$}
\newtheorem{cancel}{Cancellation}

\newtheorem{TLDiag}{}[TLDiagrams]
\newtheorem{TLDC}[TLDiag]{\{}

\newtheorem{D-ID}{Diagrammatic Identity}[section]

\newcommand{\ie}{{\it i.e.}\ }
\newcommand{\cf}{{\it cf.}\ }
\newcommand{\eg}{{\it e.g.}\ }

\newcommand{\aka}{{\it a.k.a.}\ }
\newcommand{\etc}{{\it etc.}\ }

\newcommand{\wrt}{with respect to}
\newcommand{\CC}{charge conjugation}

\newcommand{\lhs}{left-hand side}
\newcommand{\rhs}{right-hand side}

\newcommand{\role}{role}

\newcommand{\be}{\begin{equation}}
\newcommand{\ee}{\end{equation}}
\newcommand{\bea}{\begin{eqnarray}}
\newcommand{\eea}{\end{eqnarray}}
\newcommand{\beas}{\begin{eqnarray*}}
\newcommand{\eeas}{\end{eqnarray*}}

\newcommand{\bear}{\begin{array}{l}}
\newcommand{\eear}{\end{array}}

\newcommand{\bcf}{\begin{center}\begin{figure}}
\newcommand{\ecf}{\end{figure}\end{center}}

\newcommand{\bct}{\begin{center}\begin{table}}
\newcommand{\ect}{\end{table}\end{center}}

\newcommand{\ds}{\displaystyle}

\def\eq#1{(\ref{#1})}
\def\eqs#1#2{(\ref{#1},\ref{#2})}
\def\sec#1{sec.~\ref{#1}}

\def\fig#1{fig.~\ref{#1}}
\def\Fig#1{Fig.~\ref{#1}}
\def\figs#1#2{figs.~\ref{#1} and~\ref{#2}}

\newcommand{\mod}[1]{\left| #1 \right|}

\def\phi{ \varphi }
\def\A{{\cal A}}
\def\C{{\cal C}}

\newcommand{\ep}{\epsilon}

\def\hS{\hat{S}}

\def\one{\hbox{1\kern-.8mm l}}
\def\str{\mathrm{str}}

\newcommand{\Op}[1]{\mathcal{O}(p^{#1})}
\newcommand{\Oep}{\ensuremath{\mathcal{O}(\epsilon)}}
\newcommand{\Oepz}{\ensuremath{\mathcal{O}(\epsilon^0)}}
\newcommand{\OepPow}[1]{\ensuremath{\mathcal{O}\left(\epsilon^{#1}\right)}}
\newcommand{\Omom}[1]{\ensuremath{\mathcal{O}\left({\mbox{mom}^{#1}}\right)}}
\newcommand{\Oone}{\ensuremath{\mathcal{O}(1)}}
\newcommand{\Oal}{\mathcal{O}(\alpha)}
\newcommand{\OalPow}[1]{\mathcal{O}\left(\alpha^{#1}\right)}
\newcommand{\Pep}{p^{-2\epsilon}}
\newcommand{\eptoz}{\ensuremath{\epsilon \rightarrow 0}}

\newlength{\epminusone}
\newlength{\epzero}
\newcommand{\EM}{\gamma_\mathrm{EM}}
\newcommand{\altoz}{\alpha \to 0}

\newcommand{\AngVol}[1]{\not{\!\Omega_#1}}
\newcommand{\PowAngVol}[2]{\not{\! \! \Omega_#1^#2}}

\newcommand{\pder}[2]{\ensuremath{\frac{\partial #1}{\partial #2}}}

\newcommand{\pmomder}[3]{\left.\partial^{#1}_{#2}\right|_{#3}}
\newcommand{\order}[1]{\mathcal{O} (#1)}
\newcommand{\hf}{\frac{1}{2}}
\newcommand{\qt}{\frac{1}{4}}

\newlength{\PFheight}
\newcommand{\DummyKernel}{\ensuremath{\stackrel{\bullet}{\mbox{\rule{1cm}{.2mm}}}}}

\newcommand{\Square}{\ensuremath{\ensuremath{\begin{array}{c}\begin{picture}(0,0)%
\includegraphics{pstex/OLDs.pstex}%
\end{picture}%
\setlength{\unitlength}{3947sp}%
\begingroup\makeatletter\ifx\SetFigFont\undefined%
\gdef\SetFigFont#1#2#3#4#5{%
  \reset@font\fontsize{#1}{#2pt}%
  \fontfamily{#3}\fontseries{#4}\fontshape{#5}%
  \selectfont}%
\fi\endgroup%
\begin{picture}(531,249)(666,-259)
\put(889,-188){\makebox(0,0)[lb]{\smash{\SetFigFont{11}{13.2}{\rmdefault}{\mddefault}{\updefault}{\color[rgb]{0,0,0}1}%
}}}
\end{picture}
 \end{array}}}}

\newcommand{\flow}{\Lambda \partial_\Lambda}
\newcommand{\flowConstAl}{\Lambda \partial_\Lambda|_\alpha}
\newcommand{\LConstAl}{\Lambda|_\alpha}
\newcommand{\dec}[3][0]{\ensuremath{\left[ #2 \hspace{#1in} \right]^{#3}}}
\newcommand{\OLDs}{\mathcal{D}_{1\mu\nu}}
\newcommand{\TLDs}{\mathcal{D}_{2\mu\nu}}
\newcommand{\ROLDs}{\mathcal{D}_{\, 1\mu\nu}^R}

\newcommand{\NewCO}{\mathsf{B}}
\newcommand{\InvCO}{\mathsf{A}}
\newcommand{\NCOAlg}[1]{\frac{\alpha +1 + c_#1 (\alpha -1)}{\alpha c_#1}}
\newcommand{\NCTAlg}[1]{\frac{\alpha +1 + c_#1 (1-\alpha)}{\alpha c_#1}}
\newcommand{\Aone}[1]{\frac{\alpha c_#1}{\alpha+1 + c_#1(\alpha-1)}}
\newcommand{\Atwo}[1]{\frac{\alpha c_#1}{\alpha+1 + c_#1(1- \alpha)}}
\newcommand{\f}[1]{\frac{(1+\alpha) \tilde{c}_#1}{(1+\alpha)#1 \tilde{c}_#1 + 4 \alpha c_#1}}

\newcommand{\fB}[2]{\frac{(1+\alpha) \tilde{c}_#1}{(1+\alpha)#2 \tilde{c}_#1 + 4 \alpha c_#1}}
\newcommand{\gB}[2]{\frac{2 \alpha \tilde{c}_#1}{(1+\alpha)#2 \tilde{c}_#1 + 4 \alpha c_#1}}
\newcommand{\EPAone}[2]{\frac{1}{#1^2}\frac{\alpha c_#2}{(\alpha+1) + c_#2(\alpha-1)}}

\newcommand{\NUn}{\ensuremath{\mathsf{NC}}}
\newcommand{\COMP}{\mathsf{C}}
\newcommand{\Uni}[1]{\left. #1 \right|_\COMP}
\newcommand{\NUni}[1]{\left. #1 \right|_\NUn}

\newcommand{\GRk}{\rhd}
\newcommand{\GRkpr}{>}
\newcommand{\DiagDot}{\scriptstyle \bullet}

\newlength{\LabLength}
\newlength{\ProcessRefLength}
\newlength{\ProcessLength}
\newlength{\CancelRefLength}
\newlength{\CancelRefLengthB}
\newlength{\CancelLength}
\newlength{\scriptboldcurlybracket}
\newlength{\strutheight}

\newcommand{\blankstrut}{\rule{0em}{\strutheight}}

\newcommand{\cdeps}[1]{\ensuremath{\begin{array}{c}\includegraphics{./eps/#1.eps} \end{array}}} 

\newcommand{\LD}[1]{
	\settowidth{\LabLength}{\scriptsize \textbf{\ref{#1}}}
	\addtolength{\LabLength}{0.8em}
	\begin{minipage}{\LabLength}
		\scriptsize
		\begin{Diag}\label{#1}\end{Diag}
	\end{minipage}
}

\newcommand{\TLD}[1]{
	\settowidth{\LabLength}{\scriptsize \textbf{\ref{#1}}}
	\addtolength{\LabLength}{0.8em}
	\begin{minipage}{\LabLength}
		\scriptsize
		\begin{TLDiag}\label{#1}\end{TLDiag}
	\end{minipage}
}

\newcommand{\LO}[3][1]{
	\begin{array}{c}
		\LD{#3}
	\\[#1ex]
		#2
	\end{array}
}

\newcommand{\LDi}[3][1]{\LO[#1]{\ensuremath{\begin{array}{c}\input{pstex/#2.pstex_t} \end{array}}}{#3}}

\newcommand{\LHlabs}[3][0]{
\begin{array}{l}
\vspace{#1ex}
	#2\\
	#3		
\end{array}
}

\newcommand{\LDLD}[3][-0.5]{
	\LHlabs[#1]{\LD{#2}}{\LD{#3}}
}

\newcommand{\LDBl}[2][-0.5]{
	\LHlabs[#1]{\LD{#2}}{\blankstrut}
}

\newcommand{\jhep}[3]{\emph{JHEP} #1 (#2) #3}
\newcommand{\NuclPhys}[4]{\emph{Nucl.\ Phys.\ }\textbf{#1 #2} (#3) #4}
\newcommand{\PhysRev}[4]{\emph{Phys.\ Rev.\ }\textbf{#1 #2} (#3) #4}
\newcommand{\IntJModPhys}[4]{\emph{Int.\ J.\ Mod.\ Phys.\ }\textbf{#1 #2} (#3) #4}

\newcommand{\PhysLett}[4]{\emph{Phys.\ Lett.\ }\textbf{#1 #2} (#3) #4}

\newcommand{\Acta}[3]{\emph{Acta Phys.\ Slov.\ }\textbf{#1} (#2) #3}

\newcommand{\arxiv}[1]{[arXiv:#1]}
\newcommand{\hepth}[1]{hep-th/#1}

\newcommand{\http}[1]{http://#1}

\title{A Manifestly Gauge Invariant, Continuum Calculation of the $SU(N)$ Yang-Mills
two-loop $\beta$ function}

\author{
	Tim R.~Morris and Oliver J.~Rosten
\\ %[0.5cm]
	School of Physics and Astronomy,  University of Southampton,
\\ 
	Highfield, Southampton SO17 1BJ, U.K.
\\ %[0.5cm]
	E-mails: {\tt T.R.Morris@soton.ac.uk, ojr@phys.soton.ac.uk }
}

\date{}

\begin{document}

\maketitle

\begin{abstract}
	The manifestly gauge invariant exact renormalisation
	group provides a framework for performing continuum
	computations in $SU(N)$ Yang-Mills theory, without fixing the 
	gauge. We use this formalism to compute the two-loop 
	$\beta$ function in a manifestly gauge invariant way, 
	and without	specifying the details of the regularisation 
	scheme.
\end{abstract}

\vspace{-60ex}
\hfill SHEP 05-22

\newpage
\tableofcontents

\section{Introduction}
\label{sec:Intro}

In ref.~\cite{mgierg1}, a manifestly gauge invariant exact
renormalisation group (ERG) was introduced for $SU(N)$
Yang-Mills theory that can be
straightforwardly renormalised to any loop order. As such,
this formalism is suitable for general
continuum calculations, which can be performed without
fixing the gauge. 
The obvious obstruction to not fixing the gauge---namely
that the inverse of the gauge field two-point vertex
does not exist---is avoided by the particular choice
of quantities we aim to compute (just as on the lattice).
In this, and previous
works~\cite{u1,ym,ymi,ymii,Arnone:2000bv,Arnone:2001iy,sunn,Morris:2000jj,Arnone:2001iy,one,Arnone:2002ai,Arnone:2002qi,aprop,two,quarks,Thesis},
we compute vertices of the Wilsonian effective action;
in the future we aim to generalise these methods
to compute the correlators of gauge
invariant operators~\cite{InProg}.\footnote{To consider on-shell gluons, one can gauge
fix after the computation~\cite{aprop}.}

The construction of a real, gauge invariant cutoff, $\Lambda$, 
is achieved by embedding the
physical $SU(N)$ gauge theory in a spontaneously broken
$SU(N|N)$ gauge theory~\cite{Arnone:2001iy}. 
To compute the effective
action without fixing the gauge, we use the fact that there an
infinity of possible ERGs that
specify its flow as modes are integrated out~\cite{jose} (the
continuum equivalent to the infinite number of ways of blocking on
a lattice~\cite{Morris:2000jj,mgierg1}) and that out of these there 
are infinite number that
\emph{manifestly} preserve the gauge invariance. 
We then further specialise to those ERGs (still infinite in number) 
which conveniently
allow renormalisation to any loop order~\cite{Thesis,mgierg1}.

The key to doing this is first identifying 
which of the infinity of dimensionless couplings
in the regularised theory must be renormalised.
For technical reasons,  the superscalar field 
which spontaneously breaks the $SU(N|N)$ symmetry is given
zero mass dimension~\cite{aprop}, and thus is associated by the
usual dimensional reasoning with an infinite number of
dimensionless couplings. That these couplings do not
require renormalisation has been assumed in~\cite{mgierg1}
but will be proven in \sec{sec:TLD:Couplings}
(see also~\cite{Thesis}).
Thus, the only couplings requiring renormalisation
are $g(\Lambda)$, which is associated with the
physical $SU(N)$ theory, and $g_2(\Lambda)$, which 
is associated
with an unphysical copy present due to the $SU(N|N)$
regulating structure. For convenience, it is useful to define
a quantity 
\be
	\alpha := g^2_2/g^2.
\label{eq:alpha-defn}
\ee
Since the manifest preservation of gauge invariance 
ensures that the gauge fields have no
wavefunction renormalisation~\cite{ym},
$g$ and $g_2$ are the only quantities which
run. 

To facilitate calculations in our manifestly
gauge invariant ERG, a powerful set of
diagrammatic techniques have been developed~\cite{Thesis,mgierg1,oliver1}.
In this paper we complete the description of the
associated computational scheme
and demonstrate its consistency and potential
by computing the $SU(N)$ Yang-Mills two-loop
$\beta$ function, $\beta_2$, without fixing the gauge.

At the heart of the diagrammatic
calculus is an elegant representation of the flow equation. 
Its simplicity arises, counterintuitively, from 
the immense freedom in the precise construction 
of the formalism. We are able to turn this freedom
to our advantage by recognising that many of the
details of the setup are non-universal and, moreover, need
never be explicitly defined. Knowing that such details
must necessarily cancel out in the computation of
a universal quantity, we can efficiently absorb
them into diagrammatic rules. Thus, the 
diagrammatic flow equation hides a terrific amount;
whilst these hidden features must be properly understood
when constructing the formalism~\cite{Thesis,mgierg1},
they can  be essentially forgotten about when it
comes to performing actual calculations~\cite{oliver1,oliver2}.

Within our ERG,
the flow is controlled by a (generically) non-universal object,
$\hS$, the `seed
action'~\cite{Thesis,mgierg1,one,two,aprop,giqed}. 
This respects the same symmetries as the Wilsonian effective 
action, $S$, and has the same structure. However, whereas our
aim is to solve the flow for $S$, $\hS$ acts as an input.
By choosing the two-point, tree level seed action vertices equal to 
their Wilsonian effective action counterparts,
we can arrange for the ERG kernels,
integrated \wrt\ $d \ln \Lambda$, to be the inverses
of the classical two-point vertices, in the transverse
space~\cite{aprop,two,mgierg1,Thesis}; equivalently we can say
that they are inverses up to remainder terms. These 
are called `gauge remainders' and exist because
the manifest gauge invariance demands that they are there.
In recognition of both their \role\ and form,
we refer to the integrated ERG kernels as effective
propagators, mindful that they are not propagators in
the usual sense.

The diagrammatic procedure for computing
$\beta_2$ is
as follows.
We start by using the flow equations to
compute the flow of the two-point vertex corresponding
to the physical $SU(N)$ gauge field, which we suppose
carries momentum $p$. To obtain a
solvable equation for $\beta_2$,
we specialise
to the appropriate loop order and work at $\Op{2}$;
this latter step constrains the equations by allowing
the renormalisation condition for the physical coupling
$g(\Lambda)$ to feed in.

We now recognise that certain diagrams generated
by the flow comprise exclusively Wilsonian
effective action vertices joined together
with an effective propagator struck by $-\flowConstAl$.
These terms are processed by moving the $-\flowConstAl$
from the effective propagator to the diagram as a whole,
minus correction terms in which $-\flowConstAl$ strikes
the vertices. The former diagrams are called $\Lambda$-derivative
terms; the latter can be processed using the flow equation.
Amongst the terms thus generated are those which
can be simplified by applying the effective propagator
relation.
Such
terms cancel non-universal contributions up to gauge
remainders which can, themselves, be processed
diagrammatically. Iterating the diagrammatic procedure,
the expression for $\beta_2$ ultimately
reduces to the following sets of diagrams:
\begin{enumerate}
	\item $\Lambda$-derivative terms;

	\item `$\alpha$-terms', consisting of diagrams
	struck by $\partial / \partial \alpha$;

	\item `$\Op{2}$-terms', which comprise an $\Op{2}$ stub
			\ie a diagrammatic component which is manifestly
			$\Op{2}$.
\end{enumerate}

The $\Op{2}$-terms can be manipulated. In the calculation
of $\beta_1$, at any rate, the structure attaching to the stub
can be directly Taylor expanded to zeroth order in $p$---which 
can once again be done diagrammatically. The above diagrammatic
procedure is then iterated.
At two loops, as mentioned in~\cite{mgierg1},
this procedure is not so straightforward, since
na\"ive Taylor expansion can generate infra-red (IR)
divergences.

The strategy for dealing with such diagrams is to
recognise that, by considering sets of terms 
together, these IR divergences cancel out. 
Organising the calculation in this way is facilitated
by diagrammatic `subtraction techniques', which we describe in
\sec{sec:TLD:Subtractions}. Now the $\Op{2}$ can
be processed, and so $\beta_2$ can be reduced to
just $\Lambda$-derivative and $\alpha$-terms.
As anticipated in~\cite{mgierg1}, agreement
of $\beta_2$ with the standard, universal answer
is expected only in the limit that $\alpha \to 0$. In
section~\ref{sec:TLD:alpha} we demonstrate that,
subject
to some very general constraints, the $\alpha$-terms
vanish, in this limit.

It is from the $\Lambda$-derivative terms that 
the universal coefficient can be extracted, which  
is most easily done by working in $D= 4-2\epsilon$. At the one-loop level, life is
easy: each individual diagram is finite; the leading order
contribution to $\beta_1$ is finite and universal, with
all non-universal contributions
vanishing as $D\to 4$. At two-loops, as one would expect,
individual diagrams can develop divergences as
$D \to 4$. Though these must of course cancel between
terms, the
surviving finite contributions do not obviously
combine to give something universal. The trick is
to once again employ the subtraction techniques. This
allows us to 
isolate non-universal contributions,
which then cancel amongst themselves. The
remaining
contributions can be
evaluated directly, combining to yield
the expected answer.

This paper is organised as follows. In \sec{sec:Review}
we review the setup and the various diagrammatic techniques
that we will require. In \sec{ch:LambdaDerivatives:Methodology}
we describe the technology for evaluating the $\Lambda$-derivative
terms. Following a statement of the basic idea,
the principles are illustrated in the context of a computation
of $\beta_1$. The
general considerations for two-loop integrals 
are discussed and then the subtraction techniques are
explained.
Finally, we give the proof that all dimensionless
couplings besides $g$ and $g_2$ can be prevented from
running.
In \sec{sec:Numerics} we give the $\Lambda$-derivative
and $\alpha$-terms that contribute to $\beta_2$
in the $D \rightarrow 4$ limit and
extract the universal, numerical coefficient.
We conclude in
\sec{sec:Conclusion}.

\section{Review}
\label{sec:Review}

\subsection{Elements of $SU(N|N)$ Gauge Theory}	
\label{sec:Elements}

We regularise $SU(N)$ Yang-Mills theory by instead working
with $SU(N|N)$ Yang-Mills theory. The gauge field is valued in
the Lie superalgebra and thus takes the form of a Hermitian
supertraceless supermatrix:
\[
	\A_\mu = 
	\left(
		\begin{array}{cc}
			A_\mu^1 	& B_\mu 
		\\
			\bar{B}_\mu & A_\mu^2
		\end{array} 
	\right) + \A_\mu^0 \one.
\]
Here, $A^1_\mu(x)\equiv A^1_{a\mu}\tau^a_1$ is the
physical $SU(N)$ gauge field, $\tau^a_1$ being the $SU(N)$
generators orthonormalised to
$\tr(\tau^a_1\tau^b_1)=\delta^{ab}/2$, while $A^2_\mu(x)\equiv
A^2_{a\mu}\tau^a_2$ is a second unphysical $SU(N)$ gauge field.
When labelling \eg vertex coefficient functions,
we often abbreviate $A^{1,2}$ to just 1,2.
The $B$ fields are fermionic gauge fields which will gain a mass
of order $\Lambda$ from the spontaneous breaking; they play the
\role\ of gauge invariant Pauli-Villars (PV) fields, furnishing the
necessary extra regularisation to supplement the covariant higher
derivatives. 

To unambiguously define contributions which are finite
only by virtue of the PV regularisation, a preregulator must
be used in $D=4$~\cite{Arnone:2001iy}. This amounts to a prescription for 
discarding otherwise non-vanishing surface terms which
can be generated by shifting loop momenta; we use dimensional
regularisation. 
%Dimensional
%regularisation suffices, though we will see that
%there appears to be an entirely diagrammatic prescription.
%This is desirable, since 
%one might hope that it would be applicable to phenomena for which
%one must work strictly in $D=4$.

The theory is
subject to the local invariance:
\be
\label{Agauged}
\delta\A_\mu = [\nabla_\mu,\Omega(x)] +\lambda_\mu(x) \one.
\ee
The first term, in which $\nabla_\mu = \partial_\mu -i\A_\mu$, 
generates supergauge transformations. Note that the coupling, $g$,
has been scaled out of this definition. It is worth doing
this: since we do not gauge fix, the exact preservation of~\eq{Agauged}
means that none of the fields suffer wavefunction
renormalisation, even in the broken phase~\cite{aprop}.

The second term in~\eq{Agauged} divides out the centre of the algebra.
This `no $\A^0$ shift symmetry' ensures that nothing depends on $\A^0$
and that $\A^0$ has no degrees of freedom. We adopt a
prescription whereby we can effectively ignore the field $\A^0$,  altogether, 
using it  to map us into a particular diagrammatic
picture~\cite{Thesis,mgierg1}.

The spontaneous breaking is carried by a superscalar
field
\[
\C =
	\left(
		\begin{array}{cc}
			C^1		& D
		\\
			\bar{D}	& C^2
		\end{array}
	\right),
\]
which transforms covariantly:
\be
\label{Cgauged}
\delta\C = -i\,[\C,\Omega].
\ee

It can be shown that, at the classical level, the spontaneous
breaking scale (effectively the mass of $B$) tracks the covariant
higher derivative effective cutoff scale $\Lambda$, if $\C$ is
made dimensionless (by using powers of $\Lambda$) and  $\hS$ has
the minimum of its effective potential at:
\be
\label{sigma}
<\C>\ = \sigma \equiv \pmatrix{\one & 0\cr 0 & -\one}.
\ee
In this case the classical action $S_0$ also has a minimum 
at~\eq{sigma}. At the quantum level this can be imposed as a
constraint on $S$, which can be satisfied by a suitable choice of
$\hS$~\cite{aprop,Thesis}. When we shift to the broken phase, $D$ becomes
a super-Goldstone mode (eaten by $B$ in unitary gauge) whilst the
$C^i$ are Higgs bosons and can be given a running mass of order
$\Lambda$~\cite{ym,Arnone:2001iy,aprop}. Working in our manifestly
gauge invariant formalism, $B$ and $D$ gauge transform into each
other; in recognition of this, we define the composite fields
$F_M = (B_\mu, D)$,
$\bar{F}_N = (\bar{B}_\nu, -\bar{D})$, where $M$, $N$ are
five-indices~\cite{Thesis,mgierg1}.\footnote{The summation
convention for these indices is that we take each product of
components to contribute with unit weight.} 

The renormalisation conditions for the
couplings $g$ and $g_2$ are:
\bea
\label{defg}
	S[\A=A^1, \C=\sigma]	& =	& {1\over2g^2}\,\str\!\int\!\!d^D\!x\,
									\left(F^1_{\mu\nu}\right)^2+\cdots,
\\	
\label{defg2}
	S[\A=A^2, \C=\sigma] 	& =	& {1\over2g^2_2}\,\str\!\int\!\!d^D\!x\,
									\left(F^2_{\mu\nu}\right)^2+\cdots,
\eea
where the ellipses stand for higher dimension operators and the
ignored vacuum energy. 

\subsection{The Flow Equation}
\label{sec:Rev:FE}

The flow equation is most naturally phrased in its
diagrammatic form, as shown in \fig{fig:FE}. As
in our previous works, we have not drawn the improperly
regulated diagram in which the kernel bites its
own tail, having removed such terms by placing suitable
constraints on the covariantisation~\cite{ymii,aprop,mgierg1}.
\begin{center}
\begin{figure}[h]
	\be
	\label{eq:FE}
	-\flow 
	\dec{
		\ensuremath{\begin{array}{c}\input{pstex/Vertex-S.pstex_t} \end{array}}
	}{\{f\}}
	=
	\frac{1}{2}
	\dec{
		\ensuremath{\begin{array}{c}\input{pstex/Dumbbell-S-Sigma_g.pstex_t} \end{array}} - \ensuremath{\begin{array}{c}\input{pstex/Vertex-Sigma_g.pstex_t} \end{array}}
	}{\{f\}}
	\ee
\caption{A diagrammatic representation of the flow equation.}
\label{fig:FE}
\end{figure}
\end{center}

The \lhs\ depicts the flow of all independent Wilsonian effective action
vertex \emph{coefficient functions}, 
which correspond to the set of
fields, $\{f\}$. Each coefficient function has associated
with it an implied supertrace structure (and symmetry factor which,
as one would want, does not appear in the diagrammatics).
For example,
\[
	\dec{
		\ensuremath{\begin{array}{c}\begin{picture}(0,0)%
\includegraphics{pstex/Vertex-S.pstex}%
\end{picture}%
\setlength{\unitlength}{3947sp}%
\begingroup\makeatletter\ifx\SetFigFont\undefined%
\gdef\SetFigFont#1#2#3#4#5{%
  \reset@font\fontsize{#1}{#2pt}%
  \fontfamily{#3}\fontseries{#4}\fontshape{#5}%
  \selectfont}%
\fi\endgroup%
\begin{picture}(320,318)(2180,-963)
\put(2291,-859){\makebox(0,0)[lb]{\smash{\SetFigFont{11}{13.2}{\rmdefault}{\mddefault}{\updefault}{\color[rgb]{0,0,0}$S$}%
}}}
\end{picture}
 \end{array}}
	}{C^1C^1}
\]
represents both the coefficient functions $S^{C^1 C^1}$ and
$S^{C^1,C^1}$ which, respectively, are associated with the
supertrace structures $\str C^1 C^1$ and $\str C^1 \str C^1$. 

The objects on the \rhs\ of \fig{fig:FE}
have two different types of component. The lobes
represent vertices of action functionals,
where $\Sigma_g \equiv g^2S - 2 \hat{S}$. 
The object attaching
to the various lobes, \DummyKernel,  is
the sum over vertices of the covariantised ERG kernels~\cite{ymi,aprop}
and, like the action vertices, can be decorated by fields belonging to $\{f\}$.
The fields of the action vertex (vertices) to which the vertices of the kernels attach
act as labels for the ERG kernels
though, in certain circumstances, the particular decorations of
the kernel are required for unambiguous identification~\cite{Thesis,mgierg1}.
However, in actual calculations, these non-universal details are irrelevant.
We loosely refer to both individual and summed over 
vertices of the kernels simply as a kernel. 
Note that we restrict the choice of kernels such that those labelled at one end by either $A$ or $B$ and at the other
by either $C$ or $D$
do not exist~\cite{aprop}.

The rule for decorating the complete diagrams on
the \rhs\ is simple: the set of fields, $\{f\}$, are distributed in 
all independent ways between the component objects of each diagram.

Embedded within the diagrammatic rules is a prescription for evaluating the
group theory factors. 
Suppose that we wish to focus on the flow of a particular
vertex coefficient function, which necessarily has a unique
supertrace structure. On the \lhs\ of the flow equation,
we can imagine splitting the lobe up into a number lobes 
equal to the number of supertraces (traditionally, these
lobes would be joined by dotted lines, to indicate that
they are part of the same vertex~\cite{Thesis,mgierg1}).
To finish the specification of the vertex coefficient function,
the lobes should be explicitly decorated by the fields, $\{f\}$,
where the fields are, in this picture, 
to be read off each lobe in the counterclockwise sense.

On the \rhs\ of the flow equation, things are slightly
more complicated.
Each lobe is, in principle, a multi-supertrace
object and the kernel can attach to any of these supertraces.
The kernel itself is a multi-supertrace object---for more
details see~\cite{Thesis,mgierg1}---but, for our purposes, 
we need note only that the kernel is, in this more
explicit diagrammatic picture,
a double sided entity. 
Thus, whilst the dumbbell like term of \fig{fig:FE}
has at least one associated supertrace, the next diagram
has at least two, on a account of the loop. If a closed
circuit formed by a kernel is devoid
of fields then
it contributes 
a factor of $\pm N$, depending on
the flavours of the fields to which the kernel forming
the loop attaches. This is most easily appreciated by
defining the projectors
\[
	\sigma_{\pm} := \hf ( \one \pm \sigma)
\]
and noting that $\str \sigma_\pm = \pm N$. 
In the counterclockwise sense, a $\sigma_+$
can always be inserted for free after an $A^1$, $C^1$ or $\bar{F}$,
whereas a $\sigma_-$
can always be inserted for free after an $A^2$, $C^2$ or $F$.

The above prescription for evaluating the
group theory factors receives $1/N$ corrections in 
the $A^1$ and $A^2$ sectors. If a kernel
attaches to an $A^1$ or $A^2$, it comprises a direct
attachment and an indirect attachment.
In the former case, one supertrace associated
with some vertex coefficient function  is `broken
open' by an end of a kernel: the fields on
this supertrace and the single supertrace component of the
kernel are on the same circuit.
In the latter case, the kernel does not break anything open
and so the two sides of the kernel pinch together
at the end associated with the indirect attachment.
This is illustrated in
in \fig{fig:Attach}; for more detail, see~\cite{Thesis,mgierg1}.
\bcf[h]
	\[
		\ensuremath{\begin{array}{c}\begin{picture}(0,0)%
\includegraphics{pstex/Direct.pstex}%
\end{picture}%
\setlength{\unitlength}{3947sp}%
\begingroup\makeatletter\ifx\SetFigFont\undefined%
\gdef\SetFigFont#1#2#3#4#5{%
  \reset@font\fontsize{#1}{#2pt}%
  \fontfamily{#3}\fontseries{#4}\fontshape{#5}%
  \selectfont}%
\fi\endgroup%
\begin{picture}(624,477)(2089,-976)
\end{picture}
 \end{array}} \rightarrow \left.\ensuremath{\begin{array}{c}\begin{picture}(0,0)%
\includegraphics{pstex/Direct.pstex}%
\end{picture}%
\setlength{\unitlength}{3947sp}%
\begingroup\makeatletter\ifx\SetFigFont\undefined%
\gdef\SetFigFont#1#2#3#4#5{%
  \reset@font\fontsize{#1}{#2pt}%
  \fontfamily{#3}\fontseries{#4}\fontshape{#5}%
  \selectfont}%
\fi\endgroup%
\begin{picture}(624,477)(2089,-976)
\end{picture}
 \end{array}}\right|_{\mathrm{direct}} + \frac{1}{N} \left[ \ensuremath{\begin{array}{c}\begin{picture}(0,0)%
\includegraphics{pstex/Indirect-2.pstex}%
\end{picture}%
\setlength{\unitlength}{3947sp}%
\begingroup\makeatletter\ifx\SetFigFont\undefined%
\gdef\SetFigFont#1#2#3#4#5{%
  \reset@font\fontsize{#1}{#2pt}%
  \fontfamily{#3}\fontseries{#4}\fontshape{#5}%
  \selectfont}%
\fi\endgroup%
\begin{picture}(624,809)(2089,-1279)
\put(2355,-554){\makebox(0,0)[lb]{\smash{\SetFigFont{8}{9.6}{\rmdefault}{\mddefault}{\updefault}{\color[rgb]{0,0,0}$A^2$}%
}}}
\end{picture}
 \end{array}} - \ensuremath{\begin{array}{c}\begin{picture}(0,0)%
\includegraphics{pstex/Indirect-1.pstex}%
\end{picture}%
\setlength{\unitlength}{3947sp}%
\begingroup\makeatletter\ifx\SetFigFont\undefined%
\gdef\SetFigFont#1#2#3#4#5{%
  \reset@font\fontsize{#1}{#2pt}%
  \fontfamily{#3}\fontseries{#4}\fontshape{#5}%
  \selectfont}%
\fi\endgroup%
\begin{picture}(624,809)(2089,-1279)
\put(2355,-554){\makebox(0,0)[lb]{\smash{\SetFigFont{8}{9.6}{\rmdefault}{\mddefault}{\updefault}{\color[rgb]{0,0,0}$A^1$}%
}}}
\end{picture}
 \end{array}} \right]
	\]
\caption{The $1/N$ corrections to the group theory factors.}
\label{fig:Attach}
\ecf 

We can thus consider the diagram on the \lhs\ as having been unpackaged,
to give the terms on the \rhs. The dotted lines in the diagrams with indirect
attachments serve to remind us where the loose end of the kernel attaches
in the parent diagram.

\subsection{Ward Identities}
\label{sec:WIDs}

All vertices, whether they belong to
either of the actions or to the covariantised
kernels are subject to Ward Identities which, due
to the manifest gauge invariance~\eqs{Agauged}{Cgauged},
take a particularly simple form.
In this paper, we need only the Ward identities for the 
action vertices, which are 
illustrated in \fig{fig:WIDs}, though note that the
Ward identities for the vertices of the kernels are very 
similar~\cite{ymi,ymii,aprop,Thesis,oliver1}.
\bcf[h]
	\[
		\ensuremath{\begin{array}{c}\input{pstex/WID-contract.pstex_t} \end{array}} = \ensuremath{\begin{array}{c}\input{pstex/WID-PF.pstex_t} \end{array}} + \ensuremath{\begin{array}{c}\input{pstex/WID-PFb.pstex_t} \end{array}} - \ensuremath{\begin{array}{c}\input{pstex/WID-PB.pstex_t} \end{array}} - \ensuremath{\begin{array}{c}\input{pstex/WID-PBb.pstex_t} \end{array}} +\cdots
	\]
\caption{The Ward identities for action vertices.}
\label{fig:WIDs}
\ecf

On the \lhs, we contract a vertex with the momentum of
the field which carries $p$. This field---which we will
call the active field---can be either
$A^1_\rho$, $A^2_\rho$, $F_R$ or $\bar{F}_R$. In the first two cases,
the open triangle $\GRk$ represents $p_\rho$ whereas, in the
latter two cases, it represents $p_R \equiv (p_\rho, 2)$~\cite{Thesis,mgierg1}.
(Given that we often sum over all possible fields, we can take the
Feynman rule for $\GRk$ in the $C$-sector to be null.)

On the \rhs, we push the contracted momentum forward onto 
the field which directly follows the active field, in the counterclockwise
sense, and pull back (with a minus sign) onto
the field which directly precedes the active field. 
Since our diagrammatics is permutation symmetric, the struck field---which
we will call the target field---can
be either $X$, $Y$ or any of the un-drawn fields represented
by the ellipsis.
Any field(s) besides the active field and the
target field will be called spectators.

The momentum routing follows in obvious manner: for example,
in the first diagram on the \rhs, momentum $q+p$ now flows into
the vertex. In the case that the active field is fermionic,
the field pushed forward / pulled back onto is transformed
into its opposite statistic partner. There are some signs 
associated with this in the $C$ and $D$-sectors, which we
will not require here~\cite{Thesis,mgierg1}.

The half arrow which terminates the pushed forward / pulled back
active field is of no significance and can go on either side
of the active field line. It is necessary to
keep the active field line---even though the active field
is no longer part of the vertex---in order that
we can unambiguously deduce flavour changes 
and momentum routing, without reference to the parent diagram.

\subsection{Taylor Expansion of Vertices}	
\label{sec:Taylor}

For the formalism to be properly defined,
it must be the case that all vertices
are Taylor expandable to all orders
in momenta~\cite{ym,ymi,ymii}.
For the purposes of this paper, we need only
the diagrammatic rules for a particular scenario.
Consider a vertex which is part
of a complete diagram, decorated by some set of internal
fields and by a single external $A^1$ (or $A^2$).
The diagrammatic representation for the zeroth order expansion
in the momentum of the external field is all that is required
and is shown in \fig{fig:TaylorExpansion}~\cite{Thesis,mgierg1}; 
note the similarity to \fig{fig:WIDs}.
\bcf[h]
	\[
		\ensuremath{\begin{array}{c}\input{pstex/Taylor-Parent.pstex_t} \end{array}} = \cdeps{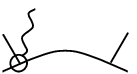} + \cdeps{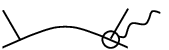} - \cdeps{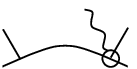} - \cdeps{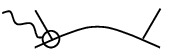} +\cdots
	\]
\caption{Diagrammatic representation of zeroth order Taylor expansion.}
\label{fig:TaylorExpansion}
\ecf

The interpretation of the diagrammatics is as follows. In the first diagram
on the \rhs, the vertex is differentiated \wrt\ the momentum carried
by the field $X$, whilst holding the momentum of the preceding field fixed
(if the preceding field carries zero momentum, it is effectively
transparent to the momentum derivative~\cite{mgierg1} 
and so we go in a clockwise sense
to the first field
which carries non-zero momentum to determine the momentum held constant).
Of course, using our current diagrammatic notation,
this latter field can be any of those
which decorate the vertex, and so we sum over all possibilities. Thus,
each cyclically ordered push forward like term has a partner,
cyclically ordered pull back like term, such that
the pair can be interpreted as
\be
	\left( \left. \partial^r_\mu \right|_s - \left. \partial^s_\mu \right|_r \right) \mathrm{Vertex},
\label{eq:Momderivs}
\ee
where $r$ and $s$ are momenta entering the vertex. 
In the case that $r=-s$, we can and will
drop either the push forward like term or pull back like term, since
the combination can be expressed as $\partial^r_\mu$; we
interpret the diagrammatic notation appropriately.

The other diagrammatics we require
is the representation of the
momentum derivative
of the effective propagators. These effective propagators are
represented simply by a line. The momentum derivative can be \wrt\
the momentum entering / leaving either end. We indicate this by
placing the momentum derivative symbol in the middle of the effective
propagator and adding an arrow, as shown in \fig{fig:dEP}.
\bcf[h]
	\[
		\cdeps{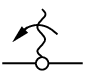}
	\]
\caption{A differentiated effective propagator.}
\label{fig:dEP}
\ecf

Now, since the left-hand end of the
effective propagator follows the momentum derivative
in the sense indicated by the arrow, we differentiate
\wrt\ the momentum \emph{entering} this end. Since equal
and opposite momentum enters the right-hand end,
it is clear that we can reverse the direction
of the arrow, at the expense of a minus sign.

In complete diagrams, Taylor expansions of various components
occur for one of two reasons. First, the expansion can
be forced: if a diagram 
contains a structure manifestly of the order in
external momentum to which we are working (in this
paper, an $\Op{2}$ stub), then other
structures carrying the external momentum
can be Taylor expanded (so long as such a step
does not generate IR divergences---see 
\sec{sec:TotalDerivativeMethods:Op2-NTE}). Alternatively,
the expansions can be unforced, but constructed for
convenience.

In both cases, diagrams contain a discontinuity in momentum
arguments. In the former case, the existence of an $\Op{2}$
stub makes it clear what the momentum routing must be. In 
the latter case, it is necessary to indicate where the
discontinuity occurs. To do this, we introduce the
`bar' notation of \fig{fig:bar}.
\bcf[h]
	\[
	\ensuremath{\begin{array}{c}\input{pstex/BarNotation.pstex_t} \end{array}}
	\]
\caption{Notation to indicate a discontinuity in momentum.}
\label{fig:bar}
\ecf

Note that there
is actually a choice of where we place the bar (to the left or right
of the top vertex),
depending on the momentum routing before Taylor
expansion. However, 
having made a choice, the location of
the bar is set, and cannot be changed.

\subsection{Charge Conjugation Invariance}

Charge conjugation invariance can be used to simplify the
diagrammatics, by allowing us to discard certain terms
and to combine others. The diagrammatic prescription for
replacing a diagram which possesses exclusively
bosonic external fields with its charge conjugate is~\cite{Thesis,mgierg1}
to reflect the diagram, picking up a sign for each
\begin{enumerate}
	\item external $A^i$,

	\item \emph{performed} gauge remainder,

	\item momentum derivative symbol (note
			that the direction of the arrow accompanying such
			symbols is reversed by the reflection of the diagram).
\end{enumerate}

\subsection{The Weak Coupling Expansion}
\label{sec:Weak}

\subsubsection{The Flow Equation}

Following~\cite{ymi,aprop}, the action has the weak
coupling expansion
\begin{equation}
	S = \sum_{i=0}^\infty \left( g^2 \right)^{i-1} S_i = \frac{1}{g^2}S_0 + S_1 + \cdots,
\label{eq:WeakCouplingExpansion-Action}
\end{equation} 
where $S_0$ is the classical effective action and the $S_{i>0}$
the $i$th-loop corrections. The seed action has a similar expansion:
\begin{equation}
	\hat{S} = \sum_{i=0}^\infty  g^{2i}\hat{S}_i.
\label{eq:WeakCouplingExpansion-SeedAction}
\end{equation} 
Note that these definitions are consistent with $\Sigma_g = g^2 S -2\hat{S}$; 
identifying powers of $g$ in the flow equation, it is clear that
$S_i$ and $\hat{S}_i$ will always appear together.
With this in mind, we now define
\begin{equation}
	\Sigma_i = S_i - 2\hat{S}_i.
\label{eq:WeakCouplingSigma}
\end{equation}

Recalling~\eq{eq:alpha-defn},
the $\beta$ functions for $g$ and $g_2$ are
\begin{eqnarray}
	\flow \frac{1}{g^2} 	& = & -2 \sum_{i=1}^\infty \beta_i(\alpha) g^{2(i-1)} 
\label{eq:WCE:flow-g}
\\
	\flow \frac{1}{g^2_2} 	& = & -2 \sum_{i=1}^\infty \tilde{\beta}_i(1/\alpha) g^{2(i-1)}_2,
\label{eq:WCE:flow-g_2}
\end{eqnarray}
where the $\beta_i(\alpha)$ are determined through the
renormalisation condition~\eq{defg} and the 
$\tilde{\beta}_i(1/\alpha)$ are determined through~\eq{defg2}.
The coefficient $\beta_1 = - \tilde{\beta}_1$ is independent of
$\alpha$, as we will explicitly demonstrate in 
\sec{sec:TLD:Methodology:1-loop}.
For generic $\alpha$, we expect
the coefficient $\beta_2(\alpha)$ to disagree with the standard
value; as we will explicitly confirm in 
section~\ref{sec:TLD:alpha}, agreement
is reached for $\beta_2(0)$.\footnote{We note that whilst
we expect $\beta_2(0) = \tilde{\beta}_2(0)$, there is no
reason to generically expect $\tilde{\beta}_2(1/\alpha) = \beta_2(\alpha)$
since $g$ and $g_2$ are not treated symmetrically in the flow equation.}

Utilising eqns.~\eqs{eq:WCE:flow-g}{eq:WCE:flow-g_2}
and rewriting $g_2^2$ in terms of $\alpha g_1^2$, it is apparent
that $\flow \alpha$ has the following weak coupling expansion:
\begin{equation}
	\flow \alpha = \sum_{i=1}^{\infty} \gamma_i g^{2i},
\label{eq:flow-alpha}
\end{equation}
where
\be
\label{eq:gamma_i}
	\gamma_i = -2 \alpha \left(\beta_i(\alpha) - \alpha^i \tilde{\beta}_i(1/\alpha) \right).
\ee

Substituting these definitions into~\eq{eq:FE},
we obtain the weak coupling expansion of the flow
equation, shown in \fig{fig:NFE:NewDiagFE}.
\begin{center}
\begin{figure}[h]
	\be
	\label{eq:WeakFlow}
	\dec{
		\ensuremath{\begin{array}{c}\input{pstex/Vertex-n-LdL.pstex_t} \end{array}} 
	}{\{f\}}
	= 
	\dec{
		\sum_{r=1}^n \left[2\left(n_r -1 \right) \beta_r +\gamma_r \pder{}{\alpha} \right]\ensuremath{\begin{array}{c}\input{pstex/Vertex-n_r-B.pstex_t} \end{array}} 
		+ \frac{1}{2} \sum_{r=0}^n \ensuremath{\begin{array}{c}\input{pstex/Dumbbell-n_r-r.pstex_t} \end{array}} -\frac{1}{2} \ensuremath{\begin{array}{c}\input{pstex/Vertex-Sigma_n_-B.pstex_t} \end{array}}
	}{\{f\}}
	\ee
\caption{The diagrammatic form for the weak coupling flow equation.}
\label{fig:NFE:NewDiagFE}
\end{figure}
\end{center}

The symbol $\bullet \equiv -\flowConstAl$. A vertex whose
argument is a letter, say $n$, represents $S_n$. We define
$n_r :=n-r$ and $n_\pm := n \pm 1$. The `bar notation' of the dumbbell
term is defined as follows:
\[
	a_0[\bar{S}_{n-r}, \bar{S}_r] 	\equiv 	a_0[S_{n-r}, S_r] - a_0[S_{n-r}, \hat{S}_r] - a_0[\hat{S}_{n-r}, S_r].
\]

\subsubsection{The Effective Propagator Relation}	\label{sec:EPReln}

The effective propagator relation~\cite{aprop} is central
to the perturbative diagrammatic approach, and arises
from examining the flow of all two-point, tree level vertices.
This is done by setting $n=0$ in~\eq{eq:WeakFlow}
and specialising $\{f\}$ to contain two fields, 
as shown in \fig{fig:TLTPs}.
We note that we can and do choose
all such vertices to be single supertrace terms~\cite{Thesis,mgierg1}.
\bcf[h]
	\be
		\ensuremath{\begin{array}{c}\input{pstex/Vertex-TLTP-LdL.pstex_t} \end{array}} = \ensuremath{\begin{array}{c}\input{pstex/Dumbbell-S_0-Sigma_0.pstex_t} \end{array}}
	\label{eq:TLTP-flow}
	\ee
\caption{Flow of all possible two-point, tree level vertices.}
\label{fig:TLTPs}
\ecf

Following~\cite{ym,ymi,ymii,aprop,Thesis,mgierg1,two},
we use the freedom inherent in $\hat{S}$ by choosing the two-point, tree
level seed action vertices equal to the corresponding Wilsonian effective
action vertices. Eqn.~\eq{eq:TLTP-flow} now simplifies.
Rearranging, integrating \wrt\ $\Lambda$ and choosing the appropriate
integration constants~\cite{Thesis,mgierg1}, we arrive at the
relationship between the integrated ERG kernels---\aka the
effective propagators---and the two-point,
tree level vertices shown in \fig{fig:EffPropReln}. Note
that we have attached the effective propagator, which only
ever appears as an internal line, to an arbitrary structure.
\bcf[h]
	\be
		\ensuremath{\begin{array}{c}\input{pstex/EffPropReln.pstex_t} \end{array}}
		\equiv \ensuremath{\begin{array}{c}\begin{picture}(0,0)%
\includegraphics{pstex/K-Delta.pstex}%
\end{picture}%
\setlength{\unitlength}{3947sp}%
\begingroup\makeatletter\ifx\SetFigFont\undefined%
\gdef\SetFigFont#1#2#3#4#5{%
  \reset@font\fontsize{#1}{#2pt}%
  \fontfamily{#3}\fontseries{#4}\fontshape{#5}%
  \selectfont}%
\fi\endgroup%
\begin{picture}(374,395)(1791,-1006)
\put(1791,-843){\makebox(0,0)[lb]{\smash{\SetFigFont{8}{9.6}{\rmdefault}{\mddefault}{\updefault}{\color[rgb]{0,0,0}$M$}%
}}}
\end{picture}
 \end{array}} - \ensuremath{\begin{array}{c}\begin{picture}(0,0)%
\includegraphics{pstex/FullGaugeRemainder.pstex}%
\end{picture}%
\setlength{\unitlength}{3947sp}%
\begingroup\makeatletter\ifx\SetFigFont\undefined%
\gdef\SetFigFont#1#2#3#4#5{%
  \reset@font\fontsize{#1}{#2pt}%
  \fontfamily{#3}\fontseries{#4}\fontshape{#5}%
  \selectfont}%
\fi\endgroup%
\begin{picture}(424,395)(2053,-930)
\put(2053,-773){\makebox(0,0)[lb]{\smash{\SetFigFont{8}{9.6}{\rmdefault}{\mddefault}{\updefault}{\color[rgb]{0,0,0}$M$}%
}}}
\end{picture}
 \end{array}} 
		\equiv \ensuremath{\begin{array}{c}\begin{picture}(0,0)%
\includegraphics{pstex/K-Delta.pstex}%
\end{picture}%
\setlength{\unitlength}{3947sp}%
\begingroup\makeatletter\ifx\SetFigFont\undefined%
\gdef\SetFigFont#1#2#3#4#5{%
  \reset@font\fontsize{#1}{#2pt}%
  \fontfamily{#3}\fontseries{#4}\fontshape{#5}%
  \selectfont}%
\fi\endgroup%
\begin{picture}(374,395)(1791,-1006)
\put(1791,-843){\makebox(0,0)[lb]{\smash{\SetFigFont{8}{9.6}{\rmdefault}{\mddefault}{\updefault}{\color[rgb]{0,0,0}$M$}%
}}}
\end{picture}
 \end{array}} - \ensuremath{\begin{array}{c}\begin{picture}(0,0)%
\includegraphics{pstex/DecomposedGR.pstex}%
\end{picture}%
\setlength{\unitlength}{3947sp}%
\begingroup\makeatletter\ifx\SetFigFont\undefined%
\gdef\SetFigFont#1#2#3#4#5{%
  \reset@font\fontsize{#1}{#2pt}%
  \fontfamily{#3}\fontseries{#4}\fontshape{#5}%
  \selectfont}%
\fi\endgroup%
\begin{picture}(540,395)(1936,-925)
\put(1936,-776){\makebox(0,0)[lb]{\smash{\SetFigFont{8}{9.6}{\rmdefault}{\mddefault}{\updefault}{\color[rgb]{0,0,0}$M$}%
}}}
\end{picture}
 \end{array}}
	\label{eq:EPReln}
	\ee
\caption{The effective propagator relation.}
\label{fig:EffPropReln}
\ecf

The structure $\GRkpr \!\! \GRk$
is called a gauge remainder~\cite{aprop}.
The individual components of
$\GRkpr \!\! \GRk$ will often be loosely
referred to as gauge remainders; where it is necessary to
 unambiguously refer to the composite structure, we will use
the terminology `full gauge remainder'.

The various components on the \rhs\ of~\eq{eq:EPReln}
can be interpreted, in the different sectors,
according to table~\ref{tab:NFE:k,k'}, where we assume
that the gauge remainder carries momentum $p$.
\renewcommand{\arraystretch}{1.5}
\begin{center}
\begin{table}[h]
	\[
	\begin{array}{c|ccc}
					& \delta_{MN}		& \GRkpr						& \GRk
	\\ \hline 
		F,\bar{F}	& \delta_{MN}		& (f_p p_\mu / \Lambda^2, g_p)	& (p_\nu, 2)
	\\
		A^i			& \delta_{\mu \nu}	& p_\mu / p^2					& p_\nu
	\\
		C^i			& \one				& \mbox{---}					& \mbox{---}
	\end{array}
	\]
\caption{Prescription for interpreting eqn.~\eq{eq:EPReln}.}
\label{tab:NFE:k,k'}
\end{table}
\end{center}
\renewcommand{\arraystretch}{1}

The functions $f(k^2/\Lambda^2)$ and $g(k^2/\Lambda^2)$ 
need never be exactly determined;
rather, they must satisfy general constraints enforced
by the requirements of
proper UV regularisation of the physical $SU(N)$ theory
and gauge invariance. We will see the effect induced by the
latter shortly. However, we will find it useful to have
concrete algebraic realisations of the two-point, tree
level vertices, and effective propagators (and hence $f$
and $g$), which we collect together in appendix~\ref{app:elements}.
The effective propagators
are denoted by $\Delta^{XY}$, where $X$ and $Y$  denote the flavours at the ends.

It is important to note that we have defined the diagrammatics
in \fig{fig:EffPropReln} such that there are no $1/N$
corrections where the effective propagator attaches
to the vertex. We do this because, when the composite
object on the \lhs\ of \fig{fig:EffPropReln} appears
in actual calculations, it always occurs inside some larger diagram.
It is straightforward to show that, in this case, the aforementioned attachment 
corrections always vanish~\cite{Thesis}. 

\subsubsection{Further Diagrammatic Identities}

We begin with a diagrammatic identity which follows
from gauge invariance and the constraint
placed on the vertices of the Wilsonian effective action by
the requirement that the minimum of the Higgs potential is
not shifted by quantum corrections:
\be
	\ensuremath{\begin{array}{c}\begin{picture}(0,0)%
\includegraphics{pstex/GR-TLTP.pstex}%
\end{picture}%
\setlength{\unitlength}{3947sp}%
\begingroup\makeatletter\ifx\SetFigFont\undefined%
\gdef\SetFigFont#1#2#3#4#5{%
  \reset@font\fontsize{#1}{#2pt}%
  \fontfamily{#3}\fontseries{#4}\fontshape{#5}%
  \selectfont}%
\fi\endgroup%
\begin{picture}(757,318)(1880,-963)
\put(2296,-857){\makebox(0,0)[lb]{\smash{\SetFigFont{11}{13.2}{\rmdefault}{\mddefault}{\updefault}{\color[rgb]{0,0,0}$0$}%
}}}
\end{picture}
 \end{array}} = 0.
\label{eq:GR-TLTP}
\ee
This follows directly in the $A$-sector, since 
one-point $A$-vertices do not exist.
In the $F$-sector, though,
we are left with one-point $C^1$ and $C^2$-vertices, but
these are constrained to be zero. 

From the effective propagator relation and~\eq{eq:GR-TLTP},
two further diagrammatic identities follow.
First, consider attaching
an effective propagator to the right-hand field in~\eq{eq:GR-TLTP}
and applying
the effective propagator before $\GRk$ has acted. Diagrammatically,
this gives
\[
	\ensuremath{\begin{array}{c}\begin{picture}(0,0)%
\includegraphics{pstex/GR-TLTP-EP.pstex}%
\end{picture}%
\setlength{\unitlength}{3947sp}%
\begingroup\makeatletter\ifx\SetFigFont\undefined%
\gdef\SetFigFont#1#2#3#4#5{%
  \reset@font\fontsize{#1}{#2pt}%
  \fontfamily{#3}\fontseries{#4}\fontshape{#5}%
  \selectfont}%
\fi\endgroup%
\begin{picture}(1081,306)(2490,-1356)
\put(2776,-1250){\makebox(0,0)[lb]{\smash{\SetFigFont{11}{13.2}{\rmdefault}{\mddefault}{\updefault}{\color[rgb]{0,0,0}0}%
}}}
\end{picture}
 \end{array}} = 0 = \cdeps{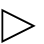} - \cdeps{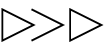},
\]
which implies the following diagrammatic identity:
\be
	\cdeps{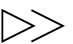} = 1.
\label{eq:GR-relation}
\ee

The effective propagator relation, together
with~\eq{eq:GR-relation}, implies that
\[
	\ensuremath{\begin{array}{c}\begin{picture}(0,0)%
\includegraphics{pstex/TLTP-EP-GR.pstex}%
\end{picture}%
\setlength{\unitlength}{3947sp}%
\begingroup\makeatletter\ifx\SetFigFont\undefined%
\gdef\SetFigFont#1#2#3#4#5{%
  \reset@font\fontsize{#1}{#2pt}%
  \fontfamily{#3}\fontseries{#4}\fontshape{#5}%
  \selectfont}%
\fi\endgroup%
\begin{picture}(1059,306)(2512,-1356)
\put(2776,-1250){\makebox(0,0)[lb]{\smash{\SetFigFont{11}{13.2}{\rmdefault}{\mddefault}{\updefault}{\color[rgb]{0,0,0}0}%
}}}
\end{picture}
 \end{array}} = \cdeps{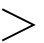} - \cdeps{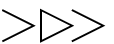} = 0.
\]
In other words, the (non-zero) structure $\ensuremath{\begin{array}{c}\begin{picture}(0,0)%
\includegraphics{pstex/EP-GR.pstex}%
\end{picture}%
\setlength{\unitlength}{3947sp}%
\begingroup\makeatletter\ifx\SetFigFont\undefined%
\gdef\SetFigFont#1#2#3#4#5{%
  \reset@font\fontsize{#1}{#2pt}%
  \fontfamily{#3}\fontseries{#4}\fontshape{#5}%
  \selectfont}%
\fi\endgroup%
\begin{picture}(406,118)(3165,-1234)
\end{picture}
 \end{array}}$ kills
a two-point, tree level vertex. But, by~\eq{eq:GR-TLTP}, 
this suggests that the structure $\ensuremath{\begin{array}{c}\begin{picture}(0,0)%
\includegraphics{pstex/EP-GR.pstex}%
\end{picture}%
\setlength{\unitlength}{3947sp}%
\begingroup\makeatletter\ifx\SetFigFont\undefined%
\gdef\SetFigFont#1#2#3#4#5{%
  \reset@font\fontsize{#1}{#2pt}%
  \fontfamily{#3}\fontseries{#4}\fontshape{#5}%
  \selectfont}%
\fi\endgroup%
\begin{picture}(406,118)(3165,-1234)
\end{picture}
 \end{array}}$
must be equal, up to some factor, to $\lhd$. Indeed,
\be
	\cdeps{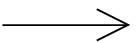} \equiv \cdeps{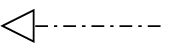},
\label{eq:PseudoEP}
\ee
where the dot-dash line represents the pseudo effective propagators 
of~\cite{Thesis,mgierg1}.

The final diagrammatic identity we require 
follows directly from
the independence of $\GRk$ on $\Lambda$:
\be
	\stackrel{\bullet}{\GRk} = 0.
\label{eq:LdL-GRk}
\ee

\subsubsection{Diagrammatic Expression For $\beta_1$}

We conclude the review by giving, in
\fig{fig:beta1}, the diagrammatic
expression of~\cite{Thesis,mgierg1} for $\beta_1$ . 
On the \rhs, we suppress 
the Lorentz indices of the external fields and work
at $\Op{2}$, as will always be done from now on.
This expression will be used throughout the rest
of this paper, but also serves to introduce the final
diagrammatic rules.
\begin{center}
\begin{figure}[h]
	\[
	4 \beta_1 \Box_{\mu \nu} (p) =
	-\frac{1}{2}
	\dec{
		\begin{array}{c}
			\begin{array}{ccccccc}
				\LD{Beta1-LdL-A}	&	& \LD{Beta1-LdL-B}	&	& \LD{Beta1-LdL-C} 	&	&\LD{Beta1-LdL-D}
			\\[1ex]
				\ensuremath{\begin{array}{c}\input{pstex/Beta1-LdL-A.pstex_t} \end{array}}	& -	& \ensuremath{\begin{array}{c}\input{pstex/Beta1-LdL-B.pstex_t} \end{array}}	&+4	& \ensuremath{\begin{array}{c}\input{pstex/Beta1-LdL-C.pstex_t} \end{array}}	& -	&\ensuremath{\begin{array}{c}\input{pstex/Beta1-LdL-D.pstex_t} \end{array}}
			\end{array}
		\\[10ex]
			\begin{array}{cccc}
					& \LD{Beta1-LdL-E}	&	& \LD{Beta1-LdL-Fc}	
			\\[1ex]
				+4	& \ensuremath{\begin{array}{c}\input{pstex/Beta1-LdL-E.pstex_t} \end{array}}	&-8	& \ensuremath{\begin{array}{c}\input{pstex/Beta1-LdL-Fc.pstex_t} \end{array}}
			\end{array}
		\end{array}
	}{\bullet}
	\]
\caption{Diagrammatic expression for $\beta_1$.}
\label{fig:beta1}
\end{figure}
\end{center}

There are a number of things to note. First is that all diagrams
are built from the following components: Wilsonian effective
action vertices, effective propagators, $\GRk$ and $\GRkpr$. There
are no seed action vertices or covariantised kernel vertices, reflecting
the universality of the $\beta_1$.

Diagrams~\ref{Beta1-LdL-A} and~\ref{Beta1-LdL-B} are straightforward,
containing just Wilsonian effective action vertices
and effective propagators.
Diagram~\ref{Beta1-LdL-C} comprises a nested gauge 
remainder~\cite{Thesis,mgierg1}. Diagrams~\ref{Beta1-LdL-D}
and~\ref{Beta1-LdL-E} possess a common bottom vertex and
effective propagator. This 
effective propagator
must be in the 
$C^i$-sector, else the diagrams vanish by \CC\ invariance. 
The structures attached to the top end
of the common effective propagator will crop up repeatedly;
for convenience we define
\[
	\cdeps{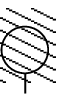} \equiv \ensuremath{\begin{array}{c}\begin{picture}(0,0)%
\includegraphics{pstex/TLThP-SimpleLoop.pstex}%
\end{picture}%
\setlength{\unitlength}{3947sp}%
\begingroup\makeatletter\ifx\SetFigFont\undefined%
\gdef\SetFigFont#1#2#3#4#5{%
  \reset@font\fontsize{#1}{#2pt}%
  \fontfamily{#3}\fontseries{#4}\fontshape{#5}%
  \selectfont}%
\fi\endgroup%
\begin{picture}(249,477)(112,329)
\put(200,483){\makebox(0,0)[lb]{\smash{\SetFigFont{11}{13.2}{\rmdefault}{\mddefault}{\updefault}{\color[rgb]{0,0,0}0}%
}}}
\end{picture}
 \end{array}} -4 \cdeps{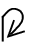}.
\]

Diagram~\ref{Beta1-LdL-Fc}
possesses an $\Op{2}$ stub: the renormalisation condition~\eq{defg} 
and gauge invariance demand that~\cite{aprop,Thesis,mgierg1}
\be
\label{eq:stub}
	S_{0 \mu \alpha}^{\ 1 \, 1}(p) = \InvCO_p \Box_{\mu \alpha}(p) = 2\Box_{\mu \alpha}(p) + \Op{4}.
\ee
In diagrams possessing an $\Op{2}$ stub and a momentum
derivative symbol, it will be understood that
the structure attaching to the stub does not carry
momentum $p$, unless explicitly indicated otherwise
(see \sec{sec:TotalDerivativeMethods:Op2-NTE}). The
structure attaching to the $\Op{2}$ stub of diagram~\ref{Beta1-LdL-Fc}
is defined in terms of (pseudo) effective propagators and
gauge remainder components:
\[
	\cdeps{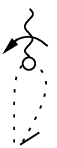} \equiv \cdeps{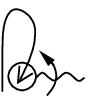} - \hf \cdeps{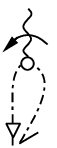}.
\]

%Diagram~\ref{Beta1-LdL-G} possesses the momentum derivative
%of a pseudo effective propagator and a $\GRk$.
%Notice that whereas diagram~\ref{Beta1-LdL-F} possesses the momentum
%derivative of an effective propagator and a $\GRkpr$, 
%diagram~\ref{Beta1-LdL-Fb} possess a plain effective propagator
%but a differentiated $\GRkpr$. Diagrams~\ref{Beta1-LdL-F}
%and~\ref{Beta1-LdL-Fb} can, in fact, be combined, as
%illustrated in \fig{fig:LittleSet-Redraw}.
%\bcf[h]
%	\[
%	\cd{Beta1-LdL-Fb} = \cd{Beta1-LdL-dF} - \cd{Beta1-LdL-F}
%	\]
%\caption{Re-expression of diagram~\ref{Beta1-LdL-Fb}.}
%\label{fig:LittleSet-Redraw}
%\ecf

%The first diagram on the \rhs\ is a total
%momentum derivative. If, for the purposes of 
%preregularisation, we work in general dimension, $D$,
%such terms are automatically discarded. In the calculation
%of $\beta_1$ and $\beta_2$, these
%are the only type of terms which vanish as a consequence of
%the preregularisation. 
%It thus seems
%that we can simply adopt the diagrammatic prescription
%to discard such terms, without having to resort to
%specifying whether or not we work strictly in $D=4$.

Finally, we introduce some nomenclature. The complete set
of diagrams inside the square brackets of
\fig{fig:beta1} will be denoted by
$\OLDs(p)$. The first three diagrams---whose sum we note
is transverse~\cite{Thesis,mgierg1}---will be referred to as
the standard set and the final diagram
as the little set. $\OLDs(p)$, modulo the little set, will
be denoted by $\ROLDs(p)$. We define
\[
	\ensuremath{\begin{array}{c}\begin{picture}(0,0)%
\includegraphics{pstex/OLDS.pstex}%
\end{picture}%
\setlength{\unitlength}{3947sp}%
\begingroup\makeatletter\ifx\SetFigFont\undefined%
\gdef\SetFigFont#1#2#3#4#5{%
  \reset@font\fontsize{#1}{#2pt}%
  \fontfamily{#3}\fontseries{#4}\fontshape{#5}%
  \selectfont}%
\fi\endgroup%
\begin{picture}(457,249)(996,-718)
\put(1184,-647){\makebox(0,0)[lb]{\smash{\SetFigFont{11}{13.2}{\rmdefault}{\mddefault}{\updefault}{\color[rgb]{0,0,0}1}%
}}}
\end{picture}
 \end{array}} \equiv \ensuremath{\begin{array}{c}\begin{picture}(0,0)%
\includegraphics{pstex/OL-A.pstex}%
\end{picture}%
\setlength{\unitlength}{3947sp}%
\begingroup\makeatletter\ifx\SetFigFont\undefined%
\gdef\SetFigFont#1#2#3#4#5{%
  \reset@font\fontsize{#1}{#2pt}%
  \fontfamily{#3}\fontseries{#4}\fontshape{#5}%
  \selectfont}%
\fi\endgroup%
\begin{picture}(326,430)(395,353)
\put(524,462){\makebox(0,0)[lb]{\smash{\SetFigFont{11}{13.2}{\rmdefault}{\mddefault}{\updefault}{\color[rgb]{0,0,0}0}%
}}}
\end{picture}
 \end{array}} - \ensuremath{\begin{array}{c}\input{pstex/OL-B.pstex_t} \end{array}} +4 \cdeps{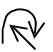} - \ensuremath{\begin{array}{c}\begin{picture}(0,0)%
\includegraphics{pstex/OL-D.pstex}%
\end{picture}%
\setlength{\unitlength}{3947sp}%
\begingroup\makeatletter\ifx\SetFigFont\undefined%
\gdef\SetFigFont#1#2#3#4#5{%
  \reset@font\fontsize{#1}{#2pt}%
  \fontfamily{#3}\fontseries{#4}\fontshape{#5}%
  \selectfont}%
\fi\endgroup%
\begin{picture}(326,754)(120,262)
\put(243,370){\makebox(0,0)[lb]{\smash{\SetFigFont{11}{13.2}{\rmdefault}{\mddefault}{\updefault}{\color[rgb]{0,0,0}0}%
}}}
\end{picture}
 \end{array}};
\]
this expression is such that, if both external legs are $A^1$s, it
reduces to $\ROLDs(p)$.

\section{$\Lambda$-Derivative Terms} 
\label{ch:LambdaDerivatives:Methodology}

In this section we detail the methodology used for treating 
$\Lambda$-derivative terms.
Following a statement of the basic idea in \sec{sec:LD:M:intro}, 
the methodology is developed in \sec{sec:TLD:Methodology:1-loop} 
by looking at the one-loop integrals which contribute to $\beta_1$. 

In \sec{sec:TLD:Methodology:2-loop}, we build on the one-loop case 
to obtain the expected general form for two-loop integrals.
As preparation for the evaluation of $\beta_2$, 
the subtraction techniques are introduced,
which will allow us to explicitly demonstrate how 
non-universal contributions can cancel between diagrams. 
The methodology for this, which also has applications to
terms which require manipulation at $\Op{2}$, is presented 
in \sec{sec:TLD:Subtractions}.

We conclude this section by showing that certain running couplings which
can spoil the universality of $\beta_2$ can always be removed by a suitable
choice of the seed action. This proves the statement made in~\cite{mgierg1}
that it is possible to ensure that the only 
couplings requiring renormalisation are $g$ and $g_2$.

\subsection{Introduction} 
\label{sec:LD:M:intro}

The simplest $\Lambda$-derivative terms we will encounter are those
contributing to $\beta_1$. From \fig{fig:beta1} we know that
we can write
\[
	4 \beta_1 \Box_{\mu \nu} (p) = -\frac{1}{2} \dec{\OLDs(p)}{\bullet}.
\]
We now want to make the integral over loop momentum (which
we will take to be $k$) to be explicit and so write
\begin{equation}
	4 \beta_1 \Box_{\mu \nu} (p) = -\frac{1}{2} \int_k \dec{\OLDs(k,p)}{\bullet}.
\label{eq:TDM:Interchange}
\end{equation}

The next step that we wish to perform is to interchange 
the order of the $\Lambda$-derivative
and the momentum integral. This step is trivial only if 
the integral is convergent, even after this
change. Temporarily ignoring this subtlety gives:
\[
	4 \beta_1 \Box_{\mu \nu} (p) = -\frac{1}{2} \dec{\int_k \OLDs(k,p)}{\bullet}.
\]

Since the \lhs\ of this equation comprises a number times $\Op{2}$, it follows
that the coefficient multiplying the $\Op{2}$ part of the \rhs\ must be
dimensionless. Consequently, we can schematically write
\[
	\beta_1 = \flowConstAl \left( \mbox{Dimensionless Quantity} \right).
\]
For the \rhs\ to survive differentiation \wrt\ $\Lambda|_\alpha$, 
it must either depend on some dimensionless running 
coupling---other than $g$ and $\alpha$---or there must be some scale, 
other than $\Lambda$,
available for the construction of dimensionless quantities; 
we show how to avoid introducing such
couplings in \sec{sec:TLD:Couplings}.

One scale which is available is $p$ and so we can 
envisage contributions to $\beta_1$ of 
the form (in $D=4$)
\[
	\flowConstAl \ln p^2 / \Lambda^2.
\]
Indeed, the standard set gives rise to contributions 
of this type (see \sec{sec:TLD-RegByp}). However,
as we will see in \sec{sec:TDM:Indep-p}, contributions
of this type cannot be formed from the little set---but we know 
from~\cite{aprop} that the little set does
contribute to $\beta_1$.

For the little set, then, what scales are there, other than $\Lambda$, available
for the construction of dimensionless quantities? Courtesy of the $SU(N|N)$ 
regularisation, there are no scales in the UV.\footnote{
Although we will find a subtle interplay between the IR and UV, 
which is commented on at
the end of 
\sec{sec:TLD-RegByp}. See also~\cite{giqed}.} 
Na\"ively, we would not expect a scale to arise
in the IR, either. However,  the key point is
that interchanging the order of differentiation \wrt\ 
$\flowConstAl$ and loop integration
in~\eq{eq:TDM:Interchange} 
has the capacity to  introduce IR divergences.
This is most easily appreciated by noting that $\Delta^{11}_k \sim 1/k^2$, 
whereas $\dot{\Delta}^{11}_k \sim 1/\Lambda^2$ (see eqn.~\eq{eq:app:NFE:EP-11}).
Thus, to legally move $\flowConstAl$ outside of the loop integral, 
we must introduce some IR regulator,
which then provides the scale necessary to form dimensionless quantities. 
After allowing 
$\flowConstAl$ to act, this unphysical scale will disappear. 
IR divergences introduced in 
this way will be called \emph{pseudo}-divergences, 
since they are an artifact of the way
in which we have chosen to perform the calculation.

Noting that in the case of the standard set it is effectively $p$ which is 
providing the IR regularisation,
our strategy for evaluating loop integrals is to look at the IR end. 
Scanning through the
list of effective propagators~\eq{eq:app:NFE:EP-11}--\eq{eq:app:NFE:EP-CiCi}, 
it is apparent that
the leading contributions occur when all effective propagators are in the $A^i$-sector;
likewise for any instances of $\GRkpr$. Gauge invariance and considerations
as to the supertrace structure will eventually determine that all contributions to
$\beta_1$ and $\beta_2$ are ultimately
limited to the lowest order momentum contributions from the 
$A^1$ sector; it is precisely this regime---and this regime alone---that is universal.

\subsection{One-loop Integrals}	
\label{sec:TLD:Methodology:1-loop}

\subsubsection{Vanishing Diagrams}

We begin our analysis by looking at diagrams~\ref{Beta1-LdL-D} and~\ref{Beta1-LdL-E} .
Recall that these two diagrams have no analogue in the computation of $\beta_1$
presented in~\cite{aprop} (as a consequence of all
actions being restricted to single supertrace terms) 
and so we had better find that they vanish.

Consider the IR end of the loop integral.
The bottom vertex does not carry the loop momentum, nor does the effective
propagator leaving it. However, we know that this effective propagator must
be in the $C$-sector. 
This immediately tells us that diagram~\ref{Beta1-LdL-E}
is IR safe, even if we interchange the order the $\Lambda$-derivative
and the momentum integral: performing this step, the loop integral just goes as
\[
	\int_k g_k.
\]

Let us now focus on diagram~\ref{Beta1-LdL-D}.
To try and find IR divergences, we take the fields involved in the 
loop integral to be in the $A^i$-sector.
To deal with the top vertex, we recall that it is 
Taylor expandable in momenta~\cite{ym,ymi,ymii}. 
Hence, to try and isolate
the most IR divergent contribution from the loop integral, we take the minimum number of
powers of momenta from the top vertex consistent with  Lorentz invariance. 
Given that the field entering
this vertex from beneath is in the $C$-sector, we can take \Omom{0}.
Diagram~\ref{Beta1-LdL-D} is thus clearly IR safe in $D=4$ since the loop integral 
looks at worst---without even taking into account the  $\Lambda$-derivative---like
\[
\int_k 1/k^2,
\]
in the IR. 

Hence, we can safely interchange the order of integration and 
differentiation \wrt\ $\LConstAl$  for both diagrams~\ref{Beta1-LdL-D} 
and~\ref{Beta1-LdL-E}.
Having done so,
we know that (the $\Op{2}$ parts of) both diagrams will vanish, 
when computed in $D=4$. In preparation for the two-loop calculation, 
where the integrals are most conveniently
evaluated using dimensional regularisation,
suppose that we now work in
$D=4-2\epsilon$ dimensions. 
Rescaling our loop momenta $k \rightarrow k/\Lambda$ we
see that, at $\Op{2}$, the diagrams acquire an overall 
factor of $\Lambda^{-2 \epsilon}$. Working in this
scheme, the $\Lambda$-derivative of the diagrams 
now $\sim \epsilon$ which, of course, vanishes in the $D \to 4$ limit.

Note that diagram~\ref{Beta1-LdL-A} (the first element of the standard set)
also vanishes in $D=4$, after differentiation
\wrt\ $\LConstAl$. However, we will always
keep this term together with the other elements 
of the standard set. This is done because
we  often exploit the fact that the set is transverse, 
and want to be able to do
this irrespective both of $D$
and whether or not the standard set is struck by a $\Lambda$-derivative.

\subsubsection{IR Regularisation Provided by $p$} \label{sec:TLD-RegByp}

In this section, we will encounter diagrams that 
survive differentiation \wrt\ $\Lambda$ in $D=4$
and for which $p$ plays a \role\ in the IR regularisation. The only diagrams
that fall into this class are diagrams~\ref{Beta1-LdL-B} and~\ref{Beta1-LdL-C}. 
However, 
as just mentioned, diagram~\ref{Beta1-LdL-A} will come along for the ride;
a full understanding of the standard set is crucial for the two-loop calculation.

Diagram~\ref{Beta1-LdL-A} is IR safe. Diagrams~\ref{Beta1-LdL-B} 
and~\ref{Beta1-LdL-C}, before
differentiation \wrt\ $\LConstAl$, 
have the IR structure
\begin{equation}
	\int_k \frac{\mathcal{O}(p^2, p.k, k^2)}{k^2 (k-p)^2}, 
\label{eq:TLD:GeneralIRStructure}
\end{equation}
where we have taken a single power of momentum from each of the 
three-point vertices, have chosen all effective propagators
to be in the $A^i$-sector and have evaluated any cutoff functions at zero
momentum.

Note that choosing the effective propagators to be in the $A^i$-sector 
constrains diagram~\ref{Beta1-LdL-B}, considerably. For three-point vertices decorated
exclusively by $A^i$-fields, it must be the case that all fields are on the same supertrace
and hence of the same flavour. Consequently, for the contributions 
to diagram~\ref{Beta1-LdL-B}
with the severest IR behaviour, all fields must be in the $A^1$-sector.

Returning to eqn.~\eq{eq:TLD:GeneralIRStructure}, 
it is clear that the presence of $p$ in the denominator is required 
to regularise the integral in the IR, at
least when we choose to take $\mathcal{O}(p^2)$ from the vertices. 
Performing the integral in $D=4$ will
then give us something of the form
\[
\mathcal{O}(p^2)(a \ln (p^2/\Lambda^2) + b).
\]
When this is hit by the $\Lambda$ derivative---which we can move 
outside the integral---only the 
first term will survive and so we will be left
with a (universal) coefficient multiplying two powers of $p$.

That the final answer is Taylor expandable in $p$ gives us an alternative way in which to 
evaluate $\Lambda$-derivative terms.
Having moved the $\Lambda$-derivative outside the integral,
we Taylor expand the denominator in $p$. Doing this, $p$ will no longer act as a regulator
and so we will then generate IR divergences, when we perform the integral. 
However, all divergences will 
be killed by the $\Lambda$-derivative. 
To parameterise these pseudo-divergences, we must introduce an IR regulator; 
it is natural to use  dimensional regularisation.

It may, at this stage, seem a little perverse to have traded one IR regulator, $p$---which
occurs naturally---for
another. However, even for diagrams which are not Taylor expandable in $p$, it will
turn out that we are often interested in the Taylor expandable part. By Taylor expanding
in $p$, the resulting integrals tend to be easier to perform. 

We now discuss how this
procedure works, in more detail:
\begin{enumerate}
\item Take $\mathcal{O}$(mom) from each of the vertices;

\item Taylor expand the denominator in $p$;

\item replace the upper limit of the radial integral with $\Lambda$, thereby cutting off modes above this scale;

\item rescale $k \rightarrow k/\Lambda$, so that the diagram acquires an overall factor of $\Lambda^{-2\epsilon}$.
\end{enumerate}

Having done the angular integral, we are left with an expression of the form:
\[
 \frac{(\Lambda^{-2\epsilon})^\bullet}{(4\pi)^{D/2}} \int_0^1 \frac{k^{D-1}}{k^4}dk  \, \mathcal{O}(p^2).
\]
Performing the integral gives us a factor of $1/\epsilon$, as expected. This is killed by a factor of $\epsilon$
arising from differentiation \wrt\ $\LConstAl$, confirming the consistency of the approach.

Before moving on, we must justify the validity of the third step. 
We know that the integral we are dealing
with has support only in the IR. However, $\int_k 1/k^4$ is not 
UV regularised and so we must incorporate the 
effects of the UV regularisation. Since the details of the 
regularisation will not affect the IR, at leading
order, we choose the simplest form that cuts off momentum modes 
above the scale $\Lambda$. The non-universal
corrections to this will necessarily remove any IR divergence, 
even before differentiation \wrt\ $\LConstAl$,
 and so vanish in the limit that $\epsilon \rightarrow 0$.
An implicit part of this step is that we now throw away all 
$F$-sector diagrams and evaluate any cutoff 
functions at zero momentum.

In readiness for the two-loop calculation, we will now extend 
our analysis of the standard set, and will give its general
form. We know the following facts about the standard set:
\begin{enumerate}
	\item 	the sum of the diagrams is transverse;

	\item 	when struck by $\flowConstAl$, the coefficient of the $\Op{2}$ 
			term is universal,
			up to \Oep\ corrections (this follows from~\cite{aprop});

	\item 	in $D=4$, the diagrams have the structure
			\[
			\Op{2} \ln (p^2/\Lambda^2) + \Op{2} +\cdots,
			\]
			where the ellipsis denotes terms which are higher order in $p$.
\end{enumerate}
From these three points and dimensions it follows that, in $D=4-2\epsilon$, 
the standard set takes the algebraic form
\[
\frac{N\Lambda^{-2\epsilon}}{(4\pi)^{D/2}} \left[ a_0 \left( 1 - \frac{p^{-2\epsilon}}{\Lambda^{-2\epsilon}} \right) 
\frac{1}{\epsilon} + \ldots \right]
\Box_{\alpha \beta}(p) + \order{p^4, p^{4-2\epsilon}},
\]
where $a_0$ is a universal coefficient and the ellipsis 
denotes terms which are higher order in $\epsilon$.

We have no further interest in the $\order{p^4, p^{4-2\epsilon}}$ 
terms and so turn to the ellipsis.
There are two ways in which we expect these terms to arise. On the one hand, 
compared to eqn.~\eq{eq:TLD:GeneralIRStructure},
we can take additional 
powers of $k$ in the numerator of the integrand, 
giving us non-universal contributions which are Taylor expandable in $p$.
On the other hand, we generically expect the coefficient 
$a_0$ to have arisen from some function of $D$ in which we have
taken $D=4$. Expanding this function in $\epsilon$ will
give rise to sub-leading contributions.
These contributions will be called computable. 
At the two-loop level, we will see how computable
contributions can combine to give universal quantities.

Just as we talk of computable parts of
some diagram, so too will we talk of the complimentary
non-computable parts. We emphasise that by non-computable
we really mean that the corresponding coefficients
cannot be computed without specifying non-universal details
of our set-up (\ie cut-off functions and the precise form 
of the covariantisation); it is not that we cannot, in principle,
calculate them.

Hence, the standard set takes the following form
\begin{equation}
\frac{N\Lambda^{-2\epsilon}}{(4\pi)^{D/2}} \left[ 
\sum_{i=0}^\infty \left(a_i + b_i\frac{p^{-2\epsilon}}{\Lambda^{-2\epsilon}} \right)\epsilon^{i-1} \right]
\Box_{\alpha \beta}(p) + \order{p^4, p^{4-2\epsilon}}, 
\label{eq:TLD:StandardSetGeneralForm}
\end{equation}
where $b_0 = -a_0$,  
the $a_{i>0}$ are a mixture of computable and non-computable 
contributions and the $b_i$ are entirely computable.

Notice that one-loop computations are insensitive to the 
$p^{-2\epsilon}/\Lambda^{-2\epsilon}$ terms: taking
into account the additional factor of $\Lambda^{-2\epsilon}$ 
sitting outside, such terms are independent of $\Lambda$
and so will be killed by the $\Lambda$-derivative. At two-loops, 
where the standard set can occur
as a sub-diagram, we expect such contributions to survive.

We now discuss how to calculate the coefficients $b_i$. 
It is convenient to begin by contracting the standard set with
$\delta_{\alpha \beta}$. The $b_i$ arise from integrals of the form
\[
	\int_k \frac{\order{p^2, p.k, k^2}}{k^2(k-p)^2}.
\]
Note that the $\order{k^2}$ term does not contribute to the $b_i$: the denominator
becomes just $1/(k-p)^2$ and so, by shifting momentum, we can remove $p$
from the denominator entirely. It is now not possible to 
generate a power of $p^{-2\epsilon}$.

The next step is to combine denominators, using Feynman parameterisation:
\[
\int_0^1 da h(a) \int_k \frac{\mathcal{O}(p^2)}{(k^2 + K^2)^2}, 
\]
where $K^2 = a(1-a)p^2$ and $h(a)$ is some function of $a$. At this stage, 
it is now tempting to proceed as before and restrict the range of the
radial integral. However, this leaves us with an unpleasant calculation 
as we cannot use standard dimensional regularisation formulae. 
Besides, there is a much simpler way
to proceed:  differentiate twice, with respect to $p^2$.

The effect on the integral is to ensure that it is UV regularised by the 
denominator of the integrand---without the need
for any cutoff regularisation. We call this 
automatic UV regularisation (which will play an important \role\ at two loops). 
Since we are  interested in the part of the integral which has support in the IR, 
there is no need for us to  restrict the range of integration, as 
doing so would only serve to
make the calculation harder.
Retaining just A-sector diagrams, we evaluate all cutoff 
functions at zero momentum, leaving us
with an integral we can do using standard dimensional 
regularisation techniques.

We perform the integral and compare it to the second derivative 
\wrt\ $p^2$ of eqn.~\eq{eq:TLD:StandardSetGeneralForm},
contracted with $\delta_{\alpha \beta}$. This is one place 
where the value of keeping the standard set together manifests itself:
because we know the standard set to be transverse, we know the 
effect of contracting with $\delta_{\alpha \beta}$.
Equating powers of $\epsilon$ allows us to determine the $b_i$. 
The first two coefficients, computed in dimensional
regularisation are:
\begin{eqnarray}
b_0 & = & -20/3   \label{eq:SS:b_0} \\
b_1 & = & -124/9 + 20 \EM /3 \label{eq:SS:b_1}
\end{eqnarray}
where $\EM$ is the Euler-Mascheroni constant. Note that $b_0$
is not just computable but also universal, being independent
of the way in which we choose to evaluate it~\cite{aprop}.

At this point it is worth pausing to consider why the coefficients
$a_0$ and $b_i$ have no dependence on $N$. We note in 
eqn.~\eq{eq:TLD:StandardSetGeneralForm} that we have extracted
an overall factor of $N$, but we might suspect that the $a_i$ and $b_i$
incorporate attachment corrections.\footnote{The overall
factor of $N$ is what we expect for the diagrams if there are no attachment 
corrections. In this case, each diagram comprises one loop
decorated by the two external fields and one empty loop. The empty loop
yields $\str \sigma_+ = N$ if the internal fields are
bosonic and $\str \sigma_- = -N$ if the internal fields are fermionic.}
 Let us look
first at diagram~\ref{Beta1-LdL-C}. Since this is formed by
the action of gauge remainders on three-point (tree level) vertices
decorated by an external field, we know from~\cite{mgierg1,Thesis} 
that we can discard all $1/N$ corrections. 

Now consider diagram~\ref{Beta1-LdL-B}. The contributions to $a_0$
and $b_i$ come when all fields are in the $A^1$-sector. If either
of the effective propagators attaches via a $1/N$ correction,  
then the external fields are always guaranteed
to be on the same supertrace, irrespective of location: the diagram vanishes by \CC,
since a three-point $A^i$ vertex changes sign under the interchange
of any pair of fields.
However, we expect the $a_{i\geq1}$, to be non-trivial functions of $1/N$,
since these coefficients receive contributions in which the vertices
of diagram~\ref{Beta1-LdL-B} are each \CC\ even (\ie $AAC$ vertices)
and from diagram~\ref{Beta1-LdL-A}.

We conclude our analysis of the standard set by
noting a beautiful interplay between the IR and the UV, illustrated 
by diagram~\ref{Beta1-LdL-C}. The strategy we have used
is to pull the $\Lambda$-derivative outside of the momentum 
integral and then focus on the IR end. 
Focusing on the IR end allows us
to throw away the regularising diagram. However, the regularising
diagram was required to
define the $A^1$-sector diagram when we interchanged the
order of differentiation \wrt\ $\Lambda$ and loop integration.

Now, suppose that we 
had left the $\Lambda$-derivative inside the integral. Then 
the $A^1$-sector diagram actually dies, since $A^1$-sector
gauge remainders are independent 
of $\Lambda$. We are left with the $B$-sector diagram, which 
provides the same leading order contribution as the $A^1$-sector 
diagram, but arising from the UV!

Interplay such as this will only arise when the components
of some diagram which gives a contribution in the IR are not
regularised by cutoff functions, alone (\cf~\cite{giqed}).

\subsubsection{Loop Integrals Independent of $p$}	\label{sec:TDM:Indep-p}

We conclude our survey of one-loop integrals by looking at the final 
diagram which contributes to $\beta_1$, \ref{Beta1-LdL-Fc}.
The difference between this diagram and the ones just analysed is that
$p$ is not involved in the IR regularisation. The integral just goes like
\[
\int_k \frac{1}{k^4}
\]
in the IR and we can use the techniques of the previous section to 
evaluate such terms. We thus expect the complete set of 
diagrams contributing to $\beta_1$ to take the following form, 
before differentiation \wrt\ $\LConstAl$:
\begin{equation}
\frac{N\Lambda^{-2\epsilon}}{(4\pi)^{D/2}} \left[
\sum_{i=0}^\infty \left( A_{i} + B_{i}\frac{p^{-2\epsilon}}{\Lambda^{-2\epsilon}} \right)\epsilon^{i-1} \right]
\Box_{\mu \nu}(p) + \order{p^4, p^{4-2\epsilon}}, 
\label{eq:TLD:1-loopGeneralForm}
\end{equation}
where the $B_i$ are computable and the $A_i$ generally 
contain both computable and non-computable parts. The universal coefficient $A_0$
yields the sole contribution to $\beta_1$, in the $\epsilon \rightarrow 0$ limit.

\subsection{Two-loop Integrals} 
\label{sec:TLD:Methodology:2-loop}

In this section, we develop the machinery of the previous 
section to deal with two-loop $\Lambda$-derivative terms.
The integrals we have to deal with fall into two classes: 
factorisable and non-factorisable.
In the former case---dealt with in 
\sec{sec:TLD:Methodology:Factorisable}---the loop-integrals 
are independent, whereas, in the 
latter case---dealt with in \sec{sec:TLD:Methodology:NFactorisable}---they are not.

Following on from the one-loop case we expect and, indeed,
find~\cite{Thesis} that we can
write\footnote{In the limit that $\alpha \rightarrow 0$.}
\[
	4\beta_2 \Box_{\mu \nu} (p) = -\frac{1}{2} \dec{\TLDs(p)}{\bullet}.
\]
One of the main sources of complication in the two-loop case is that, even
after differentiation \wrt\ $\flowConstAl$, individual elements of $\TLDs(p)$
can still possess IR divergences. It is only the sum of diagrams contributing
to $\TLDs(p)$ that we expect to give a finite (universal) contribution after
differentiation \wrt\ $\flowConstAl$.

Since we are interested in two-loop integrals which contribute to $\beta_2$ we will,
up to factors of $\Pep$, work at $\Op{2}$.

\subsubsection{The Factorisable Case} 
\label{sec:TLD:Methodology:Factorisable}

To understand the algebraic form of two-loop diagrams, 
we need first to understand their
structure. Since we are dealing with factorisable terms, 
we expect them to comprise two one-loop sub-diagrams, each of 
which carries external momentum $p$. These sub-diagrams must
be connected to each other, and so we predict that they 
will be joined together by an effective propagator
(for explicit examples, see \sec{sec:bulk}). This effective propagator
just contributes powers of the external momentum and so we take 
the general form of a factorisable two-loop integral to be:
\begin{equation}
\frac{N^2\Lambda^{-4\epsilon}}{(4\pi)^D} \sum_{i=0}^\infty \left(c_i + d_i \frac{p^{-2\epsilon}}{\Lambda^{-2\epsilon}}
+ e_i \frac{p^{-4\epsilon}}{\Lambda^{-4\epsilon}} \right) \frac{1}{\epsilon^{i-2}}\Op{2},
\label{eq:TLD:TwoLoopGenericForm}
\end{equation}
where we obtain a power of $p^{-2\epsilon}/\Lambda^{-2\epsilon}$ for 
each loop in which $p$ provides IR regularisation.

\subsubsection{The Non-Factorisable Case} 
\label{sec:TLD:Methodology:NFactorisable}

We expect  non-factorisable diagrams to possess the same
ingredients as factorisable ones, but joined together in a different way. Taking the loop momenta to be $l$
and $k$, we know that one of the effective propagators must $\sim 1/(l-k)^2$ (assuming
it to be in the $A^1$-sector).
Conservation of four-momentum at a vertex then implies that there must be at least one
effective propagator carrying $l$ and at least one carrying $k$.
Knowing that, at $\Op{2}$ the integrand must be of mass dimension $-8$, we
expect the most divergent type of diagram we can
construct (assuming no IR regularisation is provided by $p$) to take the following
form in the IR:
\[
\Op{2}\int_{l,k} \frac{\mathcal{O}(\mathrm{mom}^2)}{k^2 (l-k)^2 l^6}.
\]
In fact, we will see in \sec{sec:TLD:Subs} 
that, for the set of diagrams contributing to
$\beta_2$, gauge invariance prevents the appearance 
of $1/l^6$ and so we would find that,
taking the above form, we would be forced to have $l^2$ in the numerator. 
Hence, the most divergent type of integral we find has the following structure in the IR:
\begin{equation}
\Op{2}\int_{l,k} \frac{1}{k^2 (l-k)^2 l^4} 
\label{eq:TLD:StandardTwoLoopIntegral}.
\end{equation}
To evaluate the contribution coming from the IR, we observe 
that the $l$-integral is automatically UV regularised. Thus, using dimensional
regularisation,
we perform the $l$-integral first, with unrestricted
range of integration, and perform the $k$-integral second, 
with the range of radial integration restricted to $\Lambda$. We obtain one power of
$1/\epsilon$ from the Feynman parameter integral and a 
further power from the radial $k$-integral. 
Doing the integrals the other way around would be awkward, as we 
cannot then use an un-restricted range of integration for the inner integral.

Had there been $p$s present in the denominators, providing
regularisation, we would expect accompanying factors 
of $p^{-2\epsilon}/\Lambda^{-2\epsilon}$.
Consequently,  eqn.~\eq{eq:TLD:TwoLoopGenericForm}
is the form for a generic two-loop integral.

\subsubsection{Considerations for $\beta_2$} \label{sec:TLD:Methodology:beta2}

For the actual computation of $\beta_2$, we can
constrain some of the coefficients in eqn.~\eq{eq:TLD:TwoLoopGenericForm}. Since the terms corresponding
 to the coefficients $e_i$ are independent of $\Lambda$, we
can drop them as they will vanish after differentiation.
Given that we are comparing our final answer with the Taylor expandable, finite expression
$
\beta_2 \Box_{\mu \nu} (p),
$
it must be that the coefficients $c_0$ and $d_i$ vanish. 
The coefficients $c_0$ and $d_0$ are entirely computable. The coefficient $d_1$,
on the other hand, comprises both a purely computable part and a non-computable part
multiplied by a computable coefficient. 
The computable part and the computable coefficient must both be zero.

Ultimately, we will be left with the coefficient $c_1$ being the
only contribution to the final answer. One of the primary tasks ahead is to show that the non-universal contributions to
$c_1$ cancel between diagrams.
This problem has two sides. First, we 
must show that non-computable contributions from vertices \etc cancel out. Then
we must show that the computable contributions to $c_1$ combine to give 
the standard, universal answer.

\subsection{Subtraction Techniques} 
\label{sec:TLD:Subtractions}

\subsubsection{Basics}	\label{sec:TLD:Subtractions:Basics}

Rather than attempting to process two-loop diagrams directly, 
we perform an intermediate step whereby we 
add and subtract a set of terms designed to remove all 
non-computable contributions from the calculation.
We illustrate this technique with a simple example. 
Consider the two-loop integral arising
from the computation of the scalar two-loop $\beta$-function, 
within the ERG of~\cite{two}.
\be
	\int_{l,k} 
	\dec{
		\Delta^2_l\Delta_{l-k}\Delta_k - \frac{1}{2}\Delta^2_l\Delta^2_k 
	}{\bullet},
\label{eq:Scalarb2}
\ee
where $\Delta_l = c(l)/l^2$. We can trivially rewrite this as
\[
	\int_{l,k} 
	\dec{
		\Delta^2_l\Delta_{l-k}\Delta_k - \Delta^2_l\Delta^2_k + \Delta^2_l\Delta^2_k - \frac{1}{2}\Delta^2_l\Delta^2_k 
	}{\bullet},
\]
where we call the second term a subtraction and the 
third term its corresponding addition. The addition
trivially combines with the final term, 
though this is specific to this example
and of no particular significance. In deference to
the forthcoming gauge theory calculation, we will
leave the third and fourth terms uncombined, so that
we can illustrate the strategy generally employed.

Let us start by focusing 
on the $k$ integral, in the first term. Following~\cite{two,Bonini},
we know that we can set $c(l-k) = c(k)$, as contributions higher order in $l$ are
killed in the \eptoz\ limit.
Next, use the by now familiar prescription for the cutoff 
functions: if the integral is regularised without the cutoff
functions, then we simply evaluate them at zero momentum, 
leaving the domain of integration unrestricted. 
If the integral requires the cutoff functions for 
regularisation, then restrict the
domain of integration and evaluate the cutoff
function at zero momentum. The sub-leading
corrections to this will manifest themselves
as additional powers of momenta, in the numerator.

If we were to start
taking such sub-leading (non-computable) contributions under 
the $k$-integral, then this will allow us to
 Taylor expand the $k$-integral in $l$. 
For the $l$-integral to still diverge in the IR---and
thus survive differentiation \wrt\ $\LConstAl$ (in the limit that $\epsilon 
\rightarrow 0$)---we must take $l^0$. However, such 
non-computable contributions will be cancelled by exactly the same 
contributions coming from the second term, above. 
Hence we have:
\[
\int_{l,k} \left[ \Delta^2_l(\Delta_{l-k}\Delta_k)_\COMP - \Delta^2_l(\Delta^2_k)_\COMP + \Delta^2_l\Delta^2_k -  \frac{1}{2}\Delta^2_l\Delta^2_k \right]^\bullet
+\Oep,
\]
where $\COMP$ tells us that, when considered as a pair, the first two $k$-integrals yield a computable contribution.

For the first term not to die in the $\epsilon 
\rightarrow 0$ limit, we must set $c(l) \rightarrow 1$ and so can
 extend the `$\COMP$' to cover the whole term.
The second term, derived from the subtraction, can now
be combined with the addition. Generally, we will always
perform this step: it isolates
non-computable contributions of the first term in~\eq{eq:Scalarb2}
that survive in the $D \rightarrow 4$ limit.
We thus have:
\begin{equation}
	\int_{l,k}
	\dec{
		\left(
			\Delta^2_l\Delta_{l-k}\Delta_k
		\right)_\COMP 
		+ \Delta_l^2 \left( \Delta_k^2 \right)_\NUn - \hf \Delta_l^2 \Delta_k^2
	}{\bullet} +\Oep.
\label{eq:TLD:SubtractionSet2}
\end{equation}

The final step is to re-express each of the components of the last
term as $\COMP + \NUn$ and multiply out, collecting terms by using
the freedom to interchange $l$ and $k$. Noting that the
$\left( \Delta_l^2 \right)_\NUn \left(\Delta_k^2\right)_\NUn$ term
vanishes in the $D\to 4$ limit we are left with 
a purely computable contribution:
\[
	\int_{l,k} 
	\left[
		\Delta^2_l\Delta_{l-k}\Delta_k - \frac{1}{2}\Delta^2_l\Delta^2_k 
	\right]^\bullet_\COMP + \Oep.
\]
We have thus demonstrated that the original integral does, indeed, 
give something which is computable.

To calculate this integral, we can use a mixture of the 
techniques already discussed. In the factorisable case,
we simply restrict the ranges of the integrals. In the 
non-factorisable case, we note that the $l$-integral is automatically
UV regularised. Hence, we perform this integral first 
with an unrestricted domain of integration but then restrict the domain
for the remaining $k$-integral. It is straightforward to confirm 
that we reproduce the expected, universal answer
\[
	-\frac{17}{3} \frac{1}{(4 \pi)^4}.
\]

\subsubsection{Generalisation to the Gauge Case} 
\label{sec:TLD:Subtractions:Gauge}

Constructing subtractions in the gauge case is exactly 
analogous to the simpler case just analysed, but
the structure of the cancellations
of non-computable contributions is much richer. In the same way,
the subtractions are constructed such as to isolate  
non-computable contributions, by noting that denominators can be
Taylor expanded in momenta if sufficient powers 
of momenta are present in the numerator.

As the whole formalism is based around Taylor expansion, 
it is not surprising that we will need to use the techniques of
\sec{sec:Taylor} to Taylor expand 
vertices in momenta. We know that the lowest order terms constitute
momentum
derivatives of lower point vertices, and that the sign of this 
derivative depends on whether we have had  to push forward or
pull back. We define the subtraction to be the term which 
removes non-computable components from the parent diagram, and not by its sign.
Hence, a subtraction involving a pull back will come with a positive sign.

The real subtleties in the gauge case 
arise because the procedure of
constructing subtractions and additions
and the subsequent cancellation of all non-computable
contributions leads, generically, to a complete loss of momentum
routing invariance. As we will see, this
invariance can be partially restored, though
it is not necessarily efficient to do so.

As a first illustration, 
suppose that we have taken a non-factorisable,
two-loop diagram and constructed a factorisable
subtraction. Putting the addition to one side
for the moment, we will suppose that the sub-diagram of this latter term, 
to which we apply $\COMP$, is just the diagram shown
in \fig{fig:SubDiagram}.
\bcf[h]
	\[
		\LDi{Beta1-LdL-F}{Beta1-LdL-F}
	\]
\caption{A sub-diagram, to which we will apply $\COMP$, of some factorisable
subtraction.}
\label{fig:SubDiagram}
\ecf

The algebraic form of diagram~\ref{Beta1-LdL-F} is
\begin{equation}
\Box_{\mu \beta}(p) (\Lambda^{-2\epsilon})^\bullet \int_k \partial_\alpha^k \left[\frac{\NewCO_k}{k^2} \right] \frac{k_\beta}{k^2},
\label{eq:TLD:keepcutoff}
\end{equation}
where $\NewCO_k = \InvCO_k^{-1}$ (\cf eqn.~\eq{eq:stub}).
To compute the part of this diagram left over, after combining 
with our non-factorisable
term we do the following: evaluate the cutoff function at
zero momentum, restrict the range
of the integral, and proceed as usual.

Next, suppose that we were to move the momentum derivative from the $\NewCO_k/k^2$ term to 
the $k_\beta/k^2$ term, throwing away the total derivative, 
in the process. This yields
\begin{equation}
-\Box_{\mu \beta}(p) (\Lambda^{-2\epsilon})^\bullet \int_k \frac{\NewCO_k}{k^2}  \partial_\alpha^k \left[\frac{k_\beta}{k^2}\right] .
\label{eq:TLD:keepcutoff-B}
\end{equation}

Evaluating this integral in the usual way gives a different 
contribution, at sub-leading order,
than the integral of eqn.~\eq{eq:TLD:keepcutoff}. What is going on?
The point is, that whilst going from eqn.~\eq{eq:TLD:keepcutoff}
to eqn.~\eq{eq:TLD:keepcutoff-B} is usually a perfectly valid step, 
it breaks down when these
terms are under the influence of $\COMP$. Specifically, 
because the effect of $\COMP$ has been
to replace cutoff functions with a restricted range of integration,
we are no longer justified in throwing away what would previously have been total
momentum derivative terms. Equivalently, we have lost the freedom
to shift $k$, under the integral.

If we wish, momentum routing invariance can be retained, in this
case, by modifying what we mean by $\COMP$. We can use
the following prescription:
reinstate a term
to both the parent and subtraction (after they are under the influence of $\COMP$)
such that, at sub-leading order, we can move
momentum derivatives around with impunity. To understand what this term must be,
let us return to eqn.~\eq{eq:TLD:keepcutoff}. 
Rather than discarding the cutoff function
straight away, we will first allow the momentum derivative to act.
 
Doing so, averaging over angle and substituting
$x = k^2$ yields:
\[
\frac{N \! \AngVol{D}}{D} \Box_{\mu \alpha}(p) (\Lambda^{-2\epsilon})^\bullet \int dx x^{1-\epsilon}
\left(\frac{\NewCO'_x}{x} - \frac{\NewCO_x}{x^2} \right),
\]
where $\AngVol{D}$ is the area of the unit sphere is $D$ dimensions
divided by $(2\pi)^D$.
Integrating the first term by parts, and discarding the resulting surface term gives:
\[
\frac{N \! \AngVol{D}}{D} \Box_{\mu \alpha}(p) \int dx \frac{\NewCO_x}{x^{1+\epsilon}}(\epsilon-1).
\]
\emph{Now} if we remove the cutoff function and restrict the range of integration, 
it is apparent that the $\NewCO'$ term has provided a sub-leading contribution;
indeed, this is precisely the sub-leading contribution we are after!

We have seen how, when cutoff functions are necessary for UV regularisation,
their derivatives can supply sub-leading contributions in the IR. This can be
rephrased by saying that, under the influence of $\COMP$, total momentum derivative
contributions can no longer be discarded, at sub-leading order, 
unless we reinstate terms to parent and subtraction.

When dealing with
automatically regularised integrals, however, total momentum 
derivatives can be thrown away,
even under the influence of $\COMP$. This follows because the 
range of integration need not be restricted,
even after we have evaluated any cutoff functions at zero momentum. Equivalently,
in this case, we can move momentum derivatives around 
without the need to reinstate the derivative
of cutoff functions.

Had we combined the parent, addition and subtraction as
in the scalar example, similar considerations to those
above would apply, as follows directly from
the complementarity of $\COMP$ and $\NUn$.

Unfortunately, reinstating cutoff functions in this way
can create more problems than it solves. To understand why,
let us consider a second example, which we will
encounter in the evaluation of $\beta_2$. Our starting point
is the diagram on the \lhs\ of \fig{fig:SubEx}. On the \rhs\
is an exactly equivalent representation: we have simply taken
two copies of the diagram and reflected one of them using
\CC\ invariance~\cite{Thesis,mgierg1}. In readiness for
the construction of the subtractions, we have chosen
differing momentum routings for the two diagrams. (Note, though,
that at this stage
the two diagrams are exactly equivalent, 
since complete momentum routing invariance has not been
broken.) 
\bcf[h]
	\[
	\dec{
		\ensuremath{\begin{array}{c}\input{pstex/Diagram28.1.pstex_t} \end{array}}
	}{\bullet}
	\equiv
	\hf
	\dec{
		\begin{array}{ccc}
			\LD{ED:28.1a}		&	& \LD{ED:28.1b}
		\\[1ex]
			\ensuremath{\begin{array}{c}\input{pstex/Diagram28.1a.pstex_t} \end{array}}	& +	& \ensuremath{\begin{array}{c}\input{pstex/Diagram28.1b.pstex_t} \end{array}}
		\end{array}
	}{\bullet}
	\]
\caption{A diagram for which subtractions will be constructed.}
\label{fig:SubEx}
\ecf

As a consequence of our choices of momentum routing,
diagrams~\ref{ED:28.1a} and~\ref{ED:28.1b} will
have different subtractions. We do this for
convenience, as the subset of terms thus generated can
be manifestly combined into something useful. 
The subtractions are shown in \fig{fig:D28Ss},
where we have used \CC\ to collect terms;
each diagram possesses two labels: the top one for
the subtraction and the bottom one for the corresponding
addition.
\bcf[h]
	\[
	\dec{
		\begin{array}{cccccccc}
				& \LDLD{ED:28.1a-s1}{ED:28.1a-a1}	&		& \LDLD{ED:28.1a-s2}{ED:28.1a-a2}	& 	&\LDLD{ED:28.1a-s3}{ED:28.1a-a3}	&
		\\[1ex]
			\pm	& \ensuremath{\begin{array}{c}\input{pstex/Diagram28.1a-s1.pstex_t} \end{array}}				&\mp2	& \ensuremath{\begin{array}{c}\input{pstex/Diagram28.1a-s2.pstex_t} \end{array}} 				&\mp& \ensuremath{\begin{array}{c}\input{pstex/Diagram28.1a-s3.pstex_t} \end{array}}				&
		\\[1ex]
				& \LDLD{ED:28.1b-s1}{ED:28.1b-a1}	&		& \LDLD{ED:28.1b-s2}{ED:28.1b-a2}	& 	&\LDLD{ED:28.1b-s3}{ED:28.1b-a3}	&		&\LDLD{ED:28.1b-s4}{ED:28.1b-a4}
		\\[1ex]
			\mp	& \ensuremath{\begin{array}{c}\input{pstex/Diagram28.1b-s1.pstex_t} \end{array}}				&\pm2	& \ensuremath{\begin{array}{c}\input{pstex/Diagram28.1b-s2.pstex_t} \end{array}} 				&\pm& \ensuremath{\begin{array}{c}\input{pstex/Diagram28.1b-s3.pstex_t} \end{array}}				&\mp2	&\ensuremath{\begin{array}{c}\input{pstex/Diagram28.1b-s4.pstex_t} \end{array}}
		\end{array}
	}{\bullet}
	\]
\caption{Subtractions for diagrams~\ref{ED:28.1a} and~\ref{ED:28.1b}.}
\label{fig:D28Ss}
\ecf

Note that, in diagrams~\ref{ED:28.1a-s3}, \ref{ED:28.1b-s3} and~\ref{ED:28.1b-s4}
(and the corresponding additions), the field attached to
the circle representing the momentum derivative is implicitly
in the $A^1$-sector. This follows because the only
derivatives of vertices we consider are those which arise from a field 
in the $A^i$-sector
carrying zero momentum (and here, the $A^i$ can only be an $A^1$).

To begin the analysis of diagram~\ref{ED:28.1a} 
and its subtractions (\ref{ED:28.1a-s1},
\ref{ED:28.1a-s2} and~\ref{ED:28.1a-s3}), we focus on the 
three-legged sub-diagram carrying loop momentum $k$ (the momentum
routing of subtractions follows from the parent). 
For the diagram as a whole to have any chance of
surviving in the $\epsilon \rightarrow 0$ limit, the 
(internal) legs leaving the sub-diagram must be in
the $A^1$-sector. Thus, by Lorentz invariance, the sub-diagram 
carrying loop momentum $k$ must go as odd powers of momentum
(up to additional non-Taylor expandable functions of $l$).
Noting that, in $D=4$, the  sub-diagram carrying 
loop momentum $k$ goes as, at worst, 
$(\ln l)\times \mathcal{O}(\mathrm{mom}, \ldots)$ in the IR,
it is clear that we must take only the $\mathcal{O}$(mom) 
part of the sub-diagram.

The effect of diagrams~\ref{ED:28.1a-s1} and~\ref{ED:28.1a-s2} 
is now immediately clear: they completely remove from
diagram~\ref{ED:28.1a} all contributions in which the  
sub-diagram carrying loop momentum $k$ goes as $l$.

Let us now suppose that we take $\mathcal{O}(p)$ from 
the sub-diagram carrying loop momentum $k$. We start by noting that the only
place for this power of $p$ to come from is the 
four-point vertex. Now, if all the fields leaving the 
four-point vertex are in the $A^1$-sector,
then Lorentz invariance forces us to take an additional 
power of momentum from the four-point vertex. Recalling 
that we should not take any 
further powers of $l$ or $p$, we see that we must take 
(at least) one power of $k$ from the four-point vertex (and a  further 
power of $k$ from the three-point vertex).
The $k$-integral is now Taylor expandable in $l$. 
In the case that the $k$-dependent fields leaving the four-point vertex are not in 
the $A^1$-sector, the $k$-integral is trivially Taylor expandable in $l$.

The remaining subtraction, diagram~\ref{ED:28.1a-s3}, 
removes all these contributions; hence diagram~\ref{ED:28.1a}
turns out to be completely cancelled by its subtractions, (up to 
contributions which die in the $D \to 4$ limit).

The same cannot, however, be said for diagram~\ref{ED:28.1b}.
It is clear that whilst diagrams~\ref{ED:28.1b-s1} 
and~\ref{ED:28.1b-s2} remove from~\ref{ED:28.1b}
all contributions from the $k$-integral that are Taylor expandable in $p$,
these diagrams possess components which are not Taylor 
expandable in $p$ and so will survive.
Diagrams~\ref{ED:28.1b-s3} and~\ref{ED:28.1b-s4} remove
all contributions that are Taylor expandable in $l$ 
(as always, this statement is correct only up to contributions that
vanish anyway in the $\epsilon \rightarrow 0$ limit). 
Non-computable contributions from the $k$-integral are precisely those
which are Taylor expandable in $l$ and $p$ and so 
cancel between the parent diagram and its subtractions. 

That diagram~\ref{ED:28.1a}
is completely cancelled by its subtractions  is
in some senses a coincidence; it is certainly not generically
true that parent diagrams are thus cancelled. Indeed,
it is most efficient not to make use of this
fact. Thus, to proceed, we employ the strategy
outlined earlier: first combine parent with subtraction
and then combine the remaining components of the subtraction
with the corresponding addition. This yields the
diagrams of \fig{fig:Ex-NC}, up to $\Oep$ corrections.
\bcf[p]
	\[
	\hf
	\dec{
		\begin{array}{ccccc}		
			\LD{ED:28.1aU}			&	& \LD{ED:28.1a-NCa}				&	&\LD{ED:28.1a-NCb}
		\\[1ex]
			\Uni{\ensuremath{\begin{array}{c}\input{pstex/Diagram28.1a.pstex_t} \end{array}}}	&-2	&\NUni{\ensuremath{\begin{array}{c}\input{pstex/Diagram28.1a-s1.pstex_t} \end{array}}}	&+2	& \NUni{\ensuremath{\begin{array}{c}\input{pstex/Diagram28.1a-s2.pstex_t} \end{array}}}
		\\[1ex]
									&	&\LD{ED:28.1a-NCbi}				&	& \LD{ED:28.1a-NCc}
		\\[1ex]
									&+2	&\NUni{\ensuremath{\begin{array}{c}\input{pstex/Diagram28.1a-s2b.pstex_t} \end{array}}}	&+2	& \ensuremath{\begin{array}{c}\input{pstex/Diagram28.1a-s3NC.pstex_t} \end{array}}
		\\[2ex]					
			\LD{ED:28.1bU}			&	& \LD{ED:28.1b-NCa}				&	& \LD{ED:28.1b-NCb}
		\\[1ex]
			\Uni{\ensuremath{\begin{array}{c}\input{pstex/Diagram28.1b.pstex_t} \end{array}}}	&+2	&\NUni{\ensuremath{\begin{array}{c}\input{pstex/Diagram28.1b-s1.pstex_t} \end{array}}}	&-2	&\NUni{\ensuremath{\begin{array}{c}\input{pstex/Diagram28.1b-s2.pstex_t} \end{array}}}
		\\[1ex]
									&	& \LD{ED:28.1b-NCbi}			&	& \LD{ED:28.1b-NCc}			
		\\[1ex]
									&-2	&\NUni{\ensuremath{\begin{array}{c}\input{pstex/Diagram28.1b-s2b.pstex_t} \end{array}}}	&-2	&\ensuremath{\begin{array}{c}\input{pstex/Diagram28.1b-s3SU.pstex_t} \end{array}}		
		\\[1ex]
									&	& \LD{ED:28.1b-NCd}				&	& \LD{ED:28.1b-NCdi}	
		\\[1ex]
									&+2	& \ensuremath{\begin{array}{c}\input{pstex/Diagram28.1b-s4SU.pstex_t} \end{array}}		&+2	& \ensuremath{\begin{array}{c}\input{pstex/Diagram28.1b-s4SUb.pstex_t} \end{array}}
		\end{array}
	}{\bullet}
	\]
\caption{Result of combining diagrams~\ref{ED:28.1a} and~\ref{ED:28.1b}
with diagrams~\ref{ED:28.1a-s1}--\ref{ED:28.1b-a4}.}
\label{fig:Ex-NC}
\ecf

There are a number of important comments to make
about the diagrams of \fig{fig:Ex-NC}.
First, great care must be taken
interpreting the precise meaning of $\COMP$ and $\NUn$.
Consider evaluating diagram~\ref{ED:28.1aU}, for which
all fields must be in the $A^1$ sector.
It is most
convenient to make the $l$-integral the inner one, as this
is automatically UV regularised. Now, since we discard
all cutoff functions (and regularising diagrams),
the $k$-integral is no longer
invariant under shifts of $k$. This is crucially important.

Notice that the four-point vertex attaches to the
three-point vertex on its left via two effective propagators
carrying $k$ and $l-k$. We suppose that the dummy
indices associated with these effective propagators are
$\rho$ and $\sigma$, respectively, and that the other
index of the three-point vertex is $\alpha$.
Recall that the diagrammatic representation used for
the three-point vertex actually stands for all independent,
cyclically ordered, three-point vertices. Since three-point
$A^i$ vertices only exist as single supertrace terms,
there are two independent forms for our three-point vertex:
\[
	S^{1\, 1\, 1}_{\alpha \sigma \rho}(l,k-l,-k) \ \mbox{and}\  S^{1\, 1\, 1}_{\alpha \rho \sigma}(l,-k,k-l).
\]

Na\"ively, it should be the case that the value of
the diagram is the same in both cases; after all,
going from one cyclically ordered three-point
to the other is equivalent to relabelling
the (internal) effective propagators. However, this relabelling
is achieved via the shift in momentum $k \rightarrow -k+l$
and so we know that the value of the diagram, under
the influence of $\COMP$,  will actually be different
for each of the cyclically ordered three-point vertices.
This explains why diagrams~\ref{ED:28.1a-NCb} and~\ref{ED:28.1a-NCbi}
have been kept separate, despite the fact that these diagrams
could be directly combined if they were not under the influence
of $\NUn$ (indeed, precisely such a combination
was formed when constructing
the subtraction~\ref{ED:28.1a-s2}).

We could remove the subtleties associated
with a lack of momentum rerouting invariance, order by order in $\epsilon$,  
by re-instating the necessary
UV regularisation in the $k$ integral
for every diagram of \fig{fig:Ex-NC}.
This amounts to a different prescription for what we
mean by $\COMP$ and will generally yield
different answers for computable diagrams. 
Of course, it is not possible to unambiguously
extract the computable component of a
non-universal diagram;
only a universal
quantity formed from a sum of computable components
is independent of the computational scheme.
However, making this change to the prescription for $\COMP$ is
not generally a good idea, as it interferes with a
crucial diagrammatic step. Suppose for a moment
that the top-most sub-diagram of~\ref{ED:28.1a-NCc}
were not under the influence of $\NUn$. Then this
diagram could be immediately processed using the
effective propagator relation:
\be
	\ensuremath{\begin{array}{c}\begin{picture}(0,0)%
\includegraphics{pstex/EP-dV-EP.pstex}%
\end{picture}%
\setlength{\unitlength}{3947sp}%
\begingroup\makeatletter\ifx\SetFigFont\undefined%
\gdef\SetFigFont#1#2#3#4#5{%
  \reset@font\fontsize{#1}{#2pt}%
  \fontfamily{#3}\fontseries{#4}\fontshape{#5}%
  \selectfont}%
\fi\endgroup%
\begin{picture}(565,596)(753,40)
\put(835,352){\makebox(0,0)[lb]{\smash{\SetFigFont{10}{12.0}{\rmdefault}{\mddefault}{\updefault}{\color[rgb]{0,0,0}0}%
}}}
\end{picture}
 \end{array}} = - \cdeps{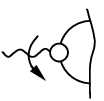} + \cdeps{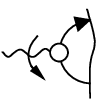} - \cdeps{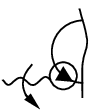}.
\label{eq:EP-dV-EP}
\ee
Can this manipulation be performed under the influence of $\NUn$?
To answer this, it is simplest to think of $\NUn$ acting on
a diagram as the full diagram, minus $\COMP$. Clearly, the
full diagram can be processed. Now, the effective propagator
relation holds if we take the lowest momentum
contribution of both the ($A^1$-sector) two-point,
tree level vertex and the corresponding effective propagator \ie
the relation works under $\COMP$ if
all cutoff functions have been thrown away.
Though by no means the only way to proceed,
it turns out that the most efficient is to completely
sacrifice momentum routing invariance, thereby allowing
the effective propagator relation 
to be straightforwardly applied. 
The processing of gauge remainders is unaffected
by $\COMP$ and hence $\NUn$. The only diagrammatic
step that cannot be made under the influence of
our current prescription for
$\COMP$, $\NUn$ is that of throwing away
total momentum derivative terms 
sitting under outer integrals. Note, though,
that such terms can always be reduced to computable
terms (at two loops). 
Consider a diagram containing a sub-diagram
both struck by a momentum derivative and under the
influence of $\NUn$.
Since a total
momentum derivative of a diagram not under
the influence of either $\COMP$ or $\NUn$ vanishes,
it follows that, in this case, the $\NUn$
can be replaced by $\COMP$, at the expense of
a minus sign. Now,
since such total momentum derivative
terms only contribute at sub-leading order in
$\epsilon$ then, in order for
the diagram as a whole to survive differentiation
\wrt\ $\LConstAl$ in the $\epsilon \to 0$ limit,
we must extend the effect of $\COMP$
to cover the entire diagram.

The second point to make about the
diagrams of \fig{fig:Ex-NC} is
that the sub-diagrams of
diagrams~\ref{ED:28.1a-NCa}--\ref{ED:28.1a-NCbi}
and~\ref{ED:28.1b-NCa}--\ref{ED:28.1b-NCbi} 	
carrying loop momentum $k$ 
can very nearly
be combined into momentum derivatives
of the second element of the 
standard set.\footnote{The external fields of
these sub-diagrams must be in the $A^1$-sector 
for the diagrams as a whole to survive in the $D \to 4$ limit.}
Had we not originally split the parent diagram
up, as in \fig{fig:SubEx}, this fact---which turns
out to be extremely useful in the evaluation of 
$\beta_2$---would not have been 
manifestly obvious. 

Let us examine diagrams~\ref{ED:28.1a-NCa}--\ref{ED:28.1a-NCbi}
more closely (an identical analysis can
be performed for~\ref{ED:28.1b-NCa}--\ref{ED:28.1b-NCbi}). 
From \sec{sec:Taylor}, the component
of diagram~\ref{ED:28.1a-NCa} in which the cyclically ordered
momentum arguments of the top-right vertex are $(k,l-k,-l)$ combines with 
diagram~\ref{ED:28.1a-NCb} such that, if the overall
sign of the diagram is taken to be positive, then the top-right
vertex is struck by
\be
	-\pmomder{-l}{\nu}{k} + \pmomder{k}{\nu}{l}.
\label{eq:momder1}
\ee
Similarly, the component
of diagram~\ref{ED:28.1a-NCa} in which the cyclically ordered
momentum arguments of the top-right vertex are $(l-k,k,-l)$ combines with 
diagram~\ref{ED:28.1a-NCb}  such that the top-right
vertex is struck by
\be
	-\pmomder{-l}{\nu}{l-k} + \pmomder{l-k}{\nu}{l} = \pmomder{l}{\nu}{k}.
\label{eq:momder2}
\ee

Eqns.~\eqs{eq:momder1}{eq:momder2} differ by the term $\pmomder{k}{\nu}{l}$.
If diagrams~\ref{ED:28.1a-NCa}--\ref{ED:28.1a-NCbi} were not under
the influence of $\NUn$, this apparent difference would vanish, as
a consequence of momentum rerouting invariance. In the current
case, it survives, of course, though it can be converted into an entirely
computable term, at the expense of the usual minus sign.

Diagrams~\ref{ED:28.1a-NCa}--\ref{ED:28.1a-NCbi}
and~\ref{ED:28.1b-NCa}--\ref{ED:28.1b-NCbi} 
naturally combine with the term 
shown in
\fig{fig:ExtraTerm}, which is
generated by the manipulation of a different parent 
diagram from that of
\fig{fig:SubEx} (this parent turns out to be diagram~\ref{diag:42.1},
which we will encounter in \sec{sec:bulk}, when we examine
the diagrams contributing to $\beta_2$). 
\bcf[h]
	\[
	2\LO{\NUni{\ensuremath{\begin{array}{c}\input{pstex/Diagram253.6.pstex_t} \end{array}} \hspace{1em}}}{d:253.6}
	\]
\caption{The partner term to diagrams~\ref{ED:28.1a-NCa}--\ref{ED:28.1a-NCbi}
and~\ref{ED:28.1b-NCa}--\ref{ED:28.1b-NCbi}.}
\label{fig:ExtraTerm}
\ecf

If it were the case that diagrams~\ref{ED:28.1a-NCa}--\ref{ED:28.1a-NCbi},
\ref{ED:28.1b-NCa}--\ref{ED:28.1b-NCbi} and~\ref{d:253.6}
were not under the influence of $\NUn$, then they could be combined
into a single term, in which the $k$-dependent sub-diagrams
sum up to give $\pmomder{l}{\nu}{k}$ of the second element
of the standard set. (Note that if the $\pmomder{l}{\nu}{k}$ 
strikes the effective propagator carrying $l-k$, then it
can be directly exchanged for $-\pmomder{k}{\nu}{l}$,
since the effective propagator depends only on $(l-k)^2$. Such
a term vanishes as a consequence of momentum rerouting invariance.)
We represent the derivative of
the second element of the
standard set and, more generally, $\ROLDs$, \wrt\ its external
momentum by
\be
	\ensuremath{\begin{array}{c}\begin{picture}(0,0)%
\includegraphics{pstex/dROLD.pstex}%
\end{picture}%
\setlength{\unitlength}{3947sp}%
\begingroup\makeatletter\ifx\SetFigFont\undefined%
\gdef\SetFigFont#1#2#3#4#5{%
  \reset@font\fontsize{#1}{#2pt}%
  \fontfamily{#3}\fontseries{#4}\fontshape{#5}%
  \selectfont}%
\fi\endgroup%
\begin{picture}(1072,353)(995,-222)
\put(1632,-92){\makebox(0,0)[lb]{\smash{\SetFigFont{11}{13.2}{\rmdefault}{\mddefault}{\updefault}{\color[rgb]{0,0,0}1}%
}}}
\end{picture}
 \end{array}}.
\label{eq:dROLDs}
\ee

In the case that $\ROLDs$ is under the influence of either
$\COMP$ or $\NUn$,
the interpretation of this notation changes slightly
for the second and third elements of the standard set
(recall that all other elements of the standard set are
entirely non-computable). For the second element of
the standard set, the interpretation is taken to be
consistent with diagrams~\ref{ED:28.1a-NCa}--\ref{ED:28.1a-NCbi},
\ref{ED:28.1b-NCa}--\ref{ED:28.1b-NCbi} and~\ref{d:253.6}
\ie eqn.~\eq{eq:dROLDs} now includes derivatives of the vertices
\wrt\ the internal momentum of the diagram, according 
to~\eqs{eq:momder1}{eq:momder2}. For the momentum
derivative of the third element of
the standard set we take:
\[
	2 \left[\ensuremath{\begin{array}{c}\input{pstex/dSS-3a.pstex_t} \end{array}} + \ensuremath{\begin{array}{c}\input{pstex/dSS-3b.pstex_t} \end{array}} \right]_{\NUn}.
\]
Were this pair not under the influence of either
$\COMP$ or $\NUn$, then
they could be combined.

\subsubsection{Application to Terms Manipulable at $\Op{2}$}		
\label{sec:TotalDerivativeMethods:Op2-NTE}

\begin{center}
\begin{figure}[h]
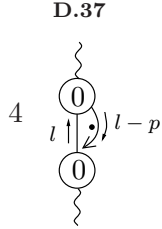

	\[
	4 \LDi{Diagram2.12-Lab}{d1:2.12-Lab}
	\]
\caption{A Taylor expandable, one-loop diagram.}
\label{fig:TLM-M:TE-Ex-A}
\end{figure}
\end{center}

The application of the subtraction techniques we have described is not limited
to $\Lambda$-derivative terms, but is useful for any class of terms in which 
we wish to perform Taylor expansions. We have already encountered such manipulations
when we dealt with the $\Op{2}$ terms in the $\beta_1$ diagrammatics
of~\cite{mgierg1,Thesis}. For example,
the elements of the little set were derived from diagrams which were Taylor expanded
in $p$. Let us return to this by considering  one of the parents of the
little set,  which is
reproduced in \fig{fig:TLM-M:TE-Ex-A}.

Due to the fact that the structure carrying momentum $l-p$ is an undecorated kernel,
rather than an effective propagator, it is clear that $p$ is not required to 
regularise this diagram in the IR; hence we can expand not only the vertex but also
the kernel to zeroth order in $p$. For this diagram, there is no need to construct
a subtraction. Indeed, at one-loop, there is never any need to construct 
subtractions for diagrams manipulable at $\Op{2}$. 

\begin{center}
\begin{figure}[h]
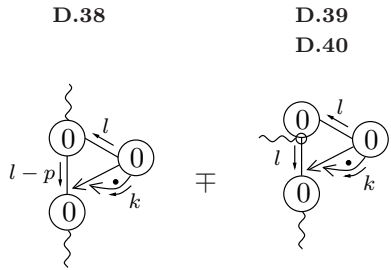

	\[
		\begin{array}{ccc}
		\vspace{0.1in}
			\LDBl{d:111b.11}	&		& \LDLD{d:111b.11-s}{d:111b.11-a}
		\\
			\ensuremath{\begin{array}{c}\input{pstex/Diagram111b.11.pstex_t} \end{array}} & \mp 	& \ensuremath{\begin{array}{c}\input{pstex/Diagram111b.11-s.pstex_t} \end{array}}
		\end{array}
	\]
\caption{A two-loop diagram with an $\Op{2}$ stub, which cannot be Taylor
expanded in $p$, and its subtraction.}
\label{fig:TLM-M:TE-Ex-B}
\end{figure}
\end{center}

At two-loops, however, the situation is different. Consider the first diagram shown 
in \fig{fig:TLM-M:TE-Ex-B}; in anticipation of what follows, 
we have constructed a subtraction.

We begin by focusing on diagram~\ref{d:111b.11}.
Since we can always Taylor expand vertices in momenta, let us suppose that
we take a power of $l$ from the top-most vertex 
(we cannot take any powers of $p$, at $\Op{2}$) and
let us choose to take a power of $k$ from the other vertex. The leading IR behaviour
of the $l$-integral is now
\[
	\int_l \frac{1}{l^2 (l-p)^2};
\]
this is not Taylor expandable in $p$. Note that had we taken a power of $l$ from
the right-hand vertex, rather than a power of $k$, then the extra power of $l$
in the integrand would render the diagram Taylor expandable in $p$ (to the
required order).

\begin{center}
\begin{figure}[h]
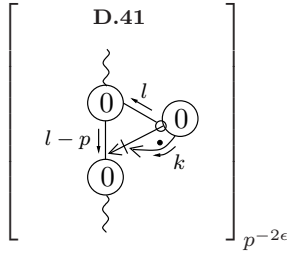

	\[
		\left[
			\LDi{Diagram111b.11-Rem}{d:111b.11-Rem}
		\right]_{\Pep}
	\]
\caption{The contribution to diagram~\ref{d:111b.11} not removed by its subtraction.}
\label{fig:TLM-M:TE-Ex-C}
\end{figure}
\end{center}

Now let us consider the subtraction and addition. The addition 
(diagram~\ref{d:111b.11-a}) is manipulated in the
usual way; this is basically what we would like to have done 
with diagram~\ref{d:111b.11},
in the first place. The effect of the subtraction on the parent is to cancel
all those components which are Taylor expandable in $p$. This immediately tells us
the following about any surviving contributions to diagram~\ref{d:111b.11}:
\begin{enumerate}
	\item	all fields carrying momentum $l$ must be in the $A^1$-sector;

	\item	we must take $\mathcal{O}(l^0)$ from the $k$-integral (note that
			the $k$-integral is Taylor expandable in $l$);

	\item	we must discard any remaining contributions to the $l$
			integral which do not $\sim \Pep$.
\end{enumerate}
The contributions to diagram~\ref{d:111b.11} not removed by
 diagram~\ref{d:111b.11-s} are shown in \fig{fig:TLM-M:TE-Ex-C}.

As required, we have taken the $\mathcal{O}(l^0)$ from the $k$-integral. 
The tag $\Pep$
demands that we take the $\Pep$ component of the diagram.
Note, of course, that this tag implicitly assumes that we are using dimensional
regularisation. However, were we to use some other means of
regularising IR divergences, diagrams such as~\ref{d:111b.11-Rem} would
still exist, but the tag would be appropriately generalised.

\subsection{Ensuring Universality} 
\label{sec:TLD:Couplings}

The central tenet of our analysis of the $\Lambda$-derivative terms
has been that the derivative \wrt\ $\LConstAl$
of a dimensionless integral must vanish, unless
there is a scale other than $\Lambda$, from which we can construct dimensionless
quantities. Implicit in this is that there are no dimensionless 
running\footnote{Where, strictly, we mean running \wrt\ $\LConstAl$.} 
couplings, hidden in the integrand. 

The most obvious candidates for dimensionless running couplings 
can immediately be discounted:
$g$ counts the loop order, and so never appears in loop 
integrals and the presence of
$\alpha$ is irrelevant, since it is held constant, when 
differentiating \wrt\ $\Lambda$.
The first question we must address is whether there are
actually any other candidates for dimensionless running couplings.

To see how they could arise, in principle, consider the flow of any
vertex, with mass dimension $\geq 0$. Now Taylor expand in momenta
and focus on the term which is the same order in momenta as the 
mass dimension of the vertex. The coefficient of this term must be
dimensionless; if this coefficient flows, then we have 
found what we are looking for.

As a first example, let us consider the flow of an $m$-loop
vertex, decorated by an arbitrary number, $q$, of $C^i$s. We take the $C^i$s to
carry momenta $r_i$. Recalling
that $C^i$s are of mass dimension zero~\cite{aprop,ymii},
all such vertices are of mass dimension four. Hence, we are interested
in the $\Omom{4}$ component of each of the vertices.

\begin{center}
\begin{figure}[h]
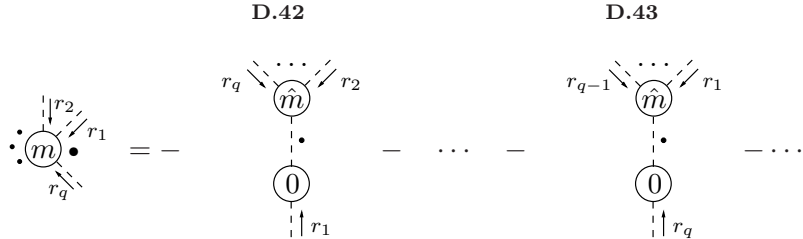

\[
	\begin{array}{ccccccc}
	\vspace{0.1in}
						&	&\LD{Dumbbell-TLTP-C-UW-Cq}	&									&\LD{Dumbbell-TLTP-C-UW-Cq-B}	&
	\\
		\ensuremath{\begin{array}{c}\input{pstex/Vertex-Cq.pstex_t} \end{array}} 	&= -&\ensuremath{\begin{array}{c}\input{pstex/Dumbbell-TLTP-C-UW-Cq.pstex_t} \end{array}}	&- \hspace{1em}\cdots \hspace{1em} -& \ensuremath{\begin{array}{c}\input{pstex/Dumbbell-TLTP-C-UW-Cq-B.pstex_t} \end{array}} 		& -\cdots
	\end{array}
\]
\caption{The flow of a vertex decorated by an arbitrary
number of $C^i$s.}
\label{fig:TLD:FlowC^q}
\end{figure}
\end{center}

The crucial point for what follows is that, no matter what the value of $m$,
the flow is guaranteed to produce a certain type of term: specifically,
we will always have a dumbbell structure consisting of a two-point, tree
level vertex, joined by an undecorated kernel to a seed action vertex.
This is illustrated in \fig{fig:TLD:FlowC^q}.

The first ellipsis denotes diagrams of the same 
structure as~\ref{Dumbbell-TLTP-C-UW-Cq} 
and~\ref{Dumbbell-TLTP-C-UW-Cq-B} but for which a different 
$C^i$ decorates the two-point,
tree level vertex. Each of these diagrams possess a 
seed action vertex. These seed action vertices are the 
highest loop vertices which appear;
moreover, all other vertices generated at this loop order possess fewer legs.
The second ellipsis denotes the remaining terms generated by the flow.

Focusing on the $\Omom{4}$ components of all diagrams generated by the flow,
we now tune  the $m$-loop, $q$-point, seed action  vertices to exactly
cancel the remaining terms. This choice of seed action is one we are entirely at
liberty to make; it ensures that there are no hidden running couplings in 
this sector of the calculation. It perhaps seems a little artificial that
we only ensure universality after some (implicit) choice for the seed action.
We must remember, though, that $\beta$-function coefficients are not strictly
universal, and that scheme dependence even at one-loop is not necessarily 
a sign of a sick formalism~\cite{aprop}. Our choice of seed action is merely done to allow
comparison of the values we compute for $\beta_1$ and $\beta_2$ with
those computed in, say, $\overline{MS}$.

In anticipation of what follows, we emphasise that the crucial ingredient in
what we have just done is
that the flow of an $m$-loop, $q$-point vertex generates an $m$-loop, $q$-point 
seed action vertex. Moreover, there are no vertices generated with higher loop
order, and for the rest of the same loop order, the number of legs is $< q$.
Consequently,
for each Wilsonian effective action vertex whose
flow we compute, it is a different seed action vertex we tune. This ensures
that we are never in the situation where we have to try and
tune the same seed action vertex in two different directions.

\begin{center}
\begin{figure}[h]
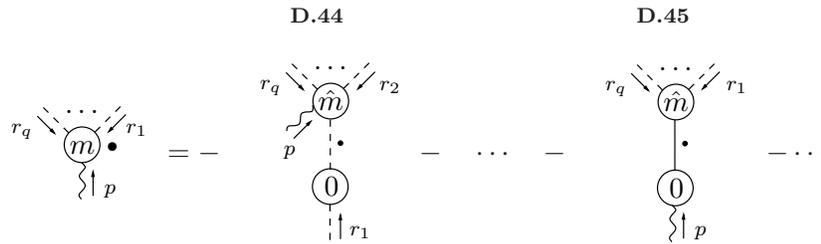

\[
	\begin{array}{ccccccc}
	\vspace{0.1in}
						&	&\LD{Dumbbell-TLTP-C-UW-A-Cq}	&									&\LD{Dumbbell-TLTP-A-UW-A-Cq}	&
	\\
		\ensuremath{\begin{array}{c}\input{pstex/Vertex-A-Cq.pstex_t} \end{array}}&= -&\ensuremath{\begin{array}{c}\input{pstex/Dumbbell-TLTP-C-UW-A-Cq.pstex_t} \end{array}}	&- \hspace{1em}\cdots \hspace{1em}-	& \ensuremath{\begin{array}{c}\input{pstex/Dumbbell-TLTP-A-UW-A-Cq.pstex_t} \end{array}}		& -\cdots
	\end{array}
\]
\caption{The flow of a vertex decorated by a single $A^i$ and an arbitrary
number of $C^i$s.}
\label{fig:TLD:FlowAC^q}
\end{figure}
\end{center}

Let us now move on to consider an $m$-loop vertex decorated by $q$ 
$C^i$s and also by a single $A^i$. The vertex is now of mass dimension three and
so it is the $\Omom{3}$ part we are interested in. This time, we are guaranteed
to generate  $m$-loop, $q+1$-point
seed action vertices, joined to a  two-point, tree-level vertex by an undecorated
kernel. Now, however, we see a difference between this case and the previous one:
the tree-level, two-point vertex can be decorated by either an $A^i$ or a $C^i$. We illustrate
this in \fig{fig:TLD:FlowAC^q}, where we take the $A^i$ to carry momentum $p$
and the $C^i$s to carry momenta $r_i$.

The first  ellipsis denotes diagrams of the same structure 
as~\ref{Dumbbell-TLTP-C-UW-A-Cq} but for which a different $C^i$ 
decorates the two-point, tree level vertex. The final
ellipsis denotes the remaining terms generated by the flow equation.

Having decorated with an $A^i$, we must now take account of gauge invariance.
The most obvious effect is that the two-point, tree level vertex of 
diagram~\ref{Dumbbell-TLTP-A-UW-A-Cq}, unlike that of 
diagram~\ref{Dumbbell-TLTP-C-UW-A-Cq}, is forced to be $\Op{2}$. Given that
we are working at \Omom{3}, this means that we must take a single power
of momentum from the corresponding $m$-loop, seed action vertex.
Now, if this single power of momentum is $p$ then, by Lorentz invariance,
its index must be contracted with the two-point, tree level vertex---killing
it. Hence, this single power of momentum must be one of the $r_i$. In turn,
this means that we can Taylor expand the $m$-loop, seed action vertex 
of diagram~\ref{Dumbbell-TLTP-A-UW-A-Cq} to zeroth
order in $p$; the effect of this is to reduce it to the derivative of
a $q$-point vertex. This means that gauge invariance has caused us to lose
one of our $m$-loop, $q+1$-point seed action vertices. We can still tune the flow
of our Wilsonian effective action vertex to zero, though, by virtue of the
presence of diagram~\ref{Dumbbell-TLTP-C-UW-A-Cq} and the diagrams represented by
the first ellipsis.

Let us examine the tuning of the seed action vertices in a little more
detail. To do this, consider contracting  the diagrams of \fig{fig:TLD:FlowAC^q}
with the momentum carried by the $A^i$. On the \lhs\ of the equation, we now have
the flow of (a set of) $m$-loop vertices, decorated by $q$ $C^i$s. Since we were
working at \Omom{3} but have contracted with a power of momentum, we should now
be looking at \Omom{4}. However, we know from our work on pure-$C^i$ vertices that
the flow of such terms has already been tuned to zero. Thus, returning
to the diagrams of \fig{fig:TLD:FlowAC^q}, gauge invariance ensures
that we need only tune the seed action vertices so as to remove those \Omom{3}\
contributions transverse in $p$! Note that this 
reproduces the conclusions of~\cite{aprop},
in which the special case of the flow of a tree level $ACC$ vertex was considered.

Next, we extend our
analysis to a vertex decorated by two $A^i$s and $q$ $C^i$s. We now work at $\Omom{2}$.
If we take $q>0$, then our analysis just mirrors what we have done:
to avoid dimensionless,  running couplings, gauge invariance ensures that we need
only tune the seed action vertices  transverse in 
the momenta of the $A^i$s. What if $q$=0? Now the 
renormalisation condition guarantees that the flow of the vertex vanishes;
this is, of course, exactly what we have been utilising to compute $\beta$-function
coefficients.

In the case of a vertex decorated by three $A^i$s and $q$ $C^i$s, we work
at \Omom{1}. Any terms involving two-point, tree level vertices decorated by
$A^i$s vanish, at the desired order in momentum. Consequently, irrespective of
the value of $q$, gauge invariance---in conjunction with what we
have just done---ensures that the flow of our vertex 
vanishes at \Omom{1}, without the need for any further tuning. Similarly, 
this result implies that gauge invariance 
can be used to demonstrate that the flow
of a vertex decorated by four $A^i$s and any number of $C^i$s vanishes,
without the need for further tuning.

Our final task is to extend this analysis to include fermionic fields.
To do this, we will treat $B$s and $D$s separately, due to their
differing mass dimensions. The point here is that neither
$S_{0 \alpha \beta}^{\ B \bar{B}}(k)$ nor $S_0^{\ D\bar{D}}(k)$ vanishes at 
zero momentum. Thus, whereas gauge invariance can force some or
all of the highest loop order seed action vertices with the maximum number of
legs
to be written as derivatives of lower point vertices when $A^i$s are
amongst the decorative fields, no such thing happens here.
Thus, both $B$s and $D$s behave essentially like the $C^i$s, of the previous analysis.

As a final point, we might worry about the vertex 
$S_{0 \alpha}^{\ B \bar{D}}(k)$---which vanishes at zero momentum. 
However, supposing that it is the $B(\bar{D})$
that is the internal field, this vertex will always be accompanied by a term
in which there is a $D(\bar{B})$ as an internal field. It is the seed action
vertex at the other end of the corresponding dumbbell which is the one we tune.

We have thus demonstrated that all dimensionless couplings, other than $g$
and $\alpha$, can be prevented from running by a suitable choice of the seed
action.

\newpage
\section{Numerical Evaluation of $\beta_2$} 
\label{sec:Numerics}

\subsection{The contributing Diagrams}
\label{sec:bulk}

In this section, we give an expression for $\beta_2$ 
in terms of the $\Lambda$-derivative and $\alpha$-terms
which survive in the $\epsilon \to 0$ limit. The complete set
of diagrams contributing to $\beta_2$ (see~\cite{Thesis})
can be derived using the techniques of~\cite{oliver1}, 
together with the subtraction techniques.
Equivalently, it can be derived directly as a special case
of the formula for arbitrary $\beta_n$~\cite{Thesis,oliver2}.
The expression for $\beta_2$ is:
	\[
		-4 \beta_2 \Box_{\mu\nu}(p) +\Oep = -\frac{1}{2} \gamma_1 \pder{}{\alpha} \OLDs(p) + \dec{\TLDs^U(p) + \TLDs^V(p)}{\bullet},
	\]
where $\TLDs^X(p) + \TLDs^Y(p)$ are given in \figs{fig:bn-X}{fig:bn-Y}.
Any diagrams contributing to $\beta_2$ are labelled \textbf{T.\#}. 	
\bcf[h]
	\[
			\begin{array}{c}
				\begin{array}{cccccc}
							& \TLD{b2-OL-A}	&	& \TLD{b2-OL-B}	&	& \TLD{Beta2-Factor-A}
				\\[1ex]
					\ds\hf	& \ensuremath{\begin{array}{c}\input{pstex/b2-OL-A.pstex_t} \end{array}}	&+2	& \ensuremath{\begin{array}{c}\input{pstex/b2-OL-B.pstex_t} \end{array}}	& + & \ensuremath{\begin{array}{c}\input{pstex/Beta2-Factor-A.pstex_t} \end{array}}
				\end{array}
			\\
				\begin{array}{cccccccc}
						& \TLD{Beta2-Factor-B}	&	& \TLD{Beta2-Factor-C}	&	& \TLD{Beta2-Factor-D}	&	& \TLD{LSSq}
				\\[1ex]
					+	& \ensuremath{\begin{array}{c}\input{pstex/Beta2-Factor-B.pstex_t} \end{array}}	&-2	& \ensuremath{\begin{array}{c}\input{pstex/Beta2-Factor-C.pstex_t} \end{array}}	&+2	& \ensuremath{\begin{array}{c}\input{pstex/Beta2-Factor-D.pstex_t} \end{array}}	&+4	& \ensuremath{\begin{array}{c}\begin{picture}(0,0)%
\includegraphics{pstex/LSSq.pstex}%
\end{picture}%
\setlength{\unitlength}{3947sp}%
\begingroup\makeatletter\ifx\SetFigFont\undefined%
\gdef\SetFigFont#1#2#3#4#5{%
  \reset@font\fontsize{#1}{#2pt}%
  \fontfamily{#3}\fontseries{#4}\fontshape{#5}%
  \selectfont}%
\fi\endgroup%
\begin{picture}(362,1596)(1150,-2216)
\put(1238,-1913){\makebox(0,0)[lb]{\smash{{\SetFigFont{11}{13.2}{\rmdefault}{\mddefault}{\updefault}{\color[rgb]{0,0,0}0}%
}}}}
\end{picture}%
 \end{array}}
				\end{array}
			\end{array}
	\]
\caption{Diagrams contributing to $\TLDs^U(p)$.}
\label{fig:bn-X}
\ecf

\bcf[tp]
	\[
		\begin{array}{c}		
			\begin{array}{cccccccc}
							&	\TLD{diag:23.1}			&	&\TLD{diag:17.1}	&   
							&\TLD{diag:31.1}	&  & \TLD{diag:28.1}
			\\[1ex]
			-\ds\frac{1}{6}	&	\ensuremath{\begin{array}{c}\input{pstex/Diagram23.1.pstex_t} \end{array}}							&-\ds\hf	& \ensuremath{\begin{array}{c}\input{pstex/Diagram17.1.pstex_t} \end{array}}					& 
					\ds+\hf	&\ensuremath{\begin{array}{c}\input{pstex/Diagram31.1.pstex_t} \end{array}}											&+& \ensuremath{\begin{array}{c}\input{pstex/Diagram28.1.pstex_t} \end{array}}
			\end{array}
		\\
			\begin{array}{cccccccc}
							&\TLD{diag:32.1}	&			&\TLD{diag:42.1}	& 	& \TLD{diag:392.4}	&	& \TLD{diag:FR26.1}
			\\[1ex]
			+\ds\frac{1}{4} &\ensuremath{\begin{array}{c}\input{pstex/Diagram32.1.pstex_t} \end{array}}	&-\ds\hf	& \ensuremath{\begin{array}{c}\input{pstex/Diagram42.1.pstex_t} \end{array}}	&+4	& \ensuremath{\begin{array}{c}\begin{picture}(0,0)%
\includegraphics{pstex/Diagram392.4.pstex}%
\end{picture}%
\setlength{\unitlength}{3947sp}%
\begingroup\makeatletter\ifx\SetFigFont\undefined%
\gdef\SetFigFont#1#2#3#4#5{%
  \reset@font\fontsize{#1}{#2pt}%
  \fontfamily{#3}\fontseries{#4}\fontshape{#5}%
  \selectfont}%
\fi\endgroup%
\begin{picture}(412,1123)(1129,-1112)
\put(1213,-818){\makebox(0,0)[lb]{\smash{\SetFigFont{11}{13.2}{\rmdefault}{\mddefault}{\updefault}{\color[rgb]{0,0,0}0}%
}}}
\end{picture}
 \end{array}}	&+8	&\ensuremath{\begin{array}{c}\begin{picture}(0,0)%
\includegraphics{pstex/DiagramFR26.1.pstex}%
\end{picture}%
\setlength{\unitlength}{3947sp}%
\begingroup\makeatletter\ifx\SetFigFont\undefined%
\gdef\SetFigFont#1#2#3#4#5{%
  \reset@font\fontsize{#1}{#2pt}%
  \fontfamily{#3}\fontseries{#4}\fontshape{#5}%
  \selectfont}%
\fi\endgroup%
\begin{picture}(455,676)(1331,-863)
\end{picture}
 \end{array}}
			\end{array}
		\\
			\begin{array}{cccccccc}
					&\TLD{diag:FR26.11}		&	&\TLD{diag:FR36.1b}		&	&\TLD{diag:FR22.3b}	&	&\TLD{diag:FR22.1b}	
					
			\\[1ex]
				+2	&\ensuremath{\begin{array}{c}\begin{picture}(0,0)%
\includegraphics{pstex/DiagramFR26.1+6.pstex}%
\end{picture}%
\setlength{\unitlength}{3947sp}%
\begingroup\makeatletter\ifx\SetFigFont\undefined%
\gdef\SetFigFont#1#2#3#4#5{%
  \reset@font\fontsize{#1}{#2pt}%
  \fontfamily{#3}\fontseries{#4}\fontshape{#5}%
  \selectfont}%
\fi\endgroup%
\begin{picture}(900,734)(1091,-791)
\end{picture}
 \end{array}}	&+4	&\ensuremath{\begin{array}{c}\input{pstex/DiagramFR36.1b.pstex_t} \end{array}}	&-4	&\ensuremath{\begin{array}{c}\begin{picture}(0,0)%
\includegraphics{pstex/DiagramFR22.3b.pstex}%
\end{picture}%
\setlength{\unitlength}{3947sp}%
\begingroup\makeatletter\ifx\SetFigFont\undefined%
\gdef\SetFigFont#1#2#3#4#5{%
  \reset@font\fontsize{#1}{#2pt}%
  \fontfamily{#3}\fontseries{#4}\fontshape{#5}%
  \selectfont}%
\fi\endgroup%
\begin{picture}(336,1251)(879,-613)
\put(1059,-325){\makebox(0,0)[lb]{\smash{{\SetFigFont{11}{13.2}{\rmdefault}{\mddefault}{\updefault}{\color[rgb]{0,0,0}0}%
}}}}
\end{picture}%
 \end{array}}&-8	&\ensuremath{\begin{array}{c}\begin{picture}(0,0)%
\includegraphics{pstex/DiagramFR22.1b.pstex}%
\end{picture}%
\setlength{\unitlength}{3947sp}%
\begingroup\makeatletter\ifx\SetFigFont\undefined%
\gdef\SetFigFont#1#2#3#4#5{%
  \reset@font\fontsize{#1}{#2pt}%
  \fontfamily{#3}\fontseries{#4}\fontshape{#5}%
  \selectfont}%
\fi\endgroup%
\begin{picture}(625,922)(560,-691)
\put(887,-395){\makebox(0,0)[lb]{\smash{{\SetFigFont{11}{13.2}{\rmdefault}{\mddefault}{\updefault}{\color[rgb]{0,0,0}0}%
}}}}
\end{picture}%
 \end{array}}
			\end{array}
		\end{array}
	\]
\caption{Diagrams contributing to $\TLDs^V(p)$.}
\label{fig:bn-Y}
\ecf

\subsection{The Universal Diagram}

\begin{center}
\begin{figure}[h]
	\[
	-\frac{1}{6}
	\dec{
		\ensuremath{\begin{array}{c}\input{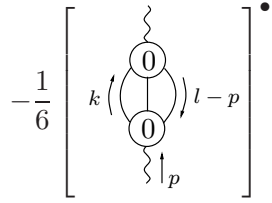} \end{array}} \hspace{1.2em}
	}{\bullet}
	\]
\caption{Reproduction of diagram~\ref{diag:23.1}.}
\label{fig:TLD:d23.1}
\end{figure}
\end{center}

Diagram~\ref{diag:23.1}, when struck by $-\flowConstAl$, 
is the sole $\Lambda$-derivative term 
which yields simply a finite,
universal contribution to $\beta_2$.\footnote{Since the leading 
order contribution to this
diagram is finite, it is not merely computable but actually universal: it
is independent of the way in which we compute it 
(see~\cite{two} for a different way of
evaluating essentially the same diagram).} We reproduce this diagram, having chosen a
particular momentum routing, in \fig{fig:TLD:d23.1}.

The requirement that we take contributions which survive in the $\eptoz$ limit
places useful constraints on the diagram. First, all 
fields must be in the $A$-sector;
given this, we are compelled to take $\Omom{0}$ from each of the vertices:
additional powers of momentum would cause the diagram to vanish.
For example, taking $\Op{2}$ from the vertices leaves us with, schematically
\[
	\Op{2} \dec{\int_{l,k} \frac{1}{l^2(l-k)^2k^2}}{\bullet};
\]
the integral is IR finite and is killed by $\flowConstAl$
in the $\eptoz$ limit.
In turn,
this forces both vertices to comprise a single supertrace: 
it is forbidden to have a single gauge field on a
supertrace; if we take two supertraces, each with two gauge fields, 
then gauge invariance demands that there is no 
$\mathcal{O}(\mathrm{mom}^0)$ contribution
to such a vertex.

Now that we know that both vertices have only a single supertrace,
all fields are forced to be in the $A^1$ sector. Temporarily ignoring
attachment corrections, the group theory factor
of the diagram must be either $N^2$ or unity. 
However, we can show that contributions
of the latter type cancel. To see this, we can
use the Ward identities to straightforwardly demonstrate~\cite{aprop,Thesis}
\begin{eqnarray*}
S^{1\, 1\, 1\, 1}_{\mu\alpha\beta\gamma}(\underline{0}) & = & -2(2\delta_{\mu \beta}\delta_{\alpha \gamma} - \delta_{\alpha \beta} \delta_{\mu \gamma}
- \delta_{\mu \alpha} \delta_{\beta \gamma}) \\
S^{1\, 1\, 1\, 1}_{\nu\gamma\beta\alpha}(\underline{0}) & = & -2(2\delta_{\nu \beta}\delta_{\alpha \gamma} - \delta_{\gamma \beta} \delta_{\nu \alpha}
- \delta_{\nu \gamma} \delta_{\beta \alpha}).
\end{eqnarray*}

Focusing on the component of diagram~\ref{diag:23.1} with a group theory
factor of unity, the locations of the external fields are independent,  
since they are always guaranteed to be on the
same supertrace. Summing over all independent locations of the external fields
yields something proportional to
\begin{equation}
	S^{1\, 1\, 1\, 1}_{\mu\alpha\beta\gamma}(\underline{0})
	+S^{1\, 1\, 1\, 1}_{\alpha\mu\beta\gamma}(\underline{0})
	+S^{1\, 1\, 1\, 1}_{\alpha\beta\mu\gamma}(\underline{0}) = 0.
\label{eq:TLD:4ptSum}
\end{equation}

Similarly, all attachment corrections can be ignored. If we suppose
that one of the effective propagators attaches via a correction
(see \fig{fig:Attach}) then the supertrace structure
of the diagram is left invariant under independently placing the ends of this
effective propagator in all independent locations. Hence the diagram vanishes
courtesy of~(\ref{eq:TLD:4ptSum}).
Increasing the number of effective propagators which attach via a
correction clearly does not change this result.

Returning to the case of direct attachment,
if the group theory goes as $N^2$,
then the locations of the external gauge fields are dependent, 
since it must be ensured that they are on the
same supertrace. Up to insertions of $A^1_{\mu,\nu}$, we can use 
charge conjugation invariance to fix the order of the three internal 
fields so long as we multiply by two. Now, there are three identical 
pairs of locations that we can place the
pair of fields $A^1_{\mu,\nu}$. 
 Including the diagram's overall factor of $-1/6$ we have:
\begin{equation}
-N^2\times S^{1\, 1\, 1\, 1}_{\mu\alpha\beta\gamma}(\underline{0}) S^{1\, 1\, 1 \, 1}_{\nu\gamma\beta\alpha}(\underline{0}) = -72N^2\delta_{\mu \nu} + \Oep.
\label{eq:diag23.1-GT}
\end{equation}

To obtain the contribution to $\beta_2$ coming from diagram~\ref{diag:23.1}, 
we must multiply
the above factor by the number obtained from the loop integral. 
Since the integral yields a finite
contribution, we simply Taylor expand the effective propagator $\Delta^{11}(l-p)$
to $\Op{2}$. Remembering to evaluate the cutoff
functions at zero momentum---which yields a factor of $1/2$ 
for each of the effective propagators---we have:
\[
	\frac{1}{8} \left[\int_{l,k} \frac{1}{k^2 (l-k)^2 l^4} \left(p.p - \frac{4(l.p)^2}{l^2} \right)\right]^\bullet,
\]
where we will define precisely what we mean by $\int_{l,k}$ in a moment.
Looking at this expression, we might worry that the 
presence of $l^4$ in the denominator
means that the integral is actually IR divergent, even after differentiation
\wrt\ $\flowConstAl$.
However,
due to the form of the $\Op{2}$ contributions, averaging over angles in the 
$l$-integral will produce a factor of 
\[
	1 - \frac{4}{D} \sim \epsilon,
\]
in addition to the power of $\epsilon$
coming from the $\Lambda$-derivative. This renders 
the contribution from diagram~\ref{diag:23.1} finite.

To evaluate the integral, we use the 
techniques of \sec{ch:LambdaDerivatives:Methodology}.
Specifically, we perform the $l$ integral first, with unrestricted
range of integration, and then perform the $k$-integral with
the radial integral cutoff at $\Lambda$.
After differentiation \wrt\ $\flowConstAl$, 
the integral gives $\PowAngVol{D}{2} p^2/32$.
Combining this with the factor coming from \eq{eq:diag23.1-GT} yields:
\be
	\mathrm{diagram} \ \ref{diag:23.1} 	= -\frac{9N^2}{(4\pi)^4 }p^2\delta_{\mu \nu} + \Oep.
\label{eq:diag23.1}
\ee

Before moving on, it is worth commenting further on the
fact that all attachment corrections in diagram~\ref{diag:23.1}
effectively vanish.
When we finally come to evaluate the numerical value of
$\beta_2$, we will be dealing with diagrams for which all
fields are in the $A^1$ sector. The highest point vertex that
we will encounter is four-point: we have already seen how
attachment corrections to such a vertex vanish. Three-point
vertices are even easier to treat. Suppose that an effective
propagator attaches via a correction to a three-point vertex,
decorated exclusively by $A$s. We can sum over the two
locations to which the effective propagator can attach,
but these two contributions cancel, by \CC.

If nested gauge remainders are in the $A$-sector, we know
from~\cite{Thesis,mgierg1} that we can ignore attachment
corrections. 

Thus, when we come to extract numerical contributions to
$\beta_2$, we will neglect attachment corrections. 
Similarly, for direct attachments, we need focus only on the cases where
the group theory goes as $N^2$.

\subsection{Finite, Non-Universal Diagrams}

\begin{center}
\begin{figure}
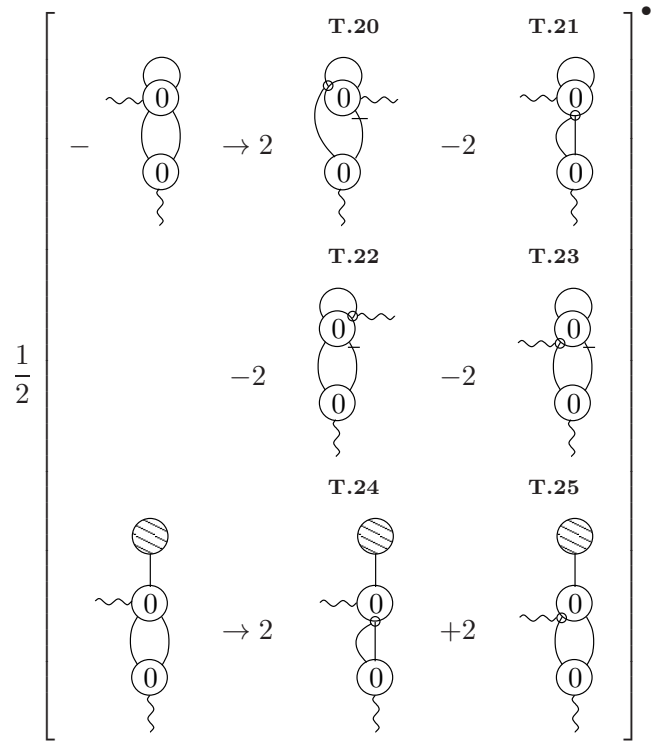

	\[
	\frac{1}{2}
	\dec{
		\begin{array}{ccccc}
								&				&\TLD{d:217.1}		&	&\TLD{d:217.2}	
		\\[1ex]
			-\ensuremath{\begin{array}{c}\input{pstex/Diagram17.1.pstex_t} \end{array}}	&\rightarrow 2	&\ensuremath{\begin{array}{c}\input{pstex/Diagram217.1.pstex_t} \end{array}}		&-2	& \ensuremath{\begin{array}{c}\input{pstex/Diagram217.2.pstex_t} \end{array}}	
		\\
								&				&\TLD{d:217.3}			&	&\TLD{d:217.4}
		\\[1ex]
								&-2				&\ensuremath{\begin{array}{c}\input{pstex/Diagram217.3.pstex_t} \end{array}}		&-2	& \ensuremath{\begin{array}{c}\input{pstex/Diagram217.4.pstex_t} \end{array}}	
		\\
								&				&\TLD{d:31.1-TE-k}	&	&\TLD{d:31.1-TE-p}
		\\[1ex]
			\ensuremath{\begin{array}{c}\input{pstex/Diagram31.1.pstex_t} \end{array}}	&\rightarrow 2	&\ensuremath{\begin{array}{c}\input{pstex/Diagram31.1-TE-k.pstex_t} \end{array}}	&+2	& \ensuremath{\begin{array}{c}\input{pstex/Diagram31.1-TE-p.pstex_t} \end{array}}
		\end{array}
	}{\bullet}
	\]
\caption{Manipulation of diagrams~\ref{diag:17.1} and~\ref{diag:31.1} under $\Lambda \partial_\Lambda|_\alpha$.
This is valid up to $\Oep$ corrections.}
\label{fig:Finite-NU:TE}
\end{figure}
\end{center}

Diagrams~\ref{diag:17.1} and~\ref{diag:31.1}
both yield finite, non-universal contributions,
when struck by $-\flowConstAl$.
All contributions which survive in the $D \to 4$
limit must pick up a $1/\epsilon$ pole (before
differentiation \wrt\ $\LConstAl$), which must come 
from the loop carrying the external momentum. 
Thus, we can take this loop to be in the $A^1$
sector and can Taylor expand the vertices which form
this loop to lowest order in momenta. The combination
of the bottom vertex and the effective propagators attached
to it yield universal factors. However, it not helpful
to convert these terms into algebra: keeping the diagrams
intact will enable us to perform cancellations at the
diagrammatic
level. Indeed, it is only useful to explicitly Taylor expand the
vertices (carrying $p$) which give a non-universal contribution, 
as shown in \fig{fig:Finite-NU:TE}. As 
usual, we have used \CC\ to collect terms~\cite{mgierg1,Thesis}.

The pattern of diagrams produced is highly suggestive:
focusing in the top-most sub-diagrams,
we can recognise terms
that contribute to momentum derivatives of $\ROLDs$.
The missing terms are buried within the more complex,
IR divergent diagrams contributing to $\beta_2$, which
we now examine.

\subsection{Subtractions}
\label{sec:TLD:Subs}

Diagrams~\ref{b2-OL-A}--\ref{LSSq}, 
\ref{diag:28.1}--\ref{diag:FR22.1b}
are all
IR divergent, even after differentiation \wrt\ $\LConstAl$. Our
strategy, as outlined in \sec{sec:TLD:Subtractions}, is 
to construct subtractions.
%To facilitate this procedure,
%we note that the following pairs of diagrams can be combined,
%up to $\Oep$ corrections:
%\ref{diag:FR26.1} and~\ref{diag:FR26.2}; \ref{diag:FR22.2} 
%and \ref{diag:FR22.1}; \ref{diag:FR24.6} and~\ref{diag:FR24.1}.
%(To see this, put all fields in the $A^1$-sector and write
%the diagrams out, algebraically.)
Specifically, we construct subtractions for
the latter set of terms, noting that this
has already been done
for diagram~\ref{diag:28.1} (see \fig{fig:D28Ss}). 
Our choice of subtractions
for diagrams~\ref{diag:32.1}--\ref{diag:FR22.1b} 
which, as noted already is not unique,
is
given in figure~\ref{fig:Subs-a}.

Following \sec{sec:TLD:Subtractions:Gauge},
we use the subtractions and additions to isolate
the computable and non-computable parts of the parent
diagrams. The latter terms are then manipulated, using
eqn.~\eq{eq:EP-dV-EP}. Processing any gauge remainders,
further progress can be made by recognising that
components of certain terms under the influence of
$\NUn$ vanish in the $\epsilon \to 0$ limit. As
an example of this, consider the
diagram shown on the \lhs\ of \fig{fig:NC-Manips}, which 
is obtained from the non-factorisable, non-computable
component of diagram~\ref{diag:42.1}.
\bcf[h]
	\[
			\NUni{\ensuremath{\begin{array}{c}\input{pstex/Diagram255b.1b.pstex_t} \end{array}} \hspace{1.2em} } \ 
			\rightarrow \ensuremath{\begin{array}{c}\input{pstex/Diagram259b.2b.pstex_t} \end{array}} \ + \Oep
	\]
\caption{Manipulation of a term under the influence of $\NUn$.}
\label{fig:NC-Manips}
\ecf

To understand how to manipulate the diagram on the \lhs,
it suffices to look at the component Taylor expandable
in $p$. We begin by temporarily ignoring the $\NUn$
and looking at the most IR divergent contribution.
In this case, the integrand goes like:
\[
	\frac{\order{l^2, l\cdot k}}{l^4(l-k)^2 k^4} 
	\Op{2}.
\]
The $\NUn$ forces additional powers of $l,k$
to appear in the numerator. The $\order{l\cdot k}$
now does not survive differentiation \wrt\ $\LConstAl$
in the $\epsilon \to 0$ limit
and the $\order{l^2}$ term survives if we
take additional powers of $l$, only.
Consequently, the $l$ integral becomes Taylor
expandable in $k$, yielding the \rhs\ of \fig{fig:NC-Manips}.
This manipulation has yielded a term which can
be processed using~\eq{eq:EP-dV-EP}.

Iterating the diagrammatic procedure yields:
\be
	-4 \beta_2 \Box_{\mu \nu}(p) +\Oep = -\frac{1}{2} \gamma_1 \pder{}{\alpha} \OLDs(p) 
	-\frac{9N^2}{(4\pi)^4}p^2 \delta_{\mu\nu}
	+\dec{
		\Uni{\TLDs^{X}(p)} + \TLDs^{Y}(p) 
	}{\bullet},
\label{eq:b2-part}
\ee
where $\TLDs^X(p)$ and $\TLDs^Y(p)$ are given,
respectively, in \figs{fig:b2-T,X}{fig:b2-T,Y}.
\bcf
	\[		
	\begin{array}{c}
		\begin{array}{cccccccc}
					&	\TLD{D:28.1a-C}	&		& \TLD{D:28.1b-C}	&		&\TLD{D:32.1-C}		&		&\TLD{D:42.1-C}
		\\[1ex]			
			\ds\hf	& \ensuremath{\begin{array}{c}\input{pstex/Diagram28.1a.pstex_t} \end{array}}	&+\ds\hf& \ensuremath{\begin{array}{c}\input{pstex/Diagram28.1b.pstex_t} \end{array}}	&+\ds\qt&\ensuremath{\begin{array}{c}\input{pstex/Diagram32.1L.pstex_t} \end{array}}\	&-\ds\hf&\ensuremath{\begin{array}{c}\input{pstex/Diagram42.1L.pstex_t} \end{array}}
		\end{array}
	\\
		\begin{array}{cccccccccc}
				&	\TLD{D:392.4a-C}	&	&\TLD{D:392.4b-C}	&	&\TLD{D:FR26.1,2a-C}		&	&\TLD{D:FR26.1,2b-C}		&	&\TLD{D:FR26.11-C}
		\\[1ex]
			+2	& \ensuremath{\begin{array}{c}\input{pstex/Diagram392.4a.pstex_t} \end{array}}\ 	&-2	&\ensuremath{\begin{array}{c}\input{pstex/Diagram392.4b.pstex_t} \end{array}}	&+4	&\hspace{-0.5em} \ensuremath{\begin{array}{c}\input{pstex/DiagramFR26.1,2-Lab.pstex_t} \end{array}}	&+4	&\hspace{-0.5em}\ensuremath{\begin{array}{c}\input{pstex/DiagramFR26.1,2-Lb.pstex_t} \end{array}}	&+2	&\hspace{-1em}\ensuremath{\begin{array}{c}\input{pstex/DiagramFR26.11-L.pstex_t} \end{array}}
		\end{array}
	\\
		\begin{array}{cccccccc}
				&\TLD{D:FR36.1-Cb}		&	& \TLD{D:FR22.3-Cb}		&	&\TLD{D:FR22.1,2-Cb}	&	& \TLD{D:FR22.1,2-Cc}
		\\[1ex]
			+4	& \ensuremath{\begin{array}{c}\input{pstex/DiagramFR36.1Lb.pstex_t} \end{array}}	&-4	& \ensuremath{\begin{array}{c}\input{pstex/DiagramFR22.3Lb.pstex_t} \end{array}}	&-4	&\ensuremath{\begin{array}{c}\input{pstex/DiagramFR22.1Lb.pstex_t} \end{array}}	&-4	&\ensuremath{\begin{array}{c}\input{pstex/DiagramFR22.1Lc.pstex_t} \end{array}}
		\end{array}
	\\
		\begin{array}{cccccc}
				& \TLD{LSSq-C}	&	&\TLD{b2-TMD-A}					&	& \TLD{b2-TMD-B}
		\\[1ex]
			 +4	& \ensuremath{\begin{array}{c}\begin{picture}(0,0)%
\includegraphics{pstex/LSSq.pstex}%
\end{picture}%
\setlength{\unitlength}{3947sp}%
\begingroup\makeatletter\ifx\SetFigFont\undefined%
\gdef\SetFigFont#1#2#3#4#5{%
  \reset@font\fontsize{#1}{#2pt}%
  \fontfamily{#3}\fontseries{#4}\fontshape{#5}%
  \selectfont}%
\fi\endgroup%
\begin{picture}(362,1596)(1150,-2216)
\put(1238,-1913){\makebox(0,0)[lb]{\smash{{\SetFigFont{11}{13.2}{\rmdefault}{\mddefault}{\updefault}{\color[rgb]{0,0,0}0}%
}}}}
\end{picture}%
 \end{array}}		&+	&\hspace{-2.1pt}\ensuremath{\begin{array}{c}\input{pstex/b2-OL-E.pstex_t} \end{array}} 	& +	&\hspace{-2.1pt}\ensuremath{\begin{array}{c}\input{pstex/b2-OL-F.pstex_t} \end{array}}
		\end{array}
	\end{array}
	\]
\caption{Diagrams contributing to $\TLDs^X(p)$.}
\label{fig:b2-T,X}
\ecf

\bcf
	\[
			\begin{array}{c}
				\begin{array}{cccccccc}
						& \TLD{b2-OL-Ab}	&	& \TLD{b2-OL-Bb}	&	& \TLD{b2-OL-C}	&	& \TLD{b2-OL-D}
				\\[1ex]
				\ds\hf	& \ensuremath{\begin{array}{c}\input{pstex/b2-OL-A.pstex_t} \end{array}}		&+2	& \ensuremath{\begin{array}{c}\input{pstex/b2-OL-B.pstex_t} \end{array}}		&-	&\ensuremath{\begin{array}{c}\input{pstex/b2-OL-C.pstex_t} \end{array}}	& -	& \ensuremath{\begin{array}{c}\input{pstex/b2-OL-D.pstex_t} \end{array}}
					\end{array}
			\\
				\begin{array}{cccccccc}
					& \TLD{Beta2-Factor-Ab}	&	& \TLD{Beta2-Factor-Bb}	&	& \TLD{Beta2-Factor-Cb}	&	& \TLD{Beta2-Factor-Db}
				\\[1ex]
				+	& \ensuremath{\begin{array}{c}\input{pstex/Beta2-Factor-A.pstex_t} \end{array}}	& +	& \ensuremath{\begin{array}{c}\input{pstex/Beta2-Factor-B.pstex_t} \end{array}}	&-2	& \ensuremath{\begin{array}{c}\input{pstex/Beta2-Factor-C.pstex_t} \end{array}}	&+2	& \ensuremath{\begin{array}{c}\input{pstex/Beta2-Factor-D.pstex_t} \end{array}}
				\end{array}
			\\
				\begin{array}{cc}
						& \TLD{LS-NC-SS}
				\\[1ex]
					+2 	& \ensuremath{\begin{array}{c}\input{pstex/LS-NC-SS.pstex_t} \end{array}}	
				\end{array}
			\end{array}
	\]
\caption{Diagrams contributing to $\TLDs^Y(p)$.}
\label{fig:b2-T,Y}
\ecf

It is thus apparent that ($\alpha$-terms aside) the expression
for $\beta_2$ is computable, up to 
diagrams~\ref{b2-OL-Ab}--\ref{LS-NC-SS}. If we
are to obtain a universal $\beta_2$ (in the
$\alpha \to 0$ limit) it must therefore
be the case that
these latter diagrams are computable also. To demonstrate
this, we need not construct further subtractions
but rather need only utilise the transversality
of \Square\ when the external legs
are in the $A^1$-sector---as they are for
all contributions to $\TLDs^Y(p)$ which
survive in the $\ep \to 0$ limit. This 
transversality can be exploited to extract the momentum
dependence and Lorentz structure of \Square:
\[
	\ensuremath{\begin{array}{c}\input{pstex/OLDs-EL.pstex_t} \end{array}} = 	\left[
						\mathcal{G}_1\frac{1}{\ep} + \mathcal{G}_2 +
						\left(
							\mathcal{H}_1\frac{1}{\ep}  + \mathcal{H}_2 
						\right)q^{-2\ep} 
					\right]2\Box_{\alpha \beta}(q) + \ldots.
\]
The ellipsis denotes terms higher order in momentum and / or $\ep$, and
the factor of two is extracted for convenience.
Notice that this transversality immediately reduces the apparent severity
of the IR divergences of diagrams~\ref{b2-OL-Ab}--\ref{b2-OL-C}
(\cf the comments above~\eq{eq:TLD:StandardTwoLoopIntegral}).

For all non-factorisable diagrams, it must be
the case that \Square\ attaches to something
computable, else the diagram as a whole vanishes in the $\ep \to 0$ limit.
This is easiest to see by working in $D=4$,
where the potentially most IR divergent contributions
to such diagrams
are obtained by taking the part of\Square\
which goes as $\order{l^2} \ln l$. Focusing, for example, on the
 part of such a diagram Taylor expandable in $p$, 
we will only find an IR divergence after integrating
over $l$ if the
$l$-integrand
picks up a factor of $1/l^6$;
such contributions are
computable.

With this in mind, and recalling that $\mathcal{G}_1$ and the $\mathcal{H}_i$ are
computable, the non-universal contributions to 
diagrams~\ref{b2-OL-Ab}--\ref{LS-NC-SS}, which survive
in the $\ep \to 0$ limit are as follows:
\begin{enumerate}
	\item	the components of diagrams~\ref{Beta2-Factor-Ab}--\ref{LS-NC-SS}
			in which we take the non-computable part of the bottom most
			sub-diagram;
	\label{it:NC}
	\item	the components of diagrams~\ref{b2-OL-Ab}--\ref{Beta2-Factor-Db}
			in which we take the $\mathcal{G}_2$ contribution to \Square\ 
			and take only the computable parts of all remaining diagrammatic
			elements.
	\label{it:fancy}
\end{enumerate}

In both cases, the sum of all contributions can
be shown to vanish, diagrammatically! In the first
case, we note that
the non-computable components of the bottom-most
sub-diagrams of \ref{Beta2-Factor-Ab}--\ref{Beta2-Factor-Bb}
are Taylor expandable in $p$ (the same cannot, in general,
be said of the computable components). Performing
the usual diagrammatic manipulations, it is straightforward
to show that the elements of item~\ref{it:NC} cancel,
amongst themselves.

To show that the diagrams of item~\ref{it:fancy} cancel,
we exploit the transversality of \Square\ noting that,
if \Square\ is contracted into an effective propagator,
we can use the effective propagator relation, up to 
terms which we do not care about:
\[
	2\Box_{\alpha \beta} (q) \Delta^{11}(q) = 1 + \order{q^2},
\]
where, if we identify $q$ with a loop momentum, the $\order{q^2}$
terms do not contribute to $\beta_2$ in the $\ep \to 0$ limit
whereas, if we identify $q$ with the external momentum, the
$\order{q^2}$ terms yield diagrams which vanish at $\Op{2}$.

Applying the effective propagator relation
in diagrams~\ref{Beta2-Factor-Ab}--\ref{Beta2-Factor-Db},
the resulting gauge remainder can be discarded: the
sub-diagram it strikes is transparent to it, by Lorentz
invariance, and so it effectively hits the $\Op{2}$ stub, killing it.
Applying the
effective propagator relation in diagram~\ref{b2-OL-Bb}, the $\mathcal{G}_2$
part of this diagram
cancels those of~\ref{Beta2-Factor-Cb} and~\ref{Beta2-Factor-Db},
up to corrections which vanish in the $\ep \to 0$ limit.
Finally, we turn to the $\mathcal{G}_2$ part of diagram~\ref{b2-OL-Ab}.
Applying the effective propagator relation, the Kronecker-$\delta$
contribution cancels the $\mathcal{G}_2$ parts of 
diagrams~\ref{b2-OL-C} and~\ref{b2-OL-D}
at leading order in $\ep$. Processing the gauge remainders
removes the $\mathcal{G}_2$ contributions from diagrams~\ref{Beta2-Factor-Ab}
and~\ref{Beta2-Factor-Bb} in the $\epsilon \to 0$ limit.

We have thus demonstrated that, up to the $\alpha$-terms, 
$\beta_2$ can be reduced to a sum of computable contributions.
The IR divergent terms and terms which go as $p^{-2\epsilon}$
(see \sec{sec:TLD:Methodology:2-loop}) cancel between diagrams. 
The following finite contributions to $-4\beta_2 \Box_{\mu \nu}(p)$
are left over, 
and can be evaluated using the techniques of
\sec{sec:TLD:Subtractions}:
\renewcommand{\arraystretch}{1.5}
\[
	\begin{array}{c|c}
		\mathrm{diagram(s)}				& N^2/(4\pi)^4
	\\ \hline
		\ref{diag:23.1}					& -9p^2 \delta_{\mu \nu}
	\\
		\mbox{\ref{D:28.1a-C}--\ref{b2-TMD-B}}	& -733/9 \Box_{\mu \nu}(p) + 9p^2 \delta_{\mu \nu}
	\\
		\mbox{\ref{b2-OL-Ab}--\ref{Beta2-Factor-Db}}	& 1141/9 \Box_{\mu \nu}(p)
	\end{array}
\]
\renewcommand{\arraystretch}{1}
Putting everything together, we have:
\be
\beta_2 \Box_{\mu \nu}(p) +\Oep = -\frac{34}{3} 
\frac{N^2}{(4\pi)^4}\Box_{\mu \nu}(p) +\frac{1}{8} \gamma_1 \pder{}{\alpha} \OLDs(p).
\ee

\subsection{The $\alpha$-terms}	
\label{sec:TLD:alpha}

\subsubsection{The problem}

To deal with the $\alpha$-terms, we must understand the
$\alpha$-dependence of $\OLDs(p)$. To make this procedure
as transparent as possible, we will use the explicit
forms for the two-point, tree level vertices, effective
propagators and gauge remainders given in appendix~\ref{app:elements}.
We note, though, that it should be possible to repeat this
analysis, using only the general properties which we
know these functions must satisfy as a consequence of
Lorentz invariance, gauge invariance, dimensions and the UV
finiteness of the theory.

In the current picture, we have an explicit algebraic realisation for four of the
members of $\OLDs(p)$: the final member of the standard set and
all of the members of the little set. The remaining diagrams
all contain either three-point or four-point vertices. It
is not our aim to choose specific forms for these vertices
through a specific choice of seed action and boundary
conditions for the flow; rather, our
aim is to show that only implicit choices
are necessary for our purposes.

Assessing the $\alpha$-dependence of $\OLDs(p)$ is complicated
by the fact that a loop integral must be performed. 
This integral must be done
before we take the $\altoz$ limit, as the two procedures do
not commute. However, whilst we cannot set $\altoz$ too soon,
we can work at small $\alpha$, and will do so henceforth.
In this limit, it follows from eqn.~\eq{eq:gamma_i}
that $\gamma_1 \sim \alpha$. Consequently, the $\alpha$-terms
will vanish, as required, only if the behaviour of
$\partial \OLDs(p) / \partial \alpha$ is better than $1/\alpha$.

From our work in \sec{ch:LambdaDerivatives:Methodology},
we know how to parameterise the $\alpha$-dependence of
$\OLDs(p)$. For our purposes, this is most easily done in $D=4$:
\begin{equation}
	\OLDs(p) = \left[ 
				4 \beta_1 \ln 
				\left(
					\frac{\left(\mbox{IR scale}\right)^2}{\Lambda^2} 
				\right) + H(\alpha) 
			\right] \Box_{\mu \nu}(p).
\label{eq:OLDs-charateristic}
\end{equation}

The non-universal function $H(\alpha)$ is independent of $\Lambda$.
We now choose to recast this equation. When constructing the (dimensionless)
argument of the logarithm, we divide the IR scale by
the only other scale available, $\Lambda$.

Let us examine this in the context of actually performing the loop integrals
to obtain~\eq{eq:OLDs-charateristic}. The appearance of the IR 
scale has been discussed,
in depth, in \sec{ch:LambdaDerivatives:Methodology}. The 
scale $\Lambda$ has
a natural interpretation as the scale at which the 
loop integrals are effectively cutoff in the UV.
However, we should not preclude the possibility that the 
loop integrals are actually cutoff
at some scale $h(\alpha) \Lambda$, where $h(\alpha)$ is a 
dimensionless function, independent of
$\Lambda$. Of course, this has no effect on the value of 
$\beta_1$ obtained by differentiating
$\OLDs(p)$ \wrt\ $\LConstAl$. With this in mind, we rewrite 
eqn.~\eq{eq:OLDs-charateristic}
as follows:
\begin{equation}
	\OLDs(p) = \left[ 
				4 \beta_1 \ln 
				\left(
					\frac{\left(\mbox{IR scale}\right)^2}{\Lambda^2 h(\alpha)} 
				\right) + \tilde{H}(\alpha) 
			\right] \Box_{\mu \nu}(p).
\label{eq:OLDs-charateristic-B}
\end{equation}

This recasting now allows us to break the problem of the $\alpha$-terms into two
parts. On the one hand, we have potential $\alpha$-dependence coming
from any non-trivial $\alpha$-dependence of the effective cutoff scale, 
parameterised by $h(\alpha)$.
On the other hand, we have $\alpha$-dependence
coming from the region of the loop integral with support, 
parameterised by $\tilde{H}(\alpha)$.
We deal with these cases in turn.

\subsubsection{Behaviour of $h(\alpha)$}

The treatment of this problem is slightly easier than one might expect.
The crucial point is that the logarithm term in 
eqn.~\eq{eq:OLDs-charateristic-B}
comes only from (UV regularised) terms with non-trivial IR behaviour.
There are five diagrams with non-trivial IR behaviour: the final two elements
of the standard set and the elements of the little set. Our strategy is
to examine these diagrams and, through a choice of cutoff functions,
ensure that the momentum integrals are cutoff at $\Lambda$; equivalently
that $h(\alpha)$ is independent of $\alpha$.

Sufficient UV regularisation can be provided by cutoff function
regularisation alone---\eg for the little set---or entirely by
the regularising sector---\eg for the final element of the standard set.
In the latter case, we are interested simply in the scale at which 
the $B$-sector diagrams regularise the $A^1$-sector diagrams. 
In the former case, we are interested
not only in this but also the scale at which the $A^1$ sector 
diagrams die off on their own.

For all that follows, we assume that the momentum of
the cutoff functions
$c(k^2/\Lambda^2)$ and $\tilde{c}(k^2/\Lambda^2)$ crosses over from large to small
for $x \equiv k^2 / \Lambda^2 \sim \order{1}$. This amounts to an implicit choice of
the non-universal details of the set-up. In this section,
we will demonstrate that, given this choice, we can
consistently arrange for all momentum integrals to 
be cutoff at $x \sim \order{1}$.

We start by looking at the
final element of the standard set
which has the algebraic form
\begin{equation}
	4N \int_k \left(\frac{(k-p)_\nu}{(k-p)^2} \frac{k_\mu}{k^2} - \frac{f_{k-p} (k-p)_\nu}{\Lambda^2} \frac{f_k k_\mu}{\Lambda^2}  \right).
\label{eq:SS-C-Alg}
\end{equation}
where
\begin{equation}
	f_k = \f{x}.
\label{eq:f_k-B}
\end{equation}
At large $x$, where $f_k = 1/x$~\cite{aprop,mgierg1}, 
we recover unbroken $SU(N|N)$, as we must,
for the theory to be regularised. Given our assumption
that $\tilde{c}_x$ crosses over at $x \sim \order{1}$,
we need
only ensure that the denominator of $f_k$ 
crosses over at $x \sim \order{1}$.
The crossover of the denominator
occurs at 
\[
	x \tilde{c}_x \sim 4 \alpha c_x.
\]
Now, since we are working at small $\alpha$ (and $x \tilde{c}_x \gg c_x$ for large
$x$), the crossover must happen for small values of $x$. Taylor expanding,
we therefore find that the crossover occurs at
\[
	x \tilde{c}_0 \sim \alpha c_0,
\]
(in the limit of small $\alpha$). $c_0$ is
fixed to be unity by the renormalisation
condition for $A^1$, eqn.~\eq{defg}; however, there is no such constraint on $\tilde{c}_0$.
In turn, this implies that the momentum integral for
the third element of the standard set is cutoff 
at $x \sim \mathcal{O}(\alpha/ \tilde{c}_0)$. Note that if
we set $\tilde{c}_0 \sim 1$, then we would indeed find the
problem that $\partial \OLDs(p) / \partial \alpha \sim 1/ \alpha$. 
Demanding that the cutoff
scale occurs at $x \sim \Oone$ then forces us to choose
\begin{equation}
	\tilde{c}_0 \sim \Oal
\label{eq:c-tilde(0)}
\end{equation}
(which is perfectly compatible with $\tilde{c}_x$ crossing over at $x \sim \order{1}$).

Let us now turn to the remaining  diagrams with non-trivial
IR behaviour. The treatment of these is somewhat different
from what we have just done, as 
we have regularisation provided not only by the $B$-sector, but
also by cutoff function regularisation. The crossover scale 
in the $B$-sector follows trivially from the 
observation that
\[
	 \Delta^{B\bar{B}}(k) = \frac{\alpha c_x f_x}{(1+\alpha) \Lambda^2}.
\]
Hence, we immediately know that, given the choice $\tilde{c}_0 \sim \Oal$, 
the crossover occurs at $x \sim \order{1}$,
in this sector.

However, we now need to show that the scale at which the
cutoff regularisation kicks in in the $A^1$-sector also 
occurs at this scale. If a diagram is sufficiently
regularised by cutoff regularisation alone, then the $B$-sector
becomes effectively redundant. If the $B$-sector is required,
in addition to cutoff regularisation, then there can be two scales
in the problem: the first is where the momentum in the $A^1$-sector
can be considered large and the second is where the momentum in the $B$-sector
can be considered large. The $A^1$ and $B$-sectors cancel
each other at the highest of these scales.

Turning now to the $A^1$-sector,
\[
	 \Delta^{11}(k) = \EPAone{k}{x},
\]
which goes as $\alpha c_x/ x$ for large $x$ and as $1/2x$ for small $x$.
The crossover occurs, for small $\alpha$, at
\[
	\alpha(c_x + 1) \sim 1-c_x.
\]
(Note that the \lhs\ dominates at sufficiently small $x$, whereas the \rhs\ dominates
at large $x$.) Again, due to the smallness of $\alpha$, we can Taylor expand in $x$
to find the crossover point, which 
occurs at
\[
	\frac{x \mod{c'_0}}{\alpha} \sim \Oone.
\]
This implies that, for the effective cutoff to be at $x \sim \Oone$,
we must choose $\mod{c'_0} \sim \Oal$.

This actually completes the analysis necessary to show that
$h(\alpha)$ can always be arranged to be independent
of $\alpha$. However, for the purposes of the next section,
it is useful to show that we can, in fact, ensure
that all momentum integrals are cutoff at $x \sim \order{1}$.
The reason that this is useful is because we expect
a one dimensional integral with an integrand of $\Oone$ but 
with support only over a range $\Delta x$ to go like $\Delta x$.
Hence, it is desirable for this range to be $\Oone$ as opposed to,
\eg, $\OalPow{-1}$.

In the $A^2$ sector, from
\[
	\Delta^{22}(k) = \frac{1}{k^2} \Atwo{x},
\]
we see that the large momentum behaviour is $\alpha c_x / x$,
whereas the small momentum behaviour is $\alpha / 2x$. Since
the $\alpha$-dependence is the same for both, the crossover
scale is clearly set by $c$, which is assumed to 
crossover at $x \sim \order{1}$. 

In the $D$ sector, given that
\[
	\Delta^{DD}(k) = \frac{\tilde{c}_x f_x}{\Lambda^4},
\]
our choice that $\tilde{c}_0 \sim \Oal$
ensures that the crossover occurs at
$x \sim \order{1}$, assuming that $\tilde{c}_x$
crosses over at $x \sim \Oone$.

In the $C^i$-sector,
\[
	\Delta^{C^i C^i}(k) = \frac{1}{\Lambda^4} \frac{\tilde{c}_x}{x + 2 \lambda \tilde{c}_x},
\]
from which it is clear that the crossover scale
is controlled by $\tilde{c}$ and $\lambda$; we
use this freedom to ensure that the crossover occurs at
$x \sim \order{1}$.

We have thus demonstrated that, by suitable choices of
the behaviours of our cutoff functions, we can guarantee that
all momentum integrals are cutoff at the scale $x \sim \order{1}$
(working at small $\alpha$); one consequence of this is that
$h(\alpha)$ is independent of $\alpha$ and so does not
generate a contribution to $\beta_2$ in the $\alpha \rightarrow 0$
limit.

We conclude this section with an interesting comment on universality.
It is clear from our analysis so far that the freedom to choose the non-universal
parts of our cutoff functions enables us to choose
$h(\alpha)$. Returning to eqn.~\eq{eq:OLDs-charateristic-B}, it thus looks
like we could generate a universal contribution to $\beta_2$ by choosing \eg
$h(\alpha) = \alpha^m$, for some $m \neq 0$. However, the universal appearance 
of this contribution is accidental,
as can be appreciated from the fact that it arises from a particular choice of
a non-universal function. Indeed, the universal $\beta_2$ will only
be obtained by arranging things so that all contributions from the
running of $\alpha$ can be removed in the $\alpha \rightarrow 0$
limit.

\subsubsection{Behaviour of $\tilde{H}(\alpha)$}

To start our analysis of $\tilde{H}(\alpha)$,
we begin by returning to the third element
of the standard set. We know that the integrand
of eqn.~\eq{eq:SS-C-Alg}
effectively has support only over the region
$0 \leq x < \Oone$. Moreover, any non-trivial 
$\alpha$-dependence of
$\tilde{H}(\alpha)$ must come from the $B$-sector,
as $A$-sector (processed) gauge remainders are 
independent of $\alpha$.

From our algebraic choice for $f$ (see equation~(\ref{eq:f_k-B})),
the most obvious  possible source of problematic $\alpha$-behaviour comes when
$x$ is small. However, this is ameliorated by our 
previous choice of $\tilde{c}_0$.
To complete our analysis of this diagram, we must now perform the loop integral.
However,
since the effective cutoff of this integral is $\Oone$ as opposed to,
say, $\OalPow{-1}$, we do not expect the loop integral to 
generate any bad $\alpha$-dependence 
(\ie dependence which diverges as $\alpha \rightarrow 0$).

Looking now at the little set, the situation is similar:
in the $B$-sector, our choice of $\tilde{c}_0$ cures
any bad $\alpha$-dependence; in the $A^1$-sector there
is not even a potential problem. In both cases, the effective cutoff
for the loop integral is $\Oone$.
This exhausts the analysis of the diagrams for which 
we have an explicit algebraic form and so now we turn to 
the diagrams possessing three and four-point
vertices.

First, we will look at the diagrams with an $A^1_\mu A^1_\nu C^i$
vertex. Neither this vertex nor the effective propagator to
which it attaches carries the loop momentum of the diagram
and so we analyse them separately. The effective propagator---which
carries zero momentum---goes as
\[
	\frac{1}{2 \lambda \Lambda^4}.
\]
We see that the $\alpha$-dependence of this can always be
controlled by a suitable choice of $\lambda$. (It can always be
ensured that this choice of $\lambda$ is compatible with the choice
which guarantees that the crossover for the $C^i$-sector effective propagator
occurs at $x \sim \order{1}$.)

What about the $A^1_\mu A^1_\nu C^i$ vertex? We have encountered this already in
section~\ref{sec:TLD:Couplings}: at $\Op{2}$ it is a dimensionless coupling
and so we tune its flow to zero. This means that the flow equations tell us
nothing about its $\alpha$-dependence: this dependence is a boundary condition. 
The solution is simply to choose the boundary condition to have sufficiently
good $\alpha$-dependence, which is something we are always at liberty to do.

Now let us examine the remaining part of these diagrams. Attached to the other end
of the $C^i$-sector effective propagator is either a hook or a three-point
vertex, decorated by a simple loop. Since the hook simply goes
as
\[
	N \int_k g_k,
\]
where $g_x= (1-xf)/2$, it is clear that the loop integral does not produce
any troublesome $\alpha$-dependence.

The case where the top part of the diagram constitutes a three-point vertex
is almost as easy to treat. Let us consider the flow of this vertex. If we
take the dimensionless part, then we know that the flow has already been tuned
to zero; in this case, we choose the boundary condition, appropriately. 
Taking the flow of the dimensionful part of the vertex we once again tune the
seed action. This time, though, we do so to ensure sufficiently good 
$\alpha$-dependence.
Note that we need not worry about constraints coming from gauge invariance. 
If the top
vertex contains two fields in the $A$-sector (it cannot contain a single one) then
gauge invariance simply tells us that the vertex is transverse; it is not related to
lower point vertices. Performing the loop integral does not generate
any bad $\alpha$-dependence.

The penultimate diagram to deal with is the second element of the
standard set, which comprises two three-point vertices, joined together
by two effective propagators. Once again, our aim is to choose
the seed action such that the $\alpha$-dependence of the three-point 
vertices is sufficiently
good. This time, however, we must worry about gauge invariance.

The first effect of gauge invariance is to relate the longitudinal
part of each of these vertices to two-point, tree level 
vertices. Referring to our list of two-point, tree level vertices
(eqns.~\eq{eq:app:TLTP-11}--\eq{eq:app:TLTP-CiCi}), it is clear that
our three-point vertices have buried in them, necessarily,  components which go
as $\OalPow{-1}$.

\begin{center}
\begin{figure}[h]
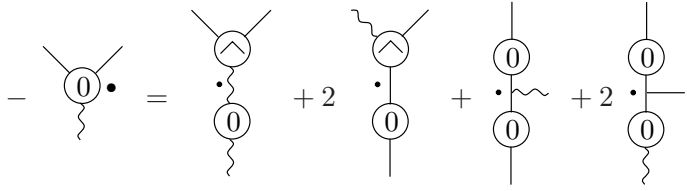

	\[
		- \ensuremath{\begin{array}{c}\input{pstex/Vertex-TLThP-E-2I.pstex_t} \end{array}} = \ensuremath{\begin{array}{c}\input{pstex/Dumbell-TLTP-E-TLThP.pstex_t} \end{array}} + 
		2\ensuremath{\begin{array}{c}\input{pstex/Dumbell-TLTP-I-TLThP.pstex_t} \end{array}} +\ensuremath{\begin{array}{c}\input{pstex/Dumbell-TLTP-DW-E-TLTP.pstex_t} \end{array}} 
		+2 \ensuremath{\begin{array}{c}\input{pstex/Dumbell-TLTP-DW-I-TLTP.pstex_t} \end{array}}
	\]
\caption{Flow of a three-point, tree level 
vertex viewed as part of a whole diagram.}
\label{fig:flow-3pt-E+2I}
\end{figure}
\end{center}

To examine the transverse components of these vertices,
we must use the flow equations. \Fig{fig:flow-3pt-E+2I}
shows the flow of a three-point vertex comprising one
external field and two identical internal fields (\ie this
vertex should be viewed as part of a whole diagram).

We expect the critical case to occur when all three fields are in 
the $A$-sector, since gauge invariance will then force us to
take $\Omom{2}$ from each two-point, tree level vertex. In particular,
this means that we will be unable to tune the three-point seed action
vertices if we take the $\Omom{3}$ part of the vertex whose flow
we are computing (the $\Omom{1}$ part is, of course, 
universal and independent of $\alpha$).

Given our choice that $c'_0 \sim \Oal$, it is straightforward to show  that
\[
	- \left. \ensuremath{\begin{array}{c}\input{pstex/Vertex-TLThP-E-2I.pstex_t} \end{array}} \right|_{\Omom{3}} \sim  \frac{\OalPow{0} \Omom{3}}{\Lambda^2}.
\]

Thus it is clear that we can always tune the seed action to ensure that
worst behaviour of the three-point, tree level vertices is the $\OalPow{-1}$
dependence forced by gauge invariance. This leading $\alpha$-dependence now cancels
between the vertices and effective propagators of the second element of the standard
set. The loop integral does not generate any bad $\alpha$-dependence

The treatment of the first element of the standard set follows, similarly:
by tuning the seed action (and choosing suitable boundary conditions
for the flow) we can ensure that the worst $\alpha$-dependence is
that forced on us by gauge invariance. This dependence is then
cancelled by the effective propagator. The loop integral does not generate any
bad $\alpha$-dependence.

We have thus demonstrated that, by suitable choice of seed action,
boundary conditions for our flow and 
non-universal behaviour of the cutoff functions, we can ensure 
that $\tilde{H}(\alpha) \sim \OalPow{0}$. It therefore follows
that we can always arrange for the $\alpha$-terms to vanish in the
$\altoz$ limit. Finally, then, we recover the standard value of
$\beta_2$:
\[
	-\frac{34}{3} \frac{N^2}{(4\pi)^4}.
\]

\section{Conclusions}	\label{sec:Conclusion}

We have performed the first manifestly gauge invariant,
continuum computation of the $SU(N)$ Yang-Mills two-loop
$\beta$ function; no gauge fixing or ghosts being
introduced at any stage.
The framework employed is the ERG
of~\cite{mgierg1,Thesis} which was constructed to allow
convenient renormalisation beyond one loop.  This
formalism is furnished with a set of powerful
diagrammatic techniques, reviewed in \sec{sec:Review}.
These techniques form the basis of the methodology used
to reduce both $\beta_1$ and $\beta_2$ to a set of
terms from which 
the universal coefficient can be readily
extracted. Let us recapitulate the basic
procedure at one loop.

From the ERG equation, a diagrammatic expression
is generated for the flow of the classical,
two-point, $A^1$ vertex.
Taking this vertex to carry
momentum $p$, we specialise to one loop
and use the renormalisation condition. This  yields
a diagrammatic expression for $\beta_1$ which,
since it contains instances of the seed action
and details of the covariantisation, is
not yet manifestly universal. To proceed, we convert
terms comprising exclusively Wilsonian effective
action vertices and undecorated kernels into 
$\Lambda$-derivative terms and corrections. A subset
of the correction terms can be simplified using the
effective propagator relation, cancelling non-universal
terms, up to gauge remainders. In turn, these remainders
can be processed diagrammatically. Iterating the procedure,
the expression for $\beta_1$ can be reduced to $\Lambda$-derivative
terms and terms possessing an $\Op{2}$ stub. At one-loop,
these latter terms are treated by Taylor expanding the
sub-diagram attached to the stub to zeroth order in $p$,
thereby allowing the $\Op{2}$ terms to be reduced to $\Lambda$-derivative
terms also.

At two loops, the procedure is much the same~\cite{mgierg1,oliver1},
though it is complicated by the fact that direct Taylor expansion of
all $\Op{2}$ terms is no longer possible. In certain diagrams,
$p$ effectively acts as an IR regulator, and so Taylor expansion in
$p$ generates spurious IR divergences. Isolation of the non-Taylor
expandable components is most easily achieved by use of subtraction
techniques, which are introduced in \sec{sec:TLD:Subtractions}. This
then allows $\beta_2$ to be reduced to a set of $\Lambda$-derivative
and $\alpha$-terms~\cite{Thesis}.

The strategy for evaluating
$\Lambda$-derivative terms is to start by interchanging the order of
differentiation \wrt\ $\LConstAl$ and loop integration. This
makes it clear that, so long as there are no dimensionless
running couplings hidden in the integrand, only those terms
which, before differentiation \wrt\ $\LConstAl$, possess some
IR scale will survive. In \sec{sec:TLD:Couplings}
it is proven that all such running couplings can be removed
from the set-up.

At one loop, the extraction of the numerical
coefficient is straightforward. Individual terms either vanish
or contribute a finite, universal number, the sum of which combine
to yield the standard answer. At two loops, extraction of
a numerical coefficient from the $\Lambda$-derivative terms
is much harder, since individual terms can now develop
IR divergences which survive even after differentiation \wrt\ $\LConstAl$.
Whilst it must be the case that the sum of diagrams is IR
finite, the sub-leading terms are no longer manifestly universal.

Though it is not possible to unambiguously
define  the universal (finite) component of a diagram which also
makes a finite, non-universal contribution
to $\beta_2$, progress
can be made. Utilising the subtraction
techniques, a prescription---which is by no means unique---can be
defined which separates a diagram into computable and non-computable
parts. By non-computable, it is meant that the corresponding
coefficient cannot be calculated without specifying the
appropriate parts of the seed action and
details of the covariantisation. The computable part can be
evaluated using only the renormalisation condition for
$A^1$, but depends
on  the precise prescription used to define what is meant
by computable.
Separating terms in this way, it can be shown, diagrammatically,
 that all non-computable contributions to $\beta_2$
cancel, amongst themselves. The sum of the computable 
contributions yields the standard, universal answer.

The final task is to treat the $\alpha$-terms.
These terms are expected to be present as a consequence of
an unphysical coupling associated with one of the
regulator fields. Since $\beta_2$ is not a physical
quantity, it is anticipated that agreement between
the value obtained in our scheme and the standard value
is found only in the limit that $\alpha$ is tuned to zero~\cite{mgierg1,Thesis}.
In \sec{sec:TLD:alpha} it is demonstrated that, by
sufficiently constraining the seed action and boundary
conditions of the flow, the $\alpha$-terms can indeed
be shown to vanish in this limit, so long as
we impose constraints on the non-universal
behaviour of the cutoff functions.

Thus, we have successfully computed $\beta_2$ without fixing
the gauge, finding agreement with the standard,
universal answer in the appropriate limit.
This demonstrates the consistency of the formalism
beyond reasonable doubt, thereby opening up the prospect
of an analytical, manifestly gauge invariant computational
scheme for $SU(N)$ Yang-Mills theory. In the future, we aim to
apply the formalism both perturbatively and non-perturbatively
and also plan to incorporate quarks.

\paragraph{Acknowledgements}
We acknowledge financial support from PPARC Rolling Grant
PPA/G/O/2002/0468.

\appendix

\section{Ingredients of the Weak Coupling Flow Equations}
\label{app:elements}

A list of our choices for all 
single supertrace seed action, two-point,
tree level vertices follows. Multi-supertrace terms
are either related to those listed,
by no-$\A^0$ symmetry, or can be
set to zero~\cite{Thesis}. The undetermined
parameter $\lambda$ can depend on $\alpha$.

\begin{eqnarray}
	\hat{S}^{\ \, 1 1}_{0 \mu \nu}(p) 		& = &   \NCOAlg{p} \Box_{\mu \nu}(p),
\label{eq:app:TLTP-11}
\\
	\hat{S}^{\ \, 2 2}_{0 \mu \nu}(p) 		& = &  	\NCTAlg{p}	\Box_{\mu \nu}(p),
\label{eq:app:TLTP-22}
\\
	\hat{S}^{\ B\bar{B}}_{0 \mu \, \nu}(p) 	& = &	\frac{\alpha + 1}{\alpha c_p}\Box_{\mu \nu}(p) 
													+ \frac{4 \Lambda^2}{\tilde{c}_p} \delta_{\mu \nu},
\label{eq:app:TLTP-BB}
\\
	\hat{S}^{\ D\bar{B}}_{0 \ \  \mu} (p)	& = & 	\frac{2 \Lambda^2 p_\mu}{\tilde{c}_p},
\label{eq:app:TLTP-BD}
\\
	\hat{S}^{\ D\bar{D}}_0(p) 				& = & 	\frac{\Lambda^2 p^2}{\tilde{c}_p},
\label{eq:app:TLTP-DD}
\\
	\hat{S}^{\ C^iC^i}_0(p)					& = &	\frac{\Lambda^2 p^2}{\tilde{c}_p} + 2 \lambda \Lambda^4.
\label{eq:app:TLTP-CiCi}
\end{eqnarray}
These choices correspond to~\cite{Thesis}
\begin{eqnarray}
	f_p & = & \fB{p}{x},
\label{eq:app:NFE-f}
\\[1ex]
	g_p	& = & \gB{p}{x},
\label{eq:app:NFE-g}
\end{eqnarray}
allowing us to write the effective propagators in the following
forms:
\begin{eqnarray}
	\Delta^{11}(p)		& = &  \frac{1}{p^2} \Aone{p},
\label{eq:app:NFE:EP-11}
\\[1ex]
	\Delta^{22}(p)		& = & \frac{1}{p^2} \Atwo{p},
\label{eq:app:NFE:EP-22}
\\[1ex]
	\Delta^{B\bar{B}}(p)	& = & \frac{1}{2 \Lambda^2} \tilde{c}_p g_p,
\label{eq:app:NFE:EP-BB}
\\[1ex]
	\Delta^{D\bar{D}}(p)	& = & \frac{1}{\Lambda^4} \tilde{c}_p f_p,
\label{eq:app:NFE:EP-DD}
\\[1ex]
	\Delta^{C^iC^i}(p)	& = & \frac{1}{\Lambda^4}	\frac{\tilde{c}_p}{x+ 2\lambda \tilde{c}_p}.
\label{eq:app:NFE:EP-CiCi}
\end{eqnarray}

\section{Subtractions for diagrams~\ref{diag:32.1}--\ref{diag:FR22.1b}}
\label{app:Subs}

\bcf
	\[
	\begin{array}{c}
		\begin{array}{cccccccccc}
%				& \LDLD{d:32.1:s1}{d:32.1:a1}	&	& \LDLD{d:32.1:s2}{d:32.1:a2}	&		& \LDLD{d:32.1:s3}{d:32.1:a3}	&
%				& \LDLD{d:42.1:s1}{d:42.1:a1}	&	& \LDLD{d:42.1:s2}{d:42.1:a2}
%		\\[1ex]
			\mp	& \ensuremath{\begin{array}{c}\input{pstex/Diagram32.1-s1.pstex_t} \end{array}}			&\mp& \ensuremath{\begin{array}{c}\input{pstex/Diagram32.1-s2.pstex_t} \end{array}}			&\pm2	& \ensuremath{\begin{array}{c}\input{pstex/Diagram32.1-s3.pstex_t} \end{array}}			&
			\pm2& \ensuremath{\begin{array}{c}\input{pstex/Diagram42.1-s1.pstex_t} \end{array}}			&\pm2	& \ensuremath{\begin{array}{c}\input{pstex/Diagram42.1-s2.pstex_t} \end{array}}
		\end{array}
	\\[1ex]
		\mp 2
		\left[
			\begin{array}{c}
				\begin{array}{ccccc}
%					\LDLD{d:392.4a:s1}{d:392.4a:a1}	&	&\LDLD{d:392.4a:s2}{d:392.4a:a2}	&	&\LDLD{d:392.4a:s3}{d:392.4a:a3}					
%				\\[1ex]
					\ensuremath{\begin{array}{c}\begin{picture}(0,0)%
\includegraphics{pstex/Diagram392.4a-s1.pstex}%
\end{picture}%
\setlength{\unitlength}{3947sp}%
\begingroup\makeatletter\ifx\SetFigFont\undefined%
\gdef\SetFigFont#1#2#3#4#5{%
  \reset@font\fontsize{#1}{#2pt}%
  \fontfamily{#3}\fontseries{#4}\fontshape{#5}%
  \selectfont}%
\fi\endgroup%
\begin{picture}(442,1107)(902,-1291)
\put(986,-997){\makebox(0,0)[lb]{\smash{\SetFigFont{11}{13.2}{\rmdefault}{\mddefault}{\updefault}{\color[rgb]{0,0,0}0}%
}}}
\end{picture}
 \end{array}}			&+	&\ensuremath{\begin{array}{c}\begin{picture}(0,0)%
\includegraphics{pstex/Diagram392.4a-s2.pstex}%
\end{picture}%
\setlength{\unitlength}{3947sp}%
\begingroup\makeatletter\ifx\SetFigFont\undefined%
\gdef\SetFigFont#1#2#3#4#5{%
  \reset@font\fontsize{#1}{#2pt}%
  \fontfamily{#3}\fontseries{#4}\fontshape{#5}%
  \selectfont}%
\fi\endgroup%
\begin{picture}(401,1100)(811,-909)
\put(895,-615){\makebox(0,0)[lb]{\smash{\SetFigFont{11}{13.2}{\rmdefault}{\mddefault}{\updefault}{\color[rgb]{0,0,0}0}%
}}}
\end{picture}
 \end{array}}				&+	&\ensuremath{\begin{array}{c}\begin{picture}(0,0)%
\includegraphics{pstex/Diagram392.4a-s3.pstex}%
\end{picture}%
\setlength{\unitlength}{3947sp}%
\begingroup\makeatletter\ifx\SetFigFont\undefined%
\gdef\SetFigFont#1#2#3#4#5{%
  \reset@font\fontsize{#1}{#2pt}%
  \fontfamily{#3}\fontseries{#4}\fontshape{#5}%
  \selectfont}%
\fi\endgroup%
\begin{picture}(474,869)(782,-243)
\put(866,-187){\makebox(0,0)[lb]{\smash{\SetFigFont{11}{13.2}{\rmdefault}{\mddefault}{\updefault}{\color[rgb]{0,0,0}0}%
}}}
\end{picture}
 \end{array}}				
				\end{array}
			\\[1ex]
				\begin{array}{cccccc}
%						&\LDLD{d:392.4b:s1}{d:392.4b:a1}&	&\LDLD{d:392.4b:s2}{d:392.4b:a2}&	&\LDLD{d:392.4b:s3}{d:392.4b:a3}
%				\\[1ex]
					-	& \ensuremath{\begin{array}{c}\begin{picture}(0,0)%
\includegraphics{pstex/Diagram392.4b-s1.pstex}%
\end{picture}%
\setlength{\unitlength}{3947sp}%
\begingroup\makeatletter\ifx\SetFigFont\undefined%
\gdef\SetFigFont#1#2#3#4#5{%
  \reset@font\fontsize{#1}{#2pt}%
  \fontfamily{#3}\fontseries{#4}\fontshape{#5}%
  \selectfont}%
\fi\endgroup%
\begin{picture}(447,1123)(851,-1097)
\put(1214,-803){\makebox(0,0)[rb]{\smash{\SetFigFont{11}{13.2}{\rmdefault}{\mddefault}{\updefault}{\color[rgb]{0,0,0}0}%
}}}
\end{picture}
 \end{array}}			&-	& \ensuremath{\begin{array}{c}\begin{picture}(0,0)%
\includegraphics{pstex/Diagram392.4b-s2.pstex}%
\end{picture}%
\setlength{\unitlength}{3947sp}%
\begingroup\makeatletter\ifx\SetFigFont\undefined%
\gdef\SetFigFont#1#2#3#4#5{%
  \reset@font\fontsize{#1}{#2pt}%
  \fontfamily{#3}\fontseries{#4}\fontshape{#5}%
  \selectfont}%
\fi\endgroup%
\begin{picture}(412,1123)(1155,-992)
\put(1483,-698){\makebox(0,0)[rb]{\smash{\SetFigFont{11}{13.2}{\rmdefault}{\mddefault}{\updefault}{\color[rgb]{0,0,0}0}%
}}}
\end{picture}
 \end{array}}			&-	& \ensuremath{\begin{array}{c}\begin{picture}(0,0)%
\includegraphics{pstex/Diagram392.4b-s3.pstex}%
\end{picture}%
\setlength{\unitlength}{3947sp}%
\begingroup\makeatletter\ifx\SetFigFont\undefined%
\gdef\SetFigFont#1#2#3#4#5{%
  \reset@font\fontsize{#1}{#2pt}%
  \fontfamily{#3}\fontseries{#4}\fontshape{#5}%
  \selectfont}%
\fi\endgroup%
\begin{picture}(563,885)(1362,-1095)
\put(1690,-1039){\makebox(0,0)[rb]{\smash{\SetFigFont{11}{13.2}{\rmdefault}{\mddefault}{\updefault}{\color[rgb]{0,0,0}0}%
}}}
\end{picture}
 \end{array}}	
				\end{array}
			\end{array}
		\right]	
	\\[1ex]
		\begin{array}{cccccc}
%					& \LDLD{d:FR26.1,2-s}{d:FR26.1,2-a}	&		&\LDLD{d:FR26.1,2-sb}{d:FR26.1,2-ab}&		& \LDLD{d:FR26.11-s}{d:FR26.11-a}
%		\\[1ex]
			\pm 4	& \cdeps{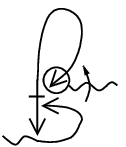}			&\mp4	&\cdeps{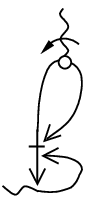} 			&\mp4	&\ensuremath{\begin{array}{c}\begin{picture}(0,0)%
\includegraphics{pstex/DiagramFR26.1+6-s.pstex}%
\end{picture}%
\setlength{\unitlength}{3947sp}%
\begingroup\makeatletter\ifx\SetFigFont\undefined%
\gdef\SetFigFont#1#2#3#4#5{%
  \reset@font\fontsize{#1}{#2pt}%
  \fontfamily{#3}\fontseries{#4}\fontshape{#5}%
  \selectfont}%
\fi\endgroup%
\begin{picture}(773,809)(3577,-569)
\end{picture}
 \end{array}}				
		\end{array}
	\\[1ex]
		\begin{array}{cccccccc}
%					& \LDLD{d:FR36.1b-s}{d:FR36.1b-a}	&		& \LDLD{d:FR22.3b-s}{d:FR22.3b-a}	&		&\LDLD{d:FR22.1b-si}{d:FR22.1b-ai}	&		& \LDLD{d:FR22.1b-sii}{d:FR22.1b-aii}
%		\\[1ex]
			\pm 4	& \ensuremath{\begin{array}{c}\input{pstex/DiagramFR36.1b-s.pstex_t} \end{array}}				&\pm 4	& \ensuremath{\begin{array}{c}\begin{picture}(0,0)%
\includegraphics{pstex/DiagramFR22.3b-s.pstex}%
\end{picture}%
\setlength{\unitlength}{3947sp}%
\begingroup\makeatletter\ifx\SetFigFont\undefined%
\gdef\SetFigFont#1#2#3#4#5{%
  \reset@font\fontsize{#1}{#2pt}%
  \fontfamily{#3}\fontseries{#4}\fontshape{#5}%
  \selectfont}%
\fi\endgroup%
\begin{picture}(342,1251)(873,-613)
\put(1059,-325){\makebox(0,0)[lb]{\smash{{\SetFigFont{11}{13.2}{\rmdefault}{\mddefault}{\updefault}{\color[rgb]{0,0,0}0}%
}}}}
\end{picture}%
 \end{array}}				&\mp 4	&\ensuremath{\begin{array}{c}\begin{picture}(0,0)%
\includegraphics{pstex/DiagramFR22.1b-si.pstex}%
\end{picture}%
\setlength{\unitlength}{3947sp}%
\begingroup\makeatletter\ifx\SetFigFont\undefined%
\gdef\SetFigFont#1#2#3#4#5{%
  \reset@font\fontsize{#1}{#2pt}%
  \fontfamily{#3}\fontseries{#4}\fontshape{#5}%
  \selectfont}%
\fi\endgroup%
\begin{picture}(362,1596)(1150,-2216)
\put(1238,-1913){\makebox(0,0)[lb]{\smash{{\SetFigFont{11}{13.2}{\rmdefault}{\mddefault}{\updefault}{\color[rgb]{0,0,0}0}%
}}}}
\end{picture}%
 \end{array}}				&\pm 4	& \ensuremath{\begin{array}{c}\begin{picture}(0,0)%
\includegraphics{pstex/DiagramFR22.1b-sii.pstex}%
\end{picture}%
\setlength{\unitlength}{3947sp}%
\begingroup\makeatletter\ifx\SetFigFont\undefined%
\gdef\SetFigFont#1#2#3#4#5{%
  \reset@font\fontsize{#1}{#2pt}%
  \fontfamily{#3}\fontseries{#4}\fontshape{#5}%
  \selectfont}%
\fi\endgroup%
\begin{picture}(928,919)(1150,-2216)
\put(1238,-1913){\makebox(0,0)[lb]{\smash{{\SetFigFont{11}{13.2}{\rmdefault}{\mddefault}{\updefault}{\color[rgb]{0,0,0}0}%
}}}}
\end{picture}%
 \end{array}}
		\end{array}
	\end{array}
	\]
\caption{Subtractions for diagrams~\ref{diag:32.1}--\ref{diag:FR22.1b}.}
\label{fig:Subs-a}
\ecf

\end{document}